\documentclass[acmsmall]{acmart}
\pdfoutput=1
\usepackage{color}
\usepackage{subfigure}
\usepackage{amsmath}
\usepackage{multirow}
\usepackage{threeparttable}
\AtBeginDocument{%
  \providecommand\BibTeX{{%
    \normalfont B\kern-0.5em{\scshape i\kern-0.25em b}\kern-0.8em\TeX}}}

\newcommand{\wadd}[1]{{\color{black}{#1}}}
\newcommand{\wdel}[1]{}

\newcommand{\xdel}[1]{}

\newcommand{\wyadd}[1]{{\color{black}{#1}}}
\newcommand{\wydel}[1]{}

\newcommand{\qadd}[1]{{\color{black}{#1}}}
\newcommand{\qdel}[1]{}

\newcommand{\hjadd}[1]{{\color{purple}{#1}}}
\newcommand{\hjdel}[1]{}

\newcommand{\A}[1]{accumulation scheme{#1}}
\begin{document}

\title{A Survey on Approximate Multiplier Designs for Energy Efficiency: From Algorithms to Circuits}


\author{Ying Wu}
\affiliation{%
  \institution{Zhejiang University}
  \city{Hangzhou}
  \country{China}
}
\email{ying.wu@zju.edu.cn}

\author{Chuangtao Chen}
\affiliation{%
  \institution{Zhejiang University}
  \city{Hangzhou}
  \country{China}}
\email{chtchen@zju.edu.cn}

\author{Weihua Xiao}
\affiliation{%
  \institution{Shanghai Jiao Tong University}
  \city{Shanghai}
  \country{China}
}
\email{019370910014@sjtu.edu.cn}

\author{Xuan Wang}
\affiliation{%
  \institution{Shanghai Jiao Tong University}
  \city{Shanghai}
  \country{China}}
\email{xuan.wang@sjtu.edu.cn}

\author{Chenyi Wen}
\affiliation{%
  \institution{Zhejiang University}
  \city{Hangzhou}
  \country{China}}
\email{wwency@zju.edu.cn}

\author{Jie Han}
\affiliation{%
  \institution{University of Alberta}
  \city{Edmonton, Alberta}
  \country{Canada}}
\email{jhan8@ualberta.ca}

\author{Xunzhao Yin}
\affiliation{%
  \institution{Zhejiang University}
  \city{Hangzhou}
  \country{China}}
\email{xzyin1@zju.edu.cn}

\author{Weikang Qian}
\affiliation{%
  \institution{Shanghai Jiao Tong University}
  \city{Shanghai}
  \country{China}}
\email{qianwk@sjtu.edu.cn}

\author{Cheng Zhuo}
\authornote{Corresponding author}
\affiliation{%
  \institution{Zhejiang University}
  \city{Hangzhou}
  \country{China}}
\email{czhuo@zju.edu.cn}

\renewcommand{\shortauthors}{Ying Wu, et al.}

\begin{abstract}
  Given the stringent requirements of \hjdel{the}\qdel{energy-efficiency requirements}\qadd{energy efficiency} for Internet-of-Things edge devices, \wydel{approximate multipliers have recently received growing attention, especially in error-resilient applications.}
  \wyadd{approximate multipliers, as a basic component of many processors and accelerators, have been constantly proposed and studied for decades, especially in error-resilient applications.}
  The computation error and energy efficiency largely depend on how and where the approximation is introduced into a design. Thus, this article aims to provide a comprehensive review of the approximation techniques in multiplier designs ranging from algorithms and architectures to circuits. We have implemented representative approximate multiplier designs in each category to understand the impact of the design techniques on accuracy and efficiency. The designs can then be effectively deployed in high-level applications, such as machine learning, to gain energy efficiency at the cost of slight accuracy loss.
\end{abstract}

\ccsdesc[500]{General and reference~Surveys and overviews}
\ccsdesc[500]{Hardware~Arithmetic and datapath circuits}
\ccsdesc[300]{Hardware~System-level fault tolerance}
\keywords{approximate computing, multiplier, algorithm, architecture, circuit}

\maketitle

\section{Introduction}
\label{sec:intro}
The rapid growth of Internet-of-Things (IoT) has made energy efficiency a critical concern for IoT devices due to the constrained resources on the edge~\cite{atzori2010internet}. Conventional processors, such as the central processing unit (CPU) and the graphics processing unit (GPU), compute with pre-determined but unnecessary full precisions for all the computational tasks, which inevitably reduces energy efficiency. \qdel{Therefore, conventional processors}However, for many error-tolerant applications, such as neural networks, computational approximation or inaccuracy can be tolerated without impacting the inference accuracy~\cite{jiang2020approximate}. Thus, it is highly desired to optimize the energy efficiency for resource-constrained IoT devices by providing sufficient instead of excessively accurate outputs. 

\qdel{Due to the error resilience of human sensory and machine learning tasks}\qadd{Given the error tolerance nature of many IoT applications}~\cite{5993675,6569370,10.5555/2971808.2972118,10.1145/2540708.2540710,liu2020retrospective}, approximate computing\qadd{ that trades numerical precision for computational efficiency} has become a promising alternative to achieving the needed energy efficiency with quality results while consuming minimum resources\qdel{ which trade\qadd{s} unnecessary precision with energy efficiency}. For \qdel{data-intensive}\qadd{computationally intensive} applications, \textit{e.g.}, neural networks and image processing, \textbf{multiplication} is probably the most frequently invoked operation~\cite{deng2019energy}, which requires significant energy and a long latency to provide accurate outputs.

As a common arithmetic component\qdel{ that}\qadd{, the multiplier} has been studied for decades~\cite{1303135,465364}\qdel{, the past focus for the multiplier}\qadd{. The focus was} mainly \qdel{placed up}on accuracy and performance in fully precise computations. Mitchell proposed an approximate logarithm-based multiplier in the 1960s~\cite{mitchell1962computer}, which, however, failed to attract significant attention for decades. Since then, the approximation technique of truncation and its variants, such as the fixed-width multiplier that enforces the same bit-width for both inputs and outputs, have been proposed for achieving area efficiency~\cite{lim1992single,schulte1993truncated}.
To reduce the truncation induced inaccuracy, around 2000, a few compensation techniques were suggested at the cost of additional complexity~\cite{jou1999design,van2000design,jon2000fixed,cho2004design,van2005generalized}. Recently, with the popularity of IoT devices and their stringent resource constraints, researchers from both academia and industry have devoted significant efforts to approximate multiplier design and optimization, which resulted in many effective and interesting approximation techniques~\cite{courbariaux2014training,deng2019energy,10.1109/TCSVT.2013.2243658,Imani_ACAM_10.1145/2934583.2934595,10.5555/2971808.2971893,imani2018rapidnn,kulkarni2011trading,7544342,6637825,7927254,10.1109/TCAD.2018.2834438, chandrakasan1995minimizing, liu2009computation, imani2019approxlp, chen2020optimally, chen2021TC, imani2019searchd, Ni_2019, vahdat2019tosam, zendegani2016roba, sarwar2016multiplier}. In general, such efforts can be categorized at multiple levels, ranging from algorithms, architectures, to circuits:\footnote{\wadd{The algorithm in this article means the computation procedure of multiplication rather than a problem-solving process. A more clear definition of the three levels is given in Section~\ref{sec:classification}.}}
\begin{itemize}
    \item At the algorithm level, the Mitchell's multiplier has been further improved to reduce the bias or hardware costs~\cite{liu2018design,ansari2020improved,ansari2019hardware,saadat2018minimally}. Additionally, a linear-fitting algorithm has been innovatively applied to transform multiplication to simpler operations such as addition and shift~\cite{imani2019approxlp,chen2020optimally}.
\item Approximate techniques \qdel{in}\qadd{at the} architecture level are more frequently revisited and explored. Various new techniques have been introduced at different stages of a binary multiplication following the conventional data flow~\cite{hashemi2015drum,vahdat2019tosam,vahdat2017letam,narayanamoorthy2014energy,kulkarni2011trading,venkatachalam2017design,yang2018low,yang2017low,esposito2018approximate,Wang20,yang2018low,ha2017multipliers,tung2019low,yang2015approximate,van2020fpga,liu2017design,jiang2015approximate}.
\item At the circuit level, various approximation techniques have been incorporated into the low-level implementations of the sub-modules of a multiplier~\cite{kulkarni2011trading,pabithra2018analysis,van2020fpga,ha2017multipliers,liu2017design,yin2018designs,liu2017design}. Such circuit-level techniques can be integrated into the designs at the other two levels.
\end{itemize}


Approximate multipliers have already been successfully applied in error-tolerant tasks, such as filtering, image processing, and machine learning~\cite{ashtaputre1985using, jou1999design, van2000design, lingamneni2013improving, brandalero2018approximate, akbari2018px, yoo20206, snigdha2016optimal, zendegani2016roba, du2014leveraging, zhang2015approxann, sarwar2016multiplier, mrazek2016design, liu2018design, hammad2018impact,ansari2019improving,hammad2019deep,leon2019cooperative,hammad2021cnn}. 
In 1985, \qdel{Savage}\qadd{Ashtaputre \textit{et al.}} designed a systolic array with approximate multipliers and demonstrated that the inaccurate multiplication can bring \qdel{an ignorable}\qadd{negligible} impact on the results~\cite{ashtaputre1985using}. 
Around 2000, fixed-width multiplier~\cite{lim1992single,schulte1993truncated} were widely used in digital signal processing such as filtering~\cite{van2000design} and wavelet transformation~\cite{jou1999design}.
Snigdha \textit{et al}. applied \hjdel{the }approximate multipliers in image compression blocks and observed over 12\% improvements in \qdel{terms of }area, power, and delay~\cite{snigdha2016optimal}.
Hammad \textit{et al}. replaced the original full-precision multipliers in VGG networks with approximate ones to support the classification tasks on CIFAR-10 and CIFAR-100, which again showed negligible accuracy losses~\cite{hammad2018impact}. After that, they further proposed to deploy \hjdel{the }approximate multipliers in training to improve the \hjdel{training }performance~\cite{hammad2019deep}.

\wyadd{In this paper, our main focus is on approximate multipliers in digital application-specific integrated circuits (ASICs). Although there are other platforms available to implement approximate multipliers, such as field programmable gate arrays (FPGAs) and in-memory multiplication, they are not within the primary scope of this paper. FPGA-based approximate multipliers are designed using the inherent FPGA basic component, Look-Up Tables (LUTs), and the conventional approximation techniques commonly used in ASICs may not yield the expected results in FPGAs~\cite{ullah2018area,guo2020small,ullah2021high,ullah2018smapproxlib,van2020fpga,prabakaran2020approxfpgas,waris2021axbms,ullah2022appaxo}. On the other hand, in-memory multiplication and accumulation is primarily implemented using analog circuits, where the computation errors are inherent to the memory devices~\cite{sebastian2020memory}.}

\textbf{Scope of the article:} Since many prior designs \hjdel{happen to }rely on hand-crafted structures or heuristics, it is then highly desired to systematically review and understand the advantages and disadvantages of the various alternatives to introducing approximation into a design. Hence, this article provides a comprehensive review of the approximate multipliers, including approximation techniques at the \hjdel{three levels, \textit{i.e.}, }algorithm, architecture, and circuit levels, the comparison of different techniques using various metrics, and the discussion of the future trend. It is noted that we re-implement various approximation techniques in Verilog to ensure the fair comparison in our evaluation, which will be released as an open-source library, \textbf{AM-Lib}~\cite{am_lib}. 
\wyadd{Since this survey paper focuses on the relative trend and differences among different approximation techniques, the experimental results and findings are based on the pre-layout simulations.}

The remainder of this article is organized as follows. In \qdel{section}\qadd{Section}~\ref{sec:back}, we review the background of approximate multipliers.
\qadd{Section~\ref{sec:classification} provides an overview of the approximate multipliers.}
Sections \ref{sec:alg} to \ref{sec:circuit} discuss the approximate multipliers at the algorithm, architecture, and circuit levels, respectively. Note that some presented designs utilize approximation techniques at multiple levels. In Section~\ref{sec:evaluate}, we evaluate the accuracy and \qdel{circuit characteristics}\qadd{hardware cost} of approximate multipliers and their applications in neural networks by considering accuracy losses and \qdel{energy-efficiency}\qadd{energy efficiency}. Then, a discussion is presented in Section \ref{sec:discussion} followed by the conclusions in Section \ref{sec:con}.

\section{Background}
\label{sec:back}
\subsection{Fixed-point vs. Floating-point Representations}

\begin{figure}[!htbp]
\centering
\includegraphics[width=.6\textwidth]{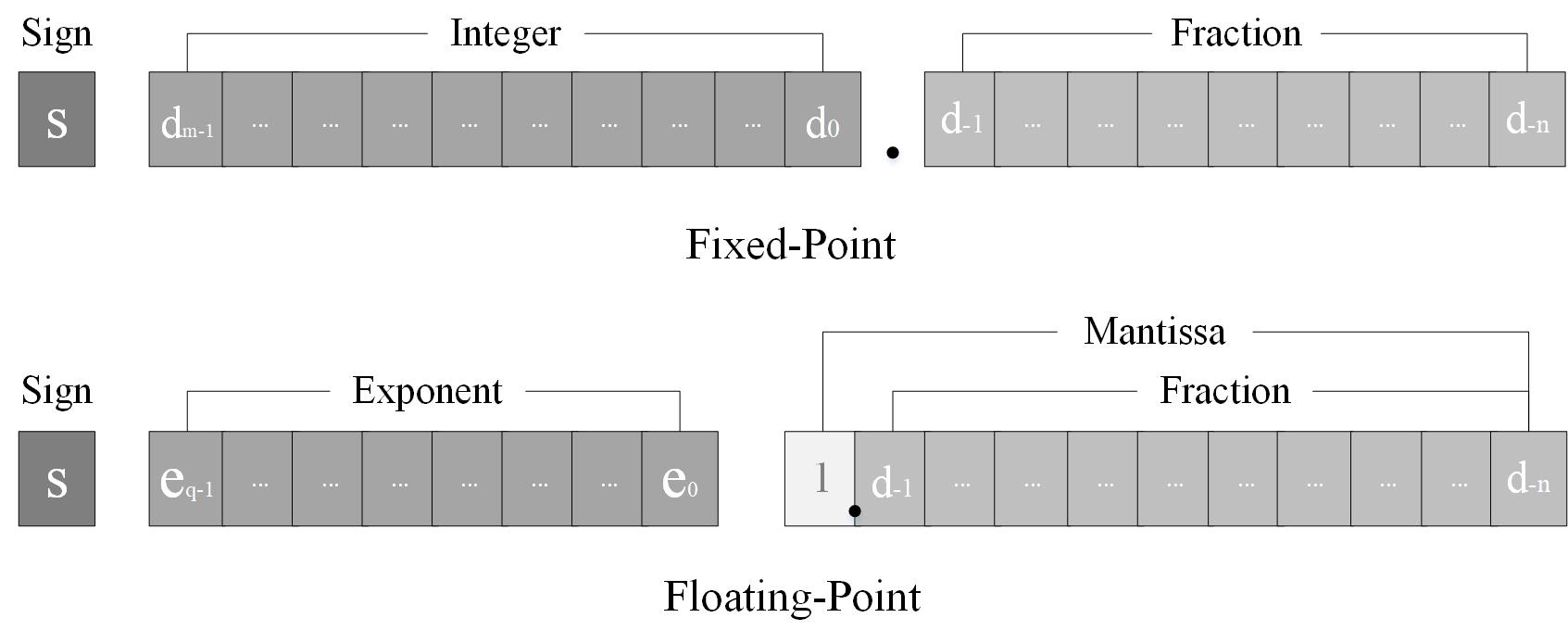}
\caption{Comparison between fixed-point and floating-point formats.}
\label{fig:background1}       
\end{figure}

\begin{figure}[!htbp]
\centering
\includegraphics[width=.6\textwidth]{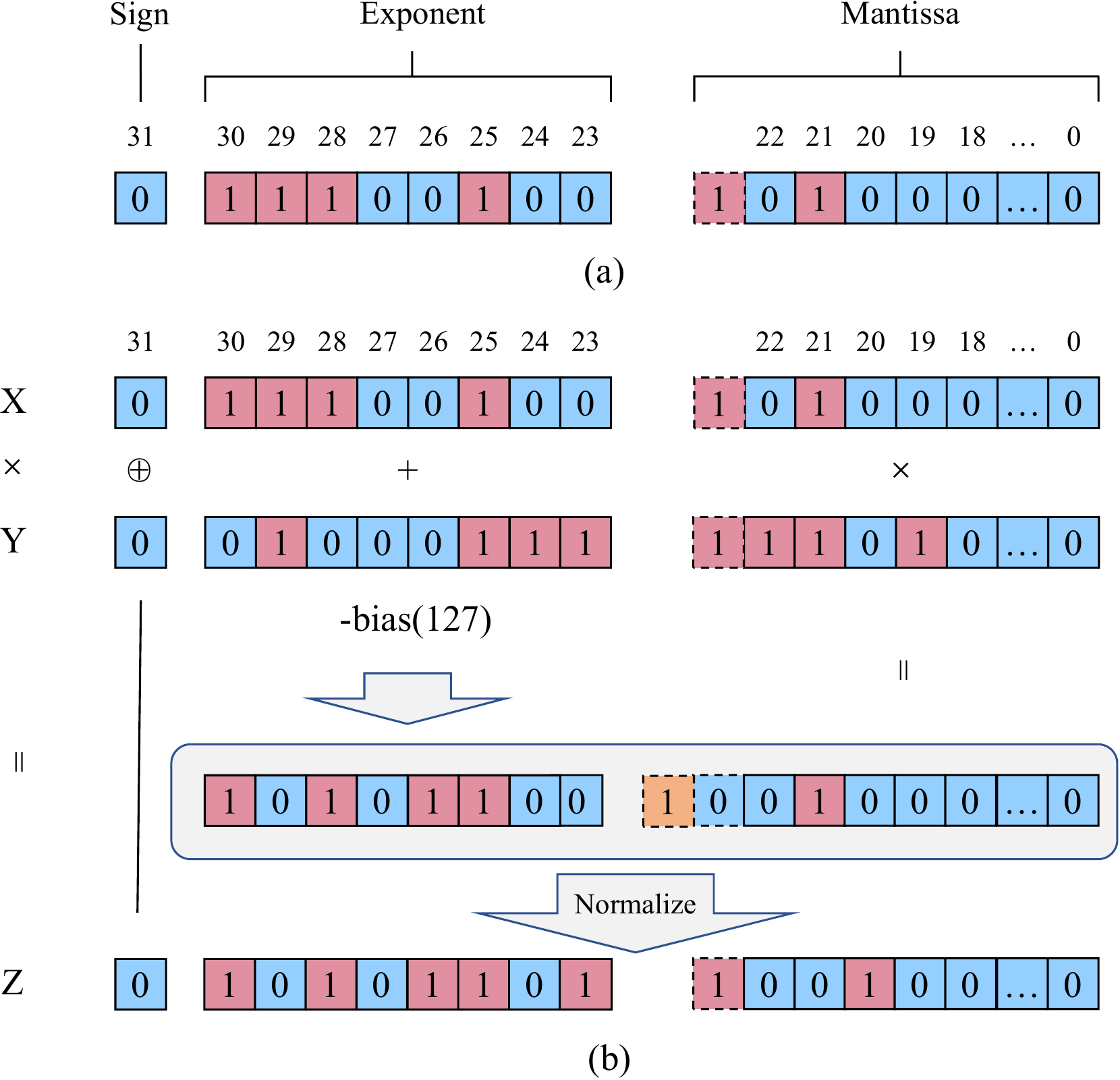}
\caption{An example of a 32-bit floating-point multiplication according to IEEE 754 standard.}
\label{fig:fp_mul_struct}       
\end{figure}

Similar to many arithmetic functions implemented in hardware, the multiplier design can be categorized into \qdel{fixed point}\qadd{fixed-point} and \qdel{floating point}\qadd{floating-point} implementations as a trade-off among accuracy, dynamic range, and cost. The major difference between fixed-point and floating-point numbers lies in whether the implementation has a specific number of digits reserved for the integer and fractional parts\hjdel{, respectively}. In other words, a fixed-point representation uses a decimal point at a fixed position. Obviously, a floating-point representation may offer a wider dynamic range and higher precision than its fixed-point counterpart, but at the cost of area, speed, and power consumption. 

Fig.~\ref{fig:background1} compares \qadd{the }fixed-point and\qadd{ the} floating-point formats in the binary number system. The fixed-point format consists of a sign bit, an integer part, and a fractional part, with a fixed binary point position. On the other hand, according to the IEEE 754 standard~\cite{IEEE_754_2019_8766229}, which is a technical standard for floating-point arithmetic, a floating-point number consists of a sign, an exponent, and \qdel{the}\wadd{a} mantissa. The mantissa of a \emph{normalized} floating-point number is a fraction between 1 and 2, where its first digit is fixed to 1 and the rest is the fraction in the range of [0,1).

With the underlying number representations, designers may use either a fixed-point or a floating-point multiplier to conduct multiplication.
For the floating-point numbers, the multiplication procedure of 32-bit floating-point numbers is shown in Fig.~\ref{fig:fp_mul_struct}.
The sign bits are XORed \hjdel{together }and the exponents are summed by an adder. Then, a bias of $127$ is subtracted from the sum to allow both negative and positive values for the exponent. Finally, the two mantissas are multiplied and shifted to the range of 1 and 2 to produce the normalized representation. The exponent will be adjusted if a shift happens. For a floating-point multiplication, the\qadd{ computation on the} mantissa part requires more energy and a longer delay than the other two parts, which is hence the focus of most research work~\cite{imani2019approxlp, chen2020optimally}. On the other hand, if no overflow\qadd{ occurs}, the fixed-point multiplication is carried out as a regular multiplication with its fractional part truncated to the specified bit-width. However, the difference between the two multiplications is actually smaller than it seems. The multiplication of the mantissa parts \qdel{for}\qadd{of} floating-point numbers can be always viewed as a special case of fixed-point multiplication, where the integer part is 1. Thus, the most critical operations in fixed- and floating-point multiplications can be considered as the same.

\begin{figure}[!htbp]
\centering
\includegraphics[width=.5\textwidth]{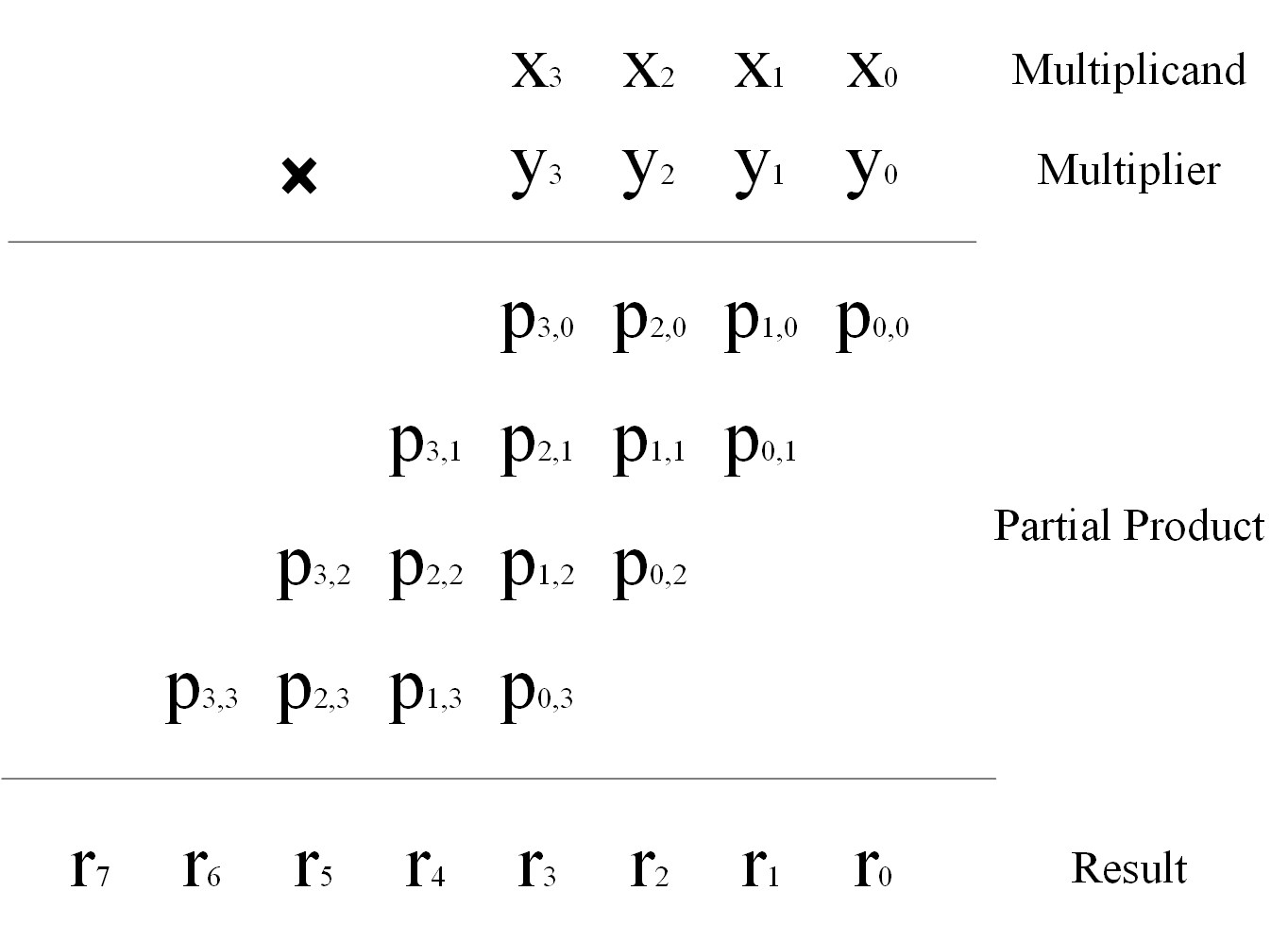}
\caption{An example of 4-bit multiplication.}
\label{fig:background2}       
\end{figure}

\begin{figure}[!htbp]
\centering
\subfigure[]{\includegraphics[width=0.6\textwidth]{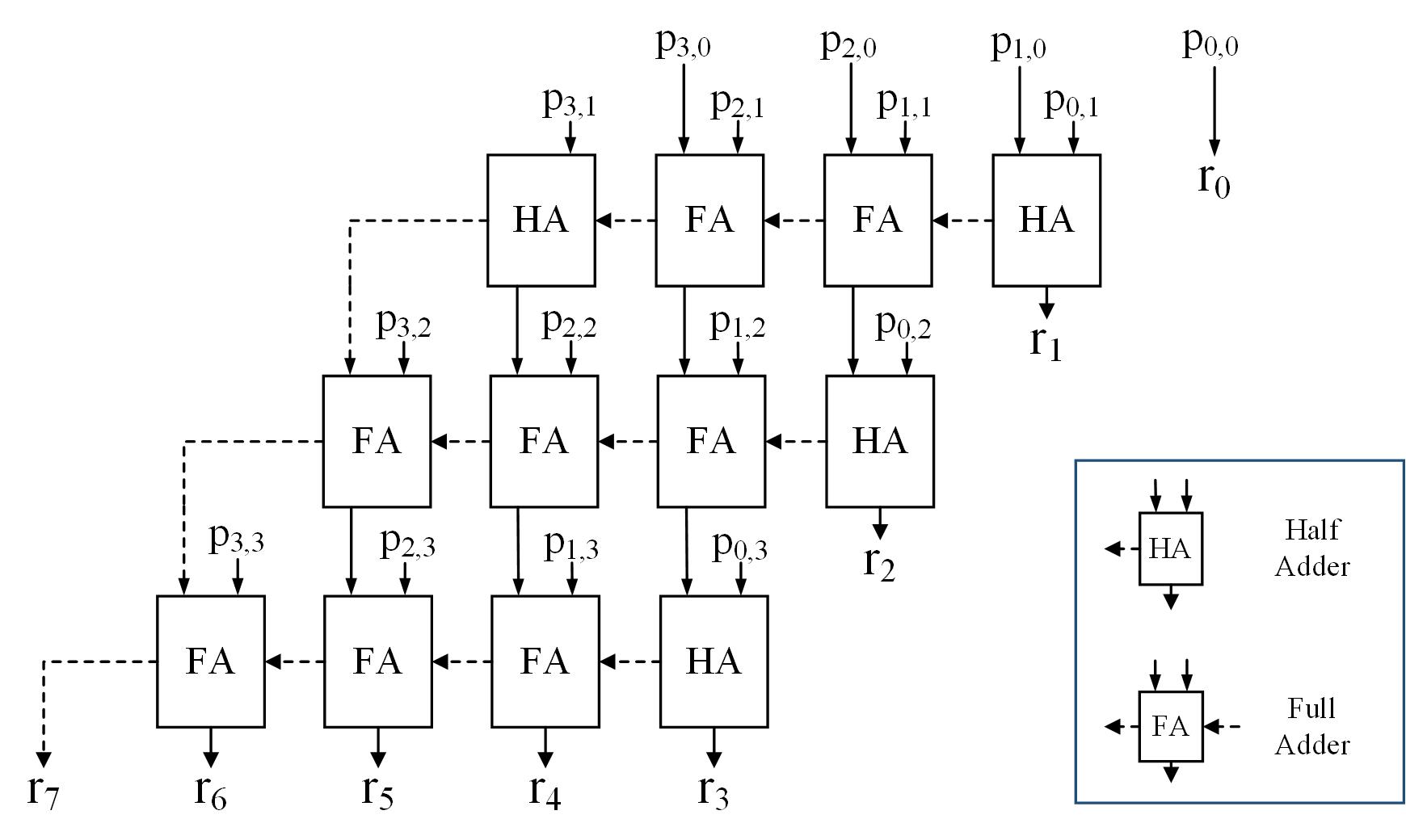}}
\subfigure[]{
\includegraphics[width=0.6\textwidth]{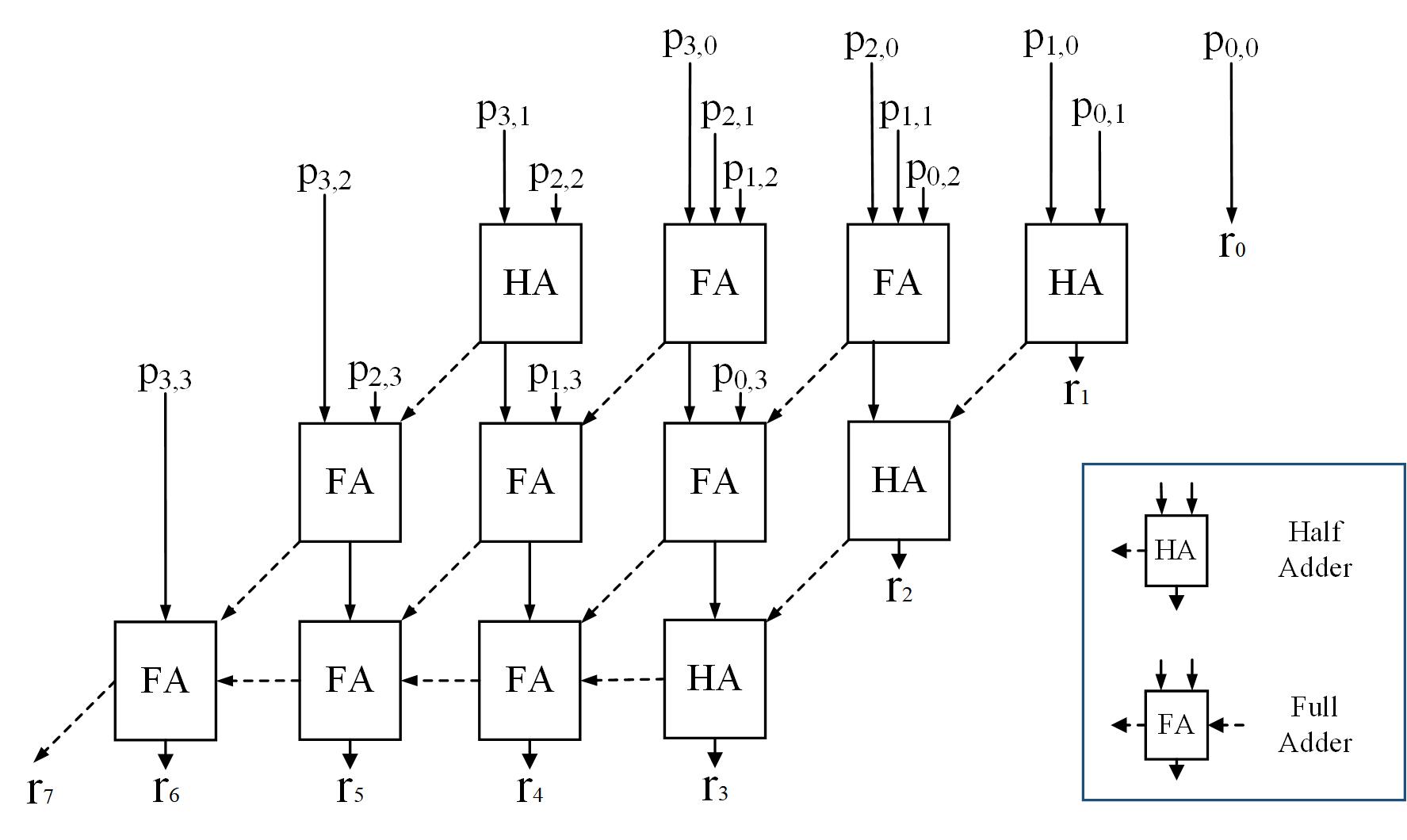}
}%
\centering
\caption{(a) An example of \qdel{RSA based}\qadd{RCA-based} accumulator; (b) An example of CSA-based accumulator.}\label{fig:background_accumulator}
\end{figure}

\subsection{Binary Multipliers}\label{sec:multiplier_binary}

Before we \qdel{go into details of}\qadd{elaborate} further on the approximate multiplier design, we \qdel{would like to }introduce the basics of a binary multiplier\qadd{ in this section}, which \hjdel{is straightforward but }sheds light on different approximation techniques introduced in the latter sections. 

Due to the nature of dealing with the two digits, $i.e.$, 0 and 1, the binary multiplication can be considered as a process of addition and shifting.
For example, consider two 4-bit operands of $x$ and $y$, where $x$ is the multiplicand and $y$ is the multiplier. As shown in Fig.~\ref{fig:background2}, similar to decimal multiplication\hjdel{ operation}, the binary multiplication is carried out for each bit of the multiplier ($i.e.$, $y_i$ for $i=0,1,2,3$) and the multiplicand ($i.e.$, $x$=\{$x_{3},x_{2},x_{1},x_{0}$\}) to generate a partial product, $e.g.$, \{$p_{3,0},p_{2,0},p_{1,0},p_{0,0}$\} for the first row. This process is then repeated for each bit of the multiplier $y$, with the partial product left-shifted by 1 bit. Finally, all the partial products are accumulated to obtain the final product as \qdel{\{$r_{7,0},r_{6,0},r_{5,0},r_{4,0},r_{3,0},r_{2,0},r_{1,0},r_{0,0}$\}}\qadd{$\{r_{7}, r_{6}, r_{5}, r_{4}, r_{3}, r_{2}, r_{1}, r_{0}\}$}. Thus, the operation of a binary multiplier can be roughly divided into three stages, data input, partial product generation, and accumulation.  

Since the partial product \wydel{does not generate a carry}is generated without a carry, the bit-wise multiplication can be calculated with AND gates. Once all the partial products are generated, we can use an array of adders to accumulate the partial products as shown in \qdel{Fig.~\ref{fig:background_accumulator}(a)}\qadd{Fig.~\ref{fig:background_accumulator}}, where HA refers to a half adder and FA refers to a full adder. Obviously, the critical path of such a structure is the \qdel{carry propagation}\qadd{carry propagation chain}. For example, Fig.~\ref{fig:background_accumulator}(a) shows a ripple-carry adder (RCA)-based accumulator, with the carries propagated horizontally from right to left, while Fig.~\ref{fig:background_accumulator}(b) shows a carry-save adder (CSA)-based accumulator with carries propagated diagonally to achieve a shorter critical path for faster speed.

\begin{figure}[!htbp]
\centering
\includegraphics[width=.7\textwidth]{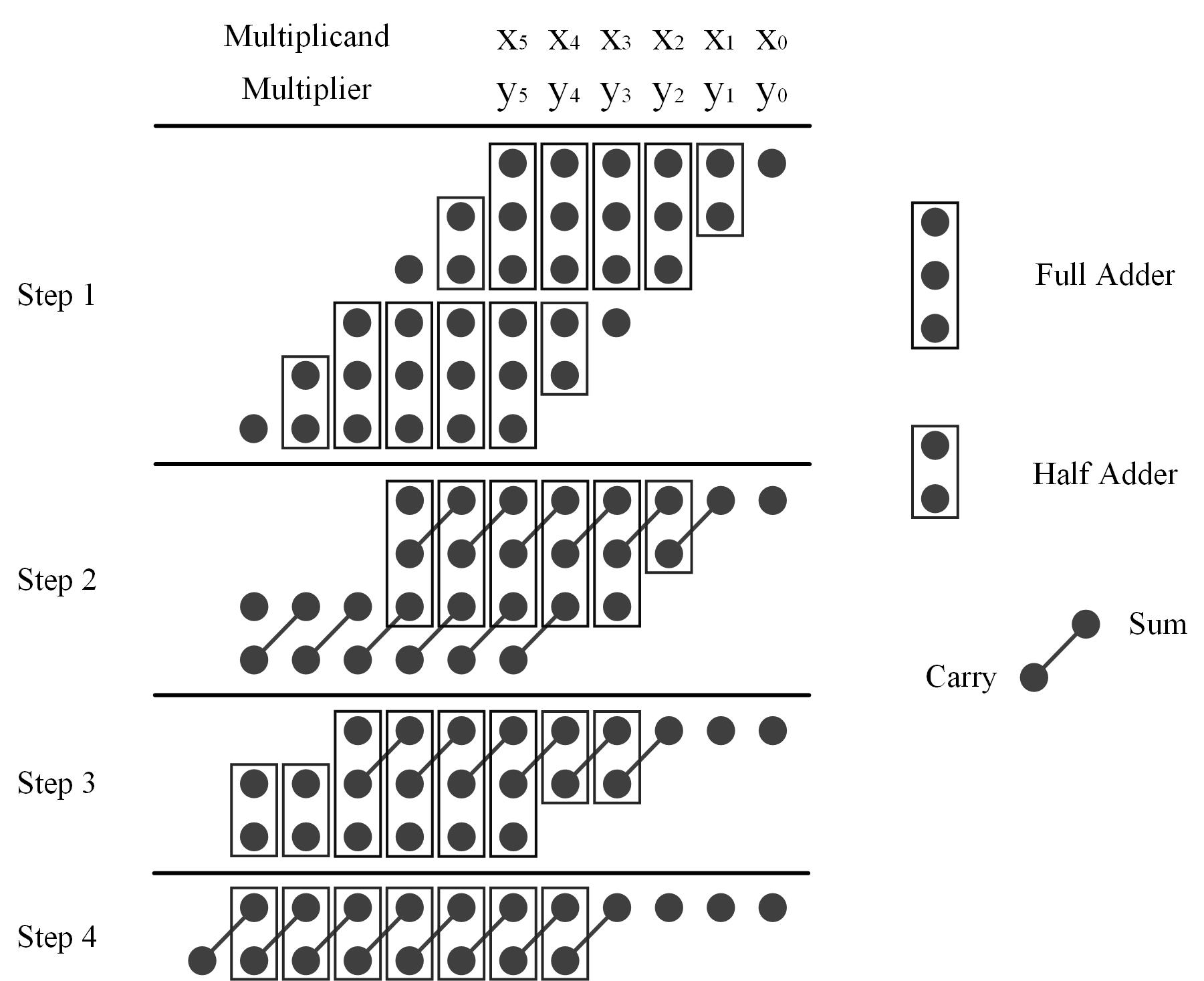}
\caption{An example of a Wallace Tree-based multiplier.}
\label{fig:background5}       
\end{figure}

In order to further accelerate the accumulation, \qdel{C.~S.~Wallace}\qadd{Wallace} proposed a tree structure in 1964~\cite{wallace1964suggestion}. As shown in Fig.~\ref{fig:background5}, the Wallace Tree groups three partial products within one column to generate two outputs, \textit{i.e.}, a sum and a carry, thereby reducing the number of partial products by a factor of approximately 1.5. The operation is repeated until only two rows are left,\qdel{ \textit{i.e.},} \qadd{\textit{e.g.}, by the first }3 steps in Fig.~\ref{fig:background5}\qdel{, which are then}\qadd{. Then, the last two rows are} added up to obtain the final result. Parallel computation and partial product compression in each stage can be utilized to speed up the accumulation process~\cite{wallace1964suggestion}.

\subsection{Exact Compressor}

\begin{figure}[!htbp]
	\centering
	\includegraphics[scale=0.98]{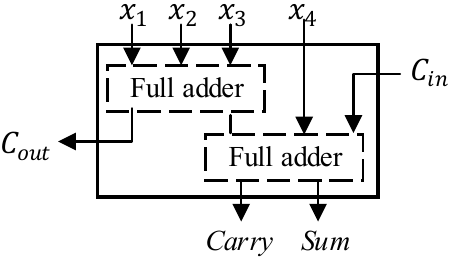}\\
	\caption{A\hjdel{n example of the} conventional exact 4-2 compressor.}
	\label{exact 4-2 compressor}
\end{figure}

In tree-based partial product accumulation, 
such as the Wallace tree~\cite{wallace1964suggestion} and the Dadda tree~\cite{dadda1965some}, compressors are used to count the number of \qdel{``ones''}\qadd{ones} \qdel{for}\qadd{within} a group of partial products.
The full adder and half adder in Fig.~\ref{fig:background5} are widely used as 3-2 compressors and 2-2 compressors, respectively.
In order to further improve the compression efficiency, higher-order compressors, such as \hjdel{exact }4-2 and 5-2 compressors~\cite{Chang2004ultra, pishvaie2012improved, baran2010energy, arasteh2018energy, veeramachaneni2007novel} or \hjdel{exact }6-3 and 7-3 compressors~\cite{fritz2017fast}, have\qadd{ also} been investigated. Fig.~\ref{exact 4-2 compressor} shows the design of a conventional 4-2 compressor. It consists of two full adders and has five primary inputs (PIs), $x_1, x_2, x_3, x_4$, and $C_{\textit{in}}$, and three primary outputs (POs), $\textit{Sum}, \textit{Carry}$, and $C_{\textit{out}}$, where $C_{\textit{in}}$ comes from the preceding less significant block, and $C_{\textit{out}}$ goes to the next more significant block. 
The outputs $\textit{Sum}$, $\textit{Carry}$, and $C_{\textit{out}}$ can be calculated as~\cite{Momeni2014}:
\begin{align*}
& \textit{Sum} = x_1\oplus x_2\oplus x_3 \oplus x_4 \oplus C_{\textit{in}}, \\
&\textit{Carry} = (x_1\oplus x_2\oplus x_3 \oplus x_4) \wedge C_{\textit{in}} 
+ \overline{(x_1\oplus x_2\oplus x_3 \oplus x_4)} \wedge x_4, \\
&C_{\textit{out}} = (x_1 \wedge(x_2 \vee x_3)) \vee (x_2 \wedge x_3).
\end{align*}

\qadd{Note that the bit significances of the \textit{Carry} and $C_{\textit{out}}$
are both \wadd{twice of}\qdel{2x higher than} that of \textit{Sum}.}

\section{Overview of Approximate Multipliers}
\label{sec:classification}

Many prior works on approximate multipliers have tackled the problem by introducing approximations at \qdel{circuit, architecture or algorithmic}\qadd{algorithm, architecture, or circuit} levels to reduce the critical path delay or improve energy efficiency.
\qadd{For example, at the algorithm level, Ahmed \textit{et al.} explored a pipelined log-based approximation using the classical Mitchell's multiplier with an iterative procedure to improve the accuracy~\cite{ahmed2016iterative}.}
\qadd{At the architecture level, many works focused on improving the conventional multiplier architecture with approximate components, such as adders, to speed up the addition or partial product generation~\cite{kulkarni2011trading, Approx_53_7598342, narayanamoorthy2014energy, AWTM_6783335, liu2014low, guo2020reconfigurable}.}
At the circuit level, references \cite{kulkarni2011trading, pabithra2018analysis, van2020fpga,ha2017multipliers,liu2017design, qian2016design,yin2018designs,lingamneni2011energy,schlachter2015automatic} proposed to approximate Boolean algebra expressions or prune out a few gates to simplify the circuit.

For all the prior works with various approximation techniques, it is actually very challenging to precisely categorize the introduced approximation to a particular design level, \textit{i.e.}, \qdel{circuit, architecture, or algorithm}\qadd{algorithm, architecture, or circuit levels}. Many of them actually involve multiple design levels, as the \qdel{high}\qadd{highest} level approximation, \textit{i.e.}, the algorithm level, may \hjdel{always }incur additional architecture changes~\cite{ahmed2016iterative, imani2017cfpu, imani2018rmac, imani2019approxlp, chen2020optimally}. Similarly, circuit-level techniques are often utilized for basic module designs applied in higher-level approximation schemes~\cite{ansari2020improved, Wang20,van2020fpga,tung2019low,ha2017multipliers,yin2018designs,esposito2018approximate,jiang2018low,venkatachalam2017design,liu2017design,qian2016design,jiang2015approximate,yang2015approximate,kulkarni2011trading,liu2009computation}.

In order to facilitate our review\hjdel{ in the following}, we introduce the following definitions to categorize the considered approximation:
\begin{itemize}
    \item Algorithm: The introduced approximation originates from a different algorithmic procedure to conduct multiplication. 
    \item Architecture: With the binary multiplier architecture in Section~\ref{sec:multiplier_binary} as a reference, the introduced approximation is intended to improve the efficiency of a particular stage in the reference architecture.
    \item Circuit: The approximation technique is not limited to a particular \hjdel{multiplier }architecture or algorithm, and it can be combined with approximation techniques at other design levels.
\end{itemize}

\wyadd{Table~\ref{tab:work_summary} summarizes the representative works reviewed in this article for digital circuit implementations. These works are listed in the reverse chronological order. We mark the types of approximate multipliers and the utilized techniques in the three considered levels.}
\wyadd{Due to the space limit, we use some abbreviations as shown in Table~\ref{tab:abbr_meaning} to represent the multiplier types and the approximation techniques or stages in the remainder of this article.}
\wyadd{For the well-explored approximation technique (acc), we provide a list of representative works that focus on dividing and applying approximate compressors~\cite{esposito2018approximate,Wang20,ha2017multipliers, tung2019low}, introducing approximation in the accumulation stage in Booth multipliers~\cite{cho2004design,song2007adaptive,wang2009high,chen2011high,farshchi2013new}, and dumping partial product rows and columns~\cite{mahdiani2009bio}. We also indicate the utilization of approximation in the accumulation stage in various works that represent other approximation techniques~\cite{yang2018low,yin2018designs,Jiang19,venkatachalam2017design,liu2017design,qian2016design,schlachter2015automatic}. As for the commonly adopted approximation technique (BR), we indicate its utilization in the listed designs without specifically listing representative designs.
Finally, for the other approximation techniques (log, linear, hy, in, enc, ECD, GP and VOS), we list the works introduced in this article, which are representative works corresponding to each respective approximation technique.
}

\begin{table}
  \caption{The meanings of abbreviations used in Table~\ref{tab:work_summary}. }
  \label{tab:abbr_meaning}
  \begin{tabular}{cl}
    \toprule
    Abbreviations & Meanings\\
    \midrule
    U & unsigned fixed-point \\
    S & signed fixed-point\\
    FP & floating-point \\
    log & logarithm-based\\
    linear &  linearization-based\\
    hy & hybrid-based\\
    in & input stage\\
    pp & partial product generation stage\\
    acc& accumulation stage\\
    enc& Booth encoding stage\\
    BR & Boolean rewriting\\
    ECD & evolutionary circuit design\\
    VOS & voltage over-scaling\\
  \bottomrule
\end{tabular}
\end{table}

\begin{table}[!htbp]
\caption{Summary of the \wydel{state-of-the-art}\wyadd{representative} designs \qdel{on}\wadd{of} approximate multipliers. }
\label{tab:work_summary} 
\begin{tabular}{|l|l|lll|lll|llll|llll|}
\hline
\multicolumn{1}{|c|}{}                       & \multicolumn{1}{c|}{}                       & \multicolumn{3}{c|}{Type}                                                   & \multicolumn{3}{c|}{Algorithm}                                              & \multicolumn{4}{c|}{Architecture}                                                                             & \multicolumn{4}{c|}{Circuit\tnote{4}}                                                             \\ \cline{3-16} 
\multicolumn{1}{|c|}{\multirow{-2}{*} {Year}} & \multicolumn{1}{c|}{\multirow{-2}{*}{Work}} & \multicolumn{1}{c|}{U}    & \multicolumn{1}{c|}{S}    & FP       & \multicolumn{1}{c|}{log}       & \multicolumn{1}{c|}{linear}    & hy    & \multicolumn{1}{c|}{in} & \multicolumn{1}{c|}{pp}        & \multicolumn{1}{c|}{acc}      & enc      & \multicolumn{1}{c|}{BR}      & \multicolumn{1}{c|}{ECD}       & \multicolumn{1}{c|}{GP} & VOS \\ \hline
2022                                                                                        & \cite{Xiao22}                                        & \multicolumn{1}{c|}{\checkmark} & \multicolumn{1}{c|}{}          &           & \multicolumn{1}{c|}{} & \multicolumn{1}{c|}{}          &           & \multicolumn{1}{c|}{}           & \multicolumn{1}{c|}{}          & \multicolumn{1}{c|}{\checkmark}          &           & \multicolumn{1}{c|}{}     & \multicolumn{1}{c|}{}          & \multicolumn{1}{c|}{}          &                    \\ \hline
2020                                                                                        & \cite{ansari2020improved}                                        & \multicolumn{1}{c|}{\checkmark} & \multicolumn{1}{c|}{}          &           & \multicolumn{1}{c|}{\checkmark} & \multicolumn{1}{c|}{}          &           & \multicolumn{1}{c|}{}           & \multicolumn{1}{c|}{}          & \multicolumn{1}{c|}{}          &           & \multicolumn{1}{c|}{\checkmark}     & \multicolumn{1}{c|}{}          & \multicolumn{1}{c|}{}          &                    \\ \hline
2020                                                                                        & \cite{chen2020optimally}                                        & \multicolumn{1}{c|}{}          & \multicolumn{1}{c|}{}          & \checkmark & \multicolumn{1}{c|}{}          & \multicolumn{1}{c|}{\checkmark} &           & \multicolumn{1}{c|}{}           & \multicolumn{1}{c|}{}          & \multicolumn{1}{c|}{}          &           & \multicolumn{1}{c|}{}              & \multicolumn{1}{c|}{}          & \multicolumn{1}{c|}{}          &                    \\ \hline
2020                                       & \cite{Wang20}                                  & \multicolumn{1}{c|}{\checkmark} &  \multicolumn{1}{c|}{}         &           &  \multicolumn{1}{c|}{}         &  \multicolumn{1}{c|}{}         &          &  \multicolumn{1}{c|}{}          &   \multicolumn{1}{c|}{}        &    \multicolumn{1}{c|}{\checkmark}       &           & \multicolumn{1}{c|}{\checkmark}             & \multicolumn{1}{c|}{}          & \multicolumn{1}{c|}{}          &                      \\ \hline
2019                                                                                        & \cite{imani2019approxlp}                                        & \multicolumn{1}{c|}{}          & \multicolumn{1}{c|}{}          & \checkmark & \multicolumn{1}{c|}{}          & \multicolumn{1}{c|}{\checkmark} &           & \multicolumn{1}{c|}{}           & \multicolumn{1}{c|}{}          & \multicolumn{1}{c|}{}          &           & \multicolumn{1}{c|}{}              & \multicolumn{1}{c|}{}          & \multicolumn{1}{c|}{}          &                    \\ \hline
2019                                                                                        & \cite{vahdat2019tosam}                                        & \multicolumn{1}{c|}{\checkmark} & \multicolumn{1}{c|}{\checkmark} &           & \multicolumn{1}{c|}{}          & \multicolumn{1}{c|}{}          &           & \multicolumn{1}{c|}{\checkmark}  & \multicolumn{1}{c|}{}          & \multicolumn{1}{c|}{\checkmark} &           & \multicolumn{1}{c|}{}              & \multicolumn{1}{c|}{}          & \multicolumn{1}{c|}{}          &                    \\ \hline
2019                                                                                        & \cite{tung2019low}                                        & \multicolumn{1}{c|}{\checkmark} & \multicolumn{1}{c|}{}          &           & \multicolumn{1}{c|}{}          & \multicolumn{1}{c|}{}          &           & \multicolumn{1}{c|}{}           & \multicolumn{1}{c|}{}          & \multicolumn{1}{c|}{\checkmark} &           & \multicolumn{1}{c|}{\checkmark}     & \multicolumn{1}{c|}{}          & \multicolumn{1}{c|}{}          &                    \\ \hline
2019                                                                                        & \cite{ansari2019hardware}                                        & \multicolumn{1}{c|}{\checkmark} & \multicolumn{1}{c|}{}          &           & \multicolumn{1}{c|}{\checkmark} & \multicolumn{1}{c|}{}          &           & \multicolumn{1}{c|}{}           & \multicolumn{1}{c|}{}          & \multicolumn{1}{c|}{}          &           & \multicolumn{1}{c|}{}              & \multicolumn{1}{c|}{}          & \multicolumn{1}{c|}{}          &                    \\ \hline
2018                                                                                        & \cite{imani2018rmac}                                        & \multicolumn{1}{c|}{}          & \multicolumn{1}{c|}{}          & \checkmark & \multicolumn{1}{c|}{}          & \multicolumn{1}{c|}{}          & \checkmark & \multicolumn{1}{c|}{}           & \multicolumn{1}{c|}{}          & \multicolumn{1}{c|}{}          &           & \multicolumn{1}{c|}{}              & \multicolumn{1}{c|}{}          &  \multicolumn{1}{c|}{}          &                   \\ \hline
2018                                                                                        & \cite{yang2018low}                                        & \multicolumn{1}{c|}{\checkmark} & \multicolumn{1}{c|}{}          &           & \multicolumn{1}{c|}{}          & \multicolumn{1}{c|}{}          &           & \multicolumn{1}{c|}{}           & \multicolumn{1}{c|}{\checkmark} & \multicolumn{1}{c|}{\checkmark} &           & \multicolumn{1}{c|}{}              & \multicolumn{1}{c|}{}          &  \multicolumn{1}{c|}{}          &                   \\ \hline
2018                                                                                        & \cite{ha2017multipliers}                                        & \multicolumn{1}{c|}{\checkmark} & \multicolumn{1}{c|}{}          &           & \multicolumn{1}{c|}{}          & \multicolumn{1}{c|}{}          &           & \multicolumn{1}{c|}{}           & \multicolumn{1}{c|}{}          & \multicolumn{1}{c|}{\checkmark} &           & \multicolumn{1}{c|}{\checkmark}     & \multicolumn{1}{c|}{}          & \multicolumn{1}{c|}{}          &                    \\ \hline
2018                                                                                        & \cite{liu2018design}                                        & \multicolumn{1}{c|}{\checkmark} & \multicolumn{1}{c|}{}          &           & \multicolumn{1}{c|}{\checkmark} & \multicolumn{1}{c|}{}          &           & \multicolumn{1}{c|}{}           & \multicolumn{1}{c|}{}          & \multicolumn{1}{c|}{}          &           & \multicolumn{1}{c|}{\checkmark} & \multicolumn{1}{c|}{}          &  \multicolumn{1}{c|}{}          &                   \\ \hline
2018                                         & \cite{saadat2018minimally}                     & \multicolumn{1}{c|}{\checkmark} & \multicolumn{1}{c|}{}          & \checkmark & \multicolumn{1}{c|}{\checkmark} & \multicolumn{1}{c|}{}          &           & \multicolumn{1}{c|}{}           & \multicolumn{1}{c|}{}          & \multicolumn{1}{c|}{}          &           & \multicolumn{1}{c|}{}              & \multicolumn{1}{c|}{}          &  \multicolumn{1}{c|}{}          &                     \\ \hline
2018                                         & \cite{yin2018designs}                          & \multicolumn{1}{c|}{}          & \multicolumn{1}{c|}{}          & \checkmark & \multicolumn{1}{c|}{}          & \multicolumn{1}{c|}{}          &           & \multicolumn{1}{c|}{}           & \multicolumn{1}{c|}{}          & \multicolumn{1}{c|}{\checkmark} & \checkmark & \multicolumn{1}{c|}{\checkmark}     & \multicolumn{1}{c|}{}          & \multicolumn{1}{c|}{}          &                      \\ \hline
2018                                                                                        & \cite{esposito2018approximate}                                        & \multicolumn{1}{c|}{\checkmark} & \multicolumn{1}{c|}{}          &           & \multicolumn{1}{c|}{}          & \multicolumn{1}{c|}{}          &           & \multicolumn{1}{c|}{}           & \multicolumn{1}{c|}{}          & \multicolumn{1}{c|}{\checkmark} &           & \multicolumn{1}{c|}{\checkmark}     & \multicolumn{1}{c|}{}          & \multicolumn{1}{c|}{}          &                    \\ \hline
2018                                                                                        & \cite{jiang2018low}                                        & \multicolumn{1}{c|}{\checkmark} & \multicolumn{1}{c|}{}          &           & \multicolumn{1}{c|}{}          & \multicolumn{1}{c|}{}          &           & \multicolumn{1}{c|}{}           & \multicolumn{1}{c|}{\checkmark} & \multicolumn{1}{c|}{\checkmark} &           & \multicolumn{1}{c|}{\checkmark}     & \multicolumn{1}{c|}{}          &  \multicolumn{1}{c|}{}          &                   \\ \hline
2017                                                                                        & \cite{venkatachalam2017design}                                        & \multicolumn{1}{c|}{\checkmark} & \multicolumn{1}{c|}{}          &           & \multicolumn{1}{c|}{}          & \multicolumn{1}{c|}{}          &           & \multicolumn{1}{c|}{}           & \multicolumn{1}{c|}{\checkmark} & \multicolumn{1}{c|}{\checkmark} &           & \multicolumn{1}{c|}{\checkmark}     & \multicolumn{1}{c|}{}          & \multicolumn{1}{c|}{}          &                    \\ \hline
2017                                         & \cite{imani2017cfpu}                           & \multicolumn{1}{c|}{}          & \multicolumn{1}{c|}{}          & \checkmark & \multicolumn{1}{c|}{}          & \multicolumn{1}{c|}{}          & \checkmark & \multicolumn{1}{c|}{}           & \multicolumn{1}{c|}{}          & \multicolumn{1}{c|}{}          &           & \multicolumn{1}{c|}{}              & \multicolumn{1}{c|}{}          &  \multicolumn{1}{c|}{}          &                     \\ \hline
2017                                         & \cite{vahdat2017letam}                         & \multicolumn{1}{c|}{\checkmark} & \multicolumn{1}{c|}{\checkmark} &           & \multicolumn{1}{c|}{}          & \multicolumn{1}{c|}{}          &           & \multicolumn{1}{c|}{\checkmark}  & \multicolumn{1}{c|}{}          & \multicolumn{1}{c|}{\checkmark} &           & \multicolumn{1}{c|}{}              & \multicolumn{1}{c|}{}          &  \multicolumn{1}{c|}{}          &                     \\ \hline
2017                                         & \cite{yang2017low}                             & \multicolumn{1}{c|}{}          & \multicolumn{1}{c|}{}          &           & \multicolumn{1}{c|}{}          & \multicolumn{1}{c|}{}          &           & \multicolumn{1}{c|}{}           & \multicolumn{1}{c|}{}          & \multicolumn{1}{c|}{}          &           & \multicolumn{1}{c|}{}              & \multicolumn{1}{c|}{}          &  \multicolumn{1}{c|}{}          &                     \\ \hline
2017                                                                                        & \cite{liu2017design}                                        & \multicolumn{1}{c|}{}          & \multicolumn{1}{c|}{\checkmark} &           & \multicolumn{1}{c|}{}          & \multicolumn{1}{c|}{}          &           & \multicolumn{1}{c|}{}           & \multicolumn{1}{c|}{}          & \multicolumn{1}{c|}{\checkmark} & \checkmark & \multicolumn{1}{c|}{\checkmark}     & \multicolumn{1}{c|}{}          &  \multicolumn{1}{c|}{}          &                   \\ \hline
2017                                                                                        & \cite{mrazek2017evoapprox8b}                                        & \multicolumn{1}{c|}{}          & \multicolumn{1}{c|}{} &           & \multicolumn{1}{c|}{}          & \multicolumn{1}{c|}{}          &           & \multicolumn{1}{c|}{}           & \multicolumn{1}{c|}{}          & \multicolumn{1}{c|}{} &  & \multicolumn{1}{c|}{}     & \multicolumn{1}{c|}{\checkmark}          &  \multicolumn{1}{c|}{}          &                   \\ \hline
2016                                                                                        & \cite{qian2016design}                                        & \multicolumn{1}{c|}{}          & \multicolumn{1}{c|}{\checkmark} &           & \multicolumn{1}{c|}{}          & \multicolumn{1}{c|}{}          &           & \multicolumn{1}{c|}{}           & \multicolumn{1}{c|}{}          & \multicolumn{1}{c|}{\checkmark} & \checkmark & \multicolumn{1}{c|}{\checkmark}     & \multicolumn{1}{c|}{}          &  \multicolumn{1}{c|}{}          &                   \\ \hline
2016                                         & \cite{ahmed2016iterative}                      & \multicolumn{1}{c|}{\checkmark} & \multicolumn{1}{c|}{}          &           & \multicolumn{1}{c|}{\checkmark} & \multicolumn{1}{c|}{}          &           & \multicolumn{1}{c|}{\checkmark}  & \multicolumn{1}{c|}{}          & \multicolumn{1}{c|}{}          &           & \multicolumn{1}{c|}{}              & \multicolumn{1}{c|}{}          &  \multicolumn{1}{c|}{}          &                     \\ \hline
2016                                                                                        & \cite{hrbacek2016automatic}                                        & \multicolumn{1}{c|}{}          & \multicolumn{1}{c|}{} &           & \multicolumn{1}{c|}{}          & \multicolumn{1}{c|}{}          &           & \multicolumn{1}{c|}{}           & \multicolumn{1}{c|}{}          & \multicolumn{1}{c|}{} &  & \multicolumn{1}{c|}{}     & \multicolumn{1}{c|}{\checkmark}          &  \multicolumn{1}{c|}{}          &                   \\ \hline
2015                                                                                        & \cite{hashemi2015drum}                                        & \multicolumn{1}{c|}{\checkmark} & \multicolumn{1}{c|}{}          &           & \multicolumn{1}{c|}{}          & \multicolumn{1}{c|}{}          &           & \multicolumn{1}{c|}{\checkmark}  & \multicolumn{1}{c|}{}          & \multicolumn{1}{c|}{}          &           & \multicolumn{1}{c|}{}              & \multicolumn{1}{c|}{}          & \multicolumn{1}{c|}{}          &                    \\ \hline
2015                                                                                        & \cite{jiang2015approximate}                                        & \multicolumn{1}{c|}{}          & \multicolumn{1}{c|}{\checkmark} &           & \multicolumn{1}{c|}{}          & \multicolumn{1}{c|}{}          &           & \multicolumn{1}{c|}{}           & \multicolumn{1}{c|}{}          & \multicolumn{1}{c|}{}          & \checkmark & \multicolumn{1}{c|}{\checkmark}     & \multicolumn{1}{c|}{}          &  \multicolumn{1}{c|}{}          &                   \\ \hline
2015                                         & \cite{schlachter2015automatic}                     & \multicolumn{1}{c|}{} & \multicolumn{1}{c|}{}          &           & \multicolumn{1}{c|}{}          & \multicolumn{1}{c|}{}          &           & \multicolumn{1}{c|}{}           & \multicolumn{1}{c|}{}          & \multicolumn{1}{c|}{\checkmark} &           & \multicolumn{1}{c|}{}     & \multicolumn{1}{c|}{}          & \multicolumn{1}{c|}{\checkmark}          &                      \\ \hline
2014                                                                                        & \cite{narayanamoorthy2014energy}                                        & \multicolumn{1}{c|}{\checkmark} & \multicolumn{1}{c|}{}          &           & \multicolumn{1}{c|}{}          & \multicolumn{1}{c|}{}          &           & \multicolumn{1}{c|}{\checkmark}  & \multicolumn{1}{c|}{}          & \multicolumn{1}{c|}{}          &           & \multicolumn{1}{c|}{}              & \multicolumn{1}{c|}{}          & \multicolumn{1}{c|}{}          &                    \\ \hline
2013                                                                                        & \cite{farshchi2013new}                                     & \multicolumn{1}{c|}{}          & \multicolumn{1}{c|}{\checkmark} &           & \multicolumn{1}{c|}{}          & \multicolumn{1}{c|}{}          &           & \multicolumn{1}{c|}{}           & \multicolumn{1}{c|}{}          & \multicolumn{1}{c|}{\checkmark} &           & \multicolumn{1}{c|}{}          & \multicolumn{1}{c|}{}          &  \multicolumn{1}{c|}{}          &                   \\ \hline

2012                                                                                        & \cite{chen2012energy}                                        & \multicolumn{1}{c|}{}          & \multicolumn{1}{c|}{}          &           & \multicolumn{1}{c|}{}          & \multicolumn{1}{c|}{}          &           & \multicolumn{1}{c|}{}           & \multicolumn{1}{c|}{}          & \multicolumn{1}{c|}{}          &           & \multicolumn{1}{c|}{}              & \multicolumn{1}{c|}{} &   \multicolumn{1}{c|}{}          &    \checkmark              \\ \hline
2011                                                                                        & \cite{kulkarni2011trading}                                        & \multicolumn{1}{c|}{\checkmark} & \multicolumn{1}{c|}{}          &           & \multicolumn{1}{c|}{}          & \multicolumn{1}{c|}{}          &           & \multicolumn{1}{c|}{}           & \multicolumn{1}{c|}{\checkmark} & \multicolumn{1}{c|}{}          &           & \multicolumn{1}{c|}{\checkmark}     & \multicolumn{1}{c|}{}          & \multicolumn{1}{c|}{}          &                    \\ \hline
2011                                                                                        & \cite{chen2011high}                                        & \multicolumn{1}{c|}{}          & \multicolumn{1}{c|}{\checkmark} &           & \multicolumn{1}{c|}{}          & \multicolumn{1}{c|}{}          &           & \multicolumn{1}{c|}{}           & \multicolumn{1}{c|}{}          & \multicolumn{1}{c|}{\checkmark} &           & \multicolumn{1}{c|}{}          & \multicolumn{1}{c|}{}          &  \multicolumn{1}{c|}{}          &                   \\ \hline

2009                                                                                        & \cite{mahdiani2009bio}                                        & \multicolumn{1}{c|}{\checkmark} & \multicolumn{1}{c|}{}          &           & \multicolumn{1}{c|}{}          & \multicolumn{1}{c|}{}          &           & \multicolumn{1}{c|}{}           & \multicolumn{1}{c|}{}          & \multicolumn{1}{c|}{\checkmark} &           & \multicolumn{1}{c|}{}              & \multicolumn{1}{c|}{}          &  \multicolumn{1}{c|}{}          &                   \\ \hline
2009                                         & \cite{lau2009energy}                           & \multicolumn{1}{c|}{\checkmark} & \multicolumn{1}{c|}{}          &           & \multicolumn{1}{c|}{}          & \multicolumn{1}{c|}{}          &           & \multicolumn{1}{c|}{}           & \multicolumn{1}{c|}{}          & \multicolumn{1}{c|}{}          &           & \multicolumn{1}{c|}{}              & \multicolumn{1}{c|}{} & \multicolumn{1}{c|}{}          &       \checkmark               \\ \hline

2009                                         & \cite{liu2009computation}                      & \multicolumn{1}{c|}{\checkmark} & \multicolumn{1}{c|}{}          &           & \multicolumn{1}{c|}{}          & \multicolumn{1}{c|}{}          &           & \multicolumn{1}{c|}{}           & \multicolumn{1}{c|}{}          & \multicolumn{1}{c|}{}          &           & \multicolumn{1}{c|}{}             & \multicolumn{1}{c|}{} &  \multicolumn{1}{c|}{}          &         \checkmark            \\ \hline
2009                                                                                        & \cite{wang2009high}                                & \multicolumn{1}{c|}{}          & \multicolumn{1}{c|}{\checkmark} &           & \multicolumn{1}{c|}{}          & \multicolumn{1}{c|}{}          &           & \multicolumn{1}{c|}{}           & \multicolumn{1}{c|}{}          & \multicolumn{1}{c|}{\checkmark} &           & \multicolumn{1}{c|}{}          & \multicolumn{1}{c|}{}          & \multicolumn{1}{c|}{}          &                    \\ \hline
2007                                                                                        & \cite{song2007adaptive}                                    & \multicolumn{1}{c|}{}          & \multicolumn{1}{c|}{\checkmark} &           & \multicolumn{1}{c|}{}          & \multicolumn{1}{c|}{}          &           & \multicolumn{1}{c|}{}           & \multicolumn{1}{c|}{}          & \multicolumn{1}{c|}{\checkmark} &           & \multicolumn{1}{c|}{}          & \multicolumn{1}{c|}{}          & \multicolumn{1}{c|}{}          &                    \\ \hline
2004                                                                                        & \cite{cho2004design}                               & \multicolumn{1}{c|}{}          & \multicolumn{1}{c|}{\checkmark} &           & \multicolumn{1}{c|}{}          & \multicolumn{1}{c|}{}          &           & \multicolumn{1}{c|}{}           & \multicolumn{1}{c|}{}          & \multicolumn{1}{c|}{\checkmark} &           & \multicolumn{1}{c|}{\checkmark} & \multicolumn{1}{c|}{}          &  \multicolumn{1}{c|}{}          &                   \\ \hline

1962                                                                                        & \cite{mitchell1962computer}                                        & \multicolumn{1}{c|}{\checkmark} & \multicolumn{1}{c|}{}          &           & \multicolumn{1}{c|}{\checkmark} & \multicolumn{1}{c|}{}          &           & \multicolumn{1}{c|}{}           & \multicolumn{1}{c|}{}          & \multicolumn{1}{c|}{}          &           & \multicolumn{1}{c|}{}              & \multicolumn{1}{c|}{}          & \multicolumn{1}{c|}{}          &                    \\ \hline
\end{tabular}
\end{table}

\section{Approximate Multiplier with Algorithm-level Approximation}\label{sec:alg}

\qadd{The algorithm-level approximate multiplier rebuilds the multiplication algorithm itself, which naturally results in a new multiplier architecture.}
In this section, we will review three different multiplier approximations at\qadd{ the} algorithm level: logarithm-based approximation, approximation with linearization, and hybrid approximation. 

\subsection{Logarithm-Based Approximation}
With logarithmic transformation, the multiplication can be converted to addition, where the two operands are the logarithms of multiplicand and multiplier, respectively. The first logarithm-based multiplier (LM) was proposed by \qdel{Mitchell \textit{et al.}}\qadd{Mitchell} in 1962~\cite{mitchell1962computer}.
For a multiplication of \qdel{$A\times B$}\qadd{two operands $A$ and $B$}, we have:
\begin{equation}
A=2^{k_1}(1+x_1)\;,
\end{equation}
\begin{equation}
\log_2 (A)=k_1+\log_2 (1+x_1)\;,
\end{equation}
where \qdel{$A$ is the input operand, }\(k_1\) indicates the position of \qadd{the }leading one\qadd{ of $A$} and \(x_1\) is the fraction part\qadd{ of $A$} that lies in \([0,1)\). The same formulation can be applied to the other operand $B$ with the parameters of \(k_2\) and \(x_2\). The logarithm of the \qdel{multiplication}\qadd{product} can be written as:
\begin{equation}\label{eq:log_multiply}
\log_2 (A\times B)=k_1+k_2+\log_2 (1+x_1)+\log_2 (1+x_2)\;.
\end{equation}

\begin{figure}[!htbp]
    \centering
    \includegraphics[scale=.5]{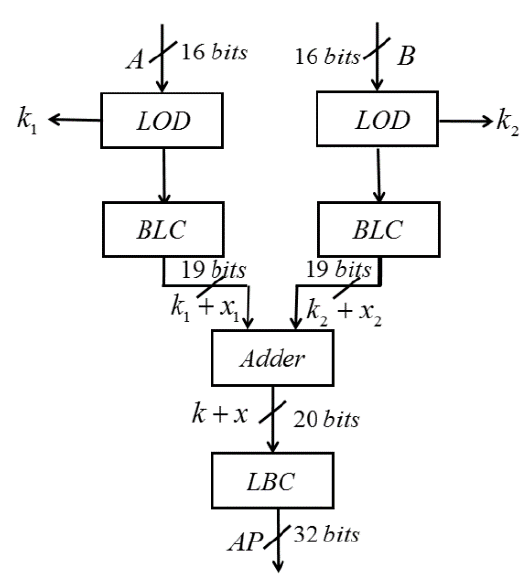}
    \caption{Procedure for the Mitchell’s algorithm~\cite{mitchell1962computer}.}
    \label{fig:algorithm1}
\end{figure}

\begin{figure}[!htbp]
    \centering
    \includegraphics[width=.7\textwidth]{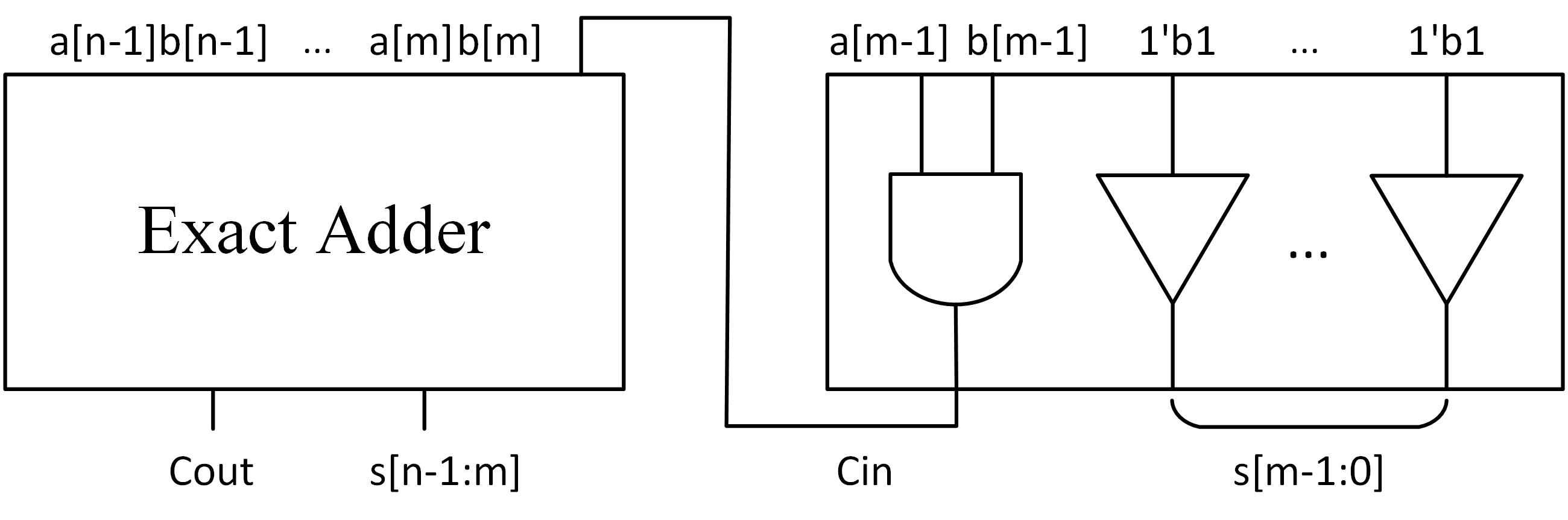}
    \caption{Architecture of an $n$-bit set-one-adder~\cite{liu2017design_log}.}
    \label{fig:algorithm2}
\end{figure}

According to Eq.~\eqref{eq:log_multiply}, the implementation based on \hjdel{the }Mitchell’s algorithm~\cite{mitchell1962computer} requires \wydel{leading one detector (LOD), binary-logarithm converter (BLC), adder, and logarithm-binary converter (LBC)}leading one detectors (LODs), binary-logarithm converters (BLCs), adders, and logarithm-binary converters (LBCs).
The procedure for Mitchell’s algorithm is illustrated in Fig.~\ref{fig:algorithm1} for a 16$\times$16 multiplier. To reduce the implementation complexity, the logarithmic expression in Eq.~\eqref{eq:log_multiply} \qdel{can be approximated by}\qadd{is approximated as}:
\begin{equation}\label{eq:log_appr}
\log_2 (x+1)\approx x, 0\leq x<1\;.
\end{equation}
Then we have: $A\times B \approx 2^{k_1+k_2+x_1+x_2}=2^{k_1+k_2}\times 2^{x_1+x_2}$.
Based on the carry of $x_1+x_2$, Eq.~\eqref{eq:log_appr} can be further approximated as:
\begin{equation}\label{eq:log_appr_2case}
A\times B\approx \begin{cases}
                    2^{k_1+k_2}(x_1+x_2+1), & \text{$x_1+x_2<1$}\;,\\
                    2^{k_1+k_2+1}(x_1+x_2), & \text{$x_1+x_2\geq 1$}\;.
                \end{cases}
\end{equation}
\wyadd{By subtracting the approximate multiplication from the exact multiplication, we can calculate the error. Taking the case of $x_1+x_2<1$ as an example, the error of Eq.~\eqref{eq:log_appr_2case} is given by:}
\begin{equation}
\begin{split}
Error &= A\times B - 2^{k_1+k_2}(x_1+x_2+1) \\ &= 2^{k_1+k_2}(1+x_1)(1+x_2)- 2^{k_1+k_2}(x_1+x_2+1)\\
          & = 2^{k_1+k_2}x_1x_2
\end{split}
\end{equation}
It is noted that the error term of $2^{k_1+k_2}x_1x_2$ has the same structure as $A\times B=2^{k_1+k_2}(1+x_1)(1+x_2)$. Then we can repeat the approximation procedure to compute $2^{k_1+k_2}x_1x_2$, which indicates an iterative process to achieve a higher accuracy using logarithm-based approximation~\cite{WOS000287716800004,ahmed2016iterative,mitchell1962computer,kim2019cost}. \wydel{In \cite{ahmed2016iterative}, the iterative approximation for $x_1+x_2\geq 1$ has been explored together with a truncation scheme}\wyadd{In \cite{ahmed2016iterative,kim2019cost,ahmed2019improved,yin2020design}, the iterative approximation has been explored together with a truncation scheme}. Liu \textit{et al.} further investigated the logarithm-based approximate multipliers using different approximate adders and find that multipliers \qdel{equipped with}\wadd{using} set-one-adders (SOAs) can achieve a higher accuracy~\cite{liu2018design}. As shown in Fig.~\ref{fig:algorithm2}, an SOA consists of one approximate adder for the lower $m$ bits and one exact adder for the higher $(n-m)$ bits. The approximate adder always sets the lower $m$ \qadd{sum }bits to logic 1 and hence, results in over-estimation. Such an over-estimation is particularly designed to compensate for the accuracy loss of a logarithm-based approximate multiplier, as Mitchell's algorithm always underestimates the multiplication result. Similar compensation schemes have been introduced in \cite{saadat2018minimally,ansari2019hardware, ansari2020improved} to improve the average error introduced by Mitchell's algorithm at the cost of area and power consumption. For example, Ansari \textit{et al.} proposed an improved logarithm transformation algorithm, which introduces double-sided errors to reduce the error bias~\cite{ansari2020improved}.

\begin{figure}[!htbp]
    \centering
    \includegraphics[scale=.35]{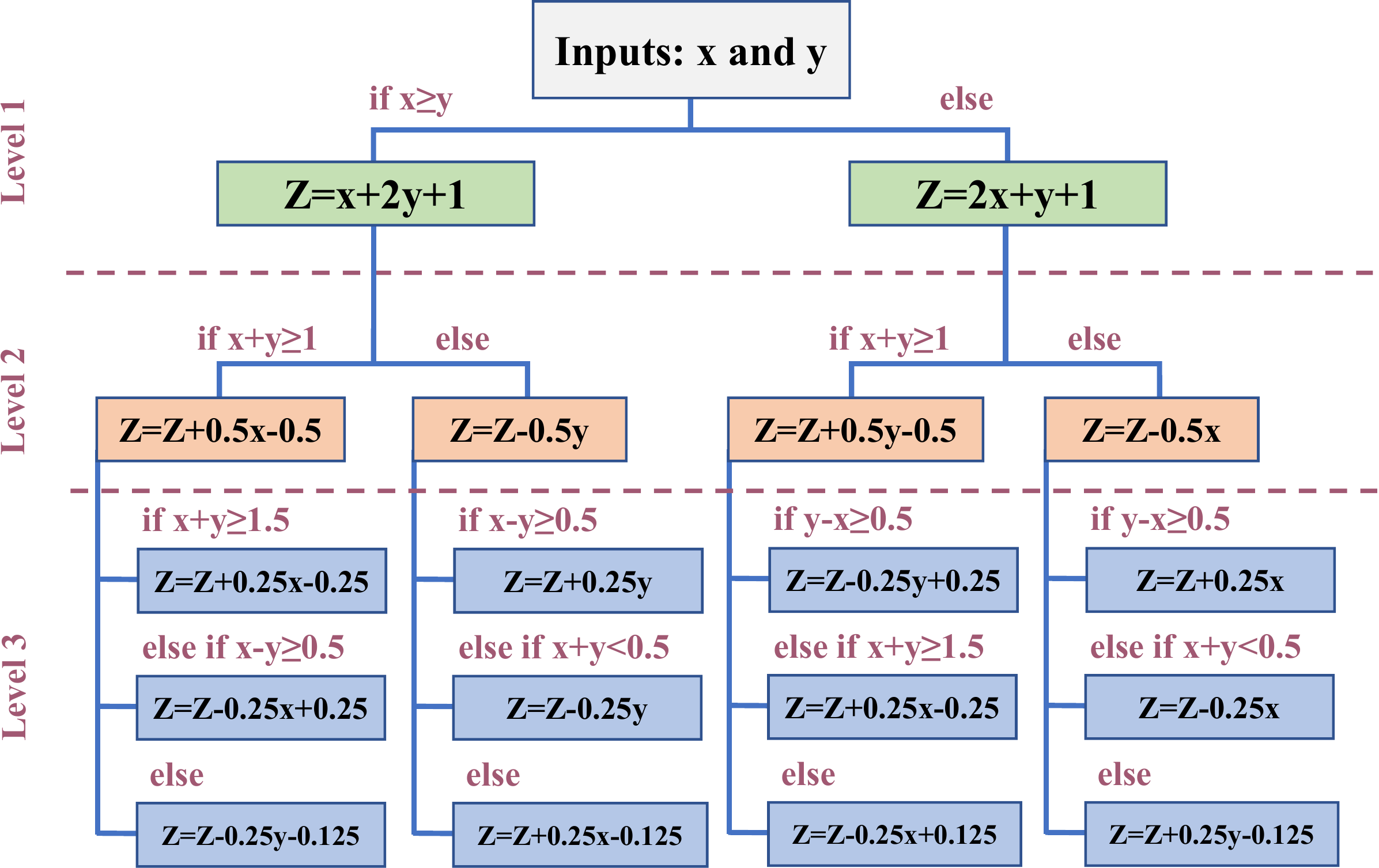}
    \caption{Computation procedure using the linear approximation in \cite{imani2019approxlp}.}
    \label{fig:algorithm3}
\end{figure}

\subsection{Approximation with Linearization}
Multiplication is a nonlinear operation that can be implemented with additions and compressions. In mathematics, it is a natural idea to approximate a nonlinear curve with a piece-wise linear function. Thus, researchers have attempted to use linear arithmetic operations to approximate the nonlinear multiplication~\cite{imani2019approxlp, chen2020optimally}. It is noted that, while logarithm-based approximate designs are based on Eq.~\eqref{eq:log_appr_2case}, it is actually a special case of linearization.

\begin{figure}[!htbp]
    \centering
    \includegraphics[scale=.07]{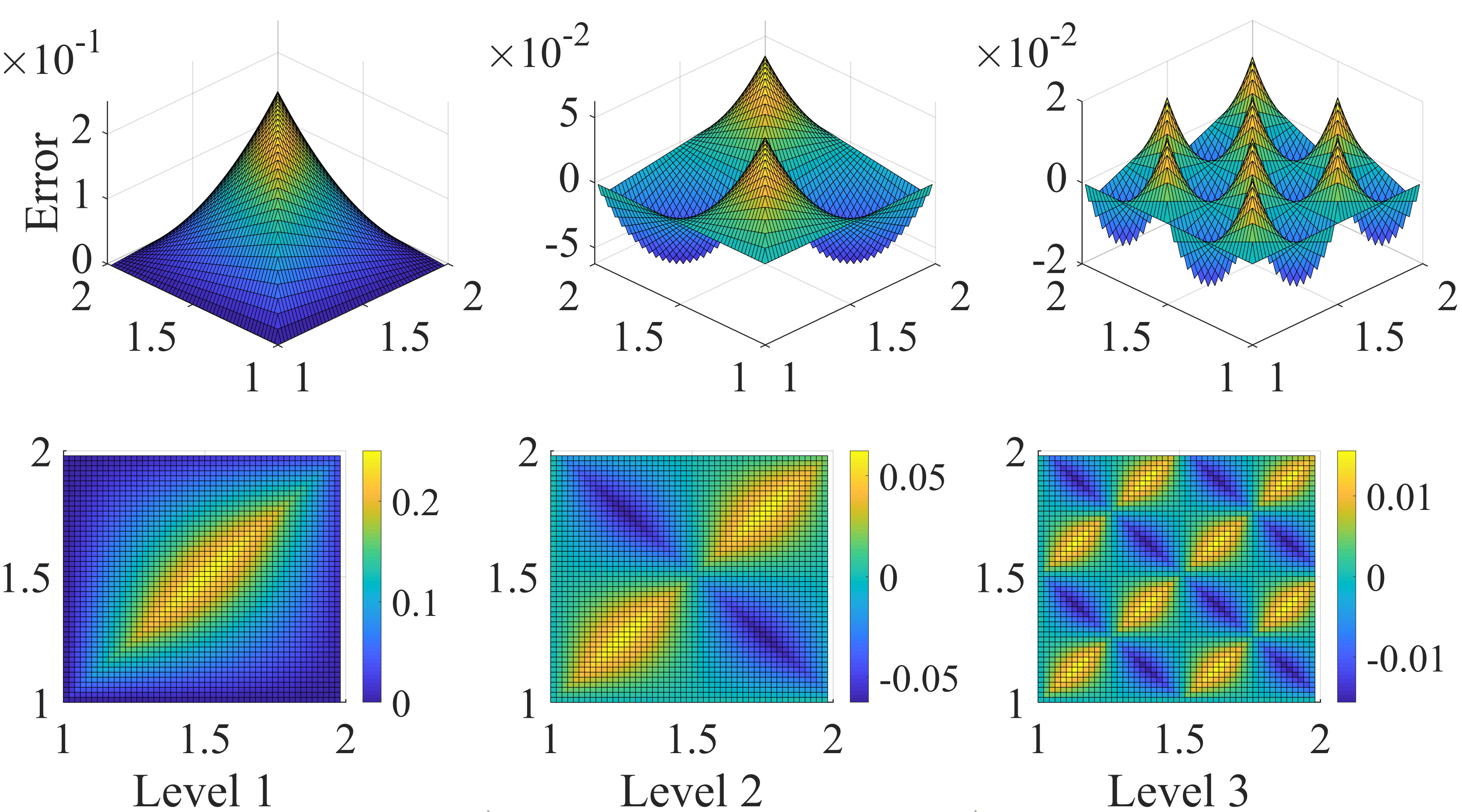}
    \caption{Error distributions of ApproxLP at different approximation levels~\cite{imani2019approxlp}.}
    \label{fig:approxlp2d}
\end{figure}

Without loss of generality, the multiplication can be considered as a function of two variables, whose linear approximation can be \hjdel{always }expressed as:
\begin{equation}\label{eq:linear_approx}
f=xy \approx f_{approx}=ax+by+c;,
\end{equation}
where $x$ and $y$ are the input operands\qdel{;}\qadd{, and} $a$, $b$, and $c$ are \hjdel{the }coefficients. In \cite{imani2019approxlp}, an iterative linear approximation for floating-point multiplication is proposed to approximate the multiplication according to Eq.~\eqref{eq:linear_approx}.
For the mantissas of normalized floating-point numbers, the range of the product is $[1,2)\times[1,2)$, which is a square domain. By appropriately partitioning the domain into smaller sub-domains and assigning a proper linear function to each, the original nonlinear surface for the multiplication can be approximated by a series of piece-wise linear functions, one for each sub-domain. Fig.~\ref{fig:algorithm3} summarizes the computation procedure called ApproxLP using the linear approximation in \cite{imani2019approxlp}. It is clear that the accuracy can be improved by partitioning more sub-domains, the number of which grows exponentially with the \qdel{approximate}\qadd{approximation} level. Thus, the efficiency of ApproxLP in \cite{imani2019approxlp} actually quickly degrades with a larger approximation level\qdel{s}. Moreover, the comparators used for each level in Fig.~\ref{fig:algorithm3} also introduce non-trivial delay overhead. Fig.~\ref{fig:approxlp2d} plots the error distributions of mantissa multiplications in ApproxLP for different \qdel{approximate}\qadd{approximation} levels with $x$-axis and $y$-axis representing the mantissa range\wydel{, which are symmetric over 0 and hence result in a zero average error}.

To reduce the number of comparators, Chen  \textit{et al.} proposed to partition the input domain into identical smaller square sub-domains~\cite{chen2020optimally}. For one level higher, each domain (or sub-domain\qdel{s}) is further partitioned into four identical smaller ones. With such an iterative process, there are $4^n$ sub-domains for \qdel{level $n$}\qadd{level-$n$} approximation. For a rectangular domain $[x_1,x_2 ]\times [y_1,y_2]$, the optimal coefficients to minimize the mean square error (MSE) between $f_{approx}=ax+by+c$ and $f=xy$ are~\cite{chen2020optimally}:
\begin{equation}\label{eq:optimal_coeffs}
    a = \frac{y_1+y_2}{2}, b = \frac{x_1+x_2}{2}, c = -ab.
\end{equation}

\begin{figure}[!htbp]
    \centering
    \includegraphics[width=0.7\textwidth]{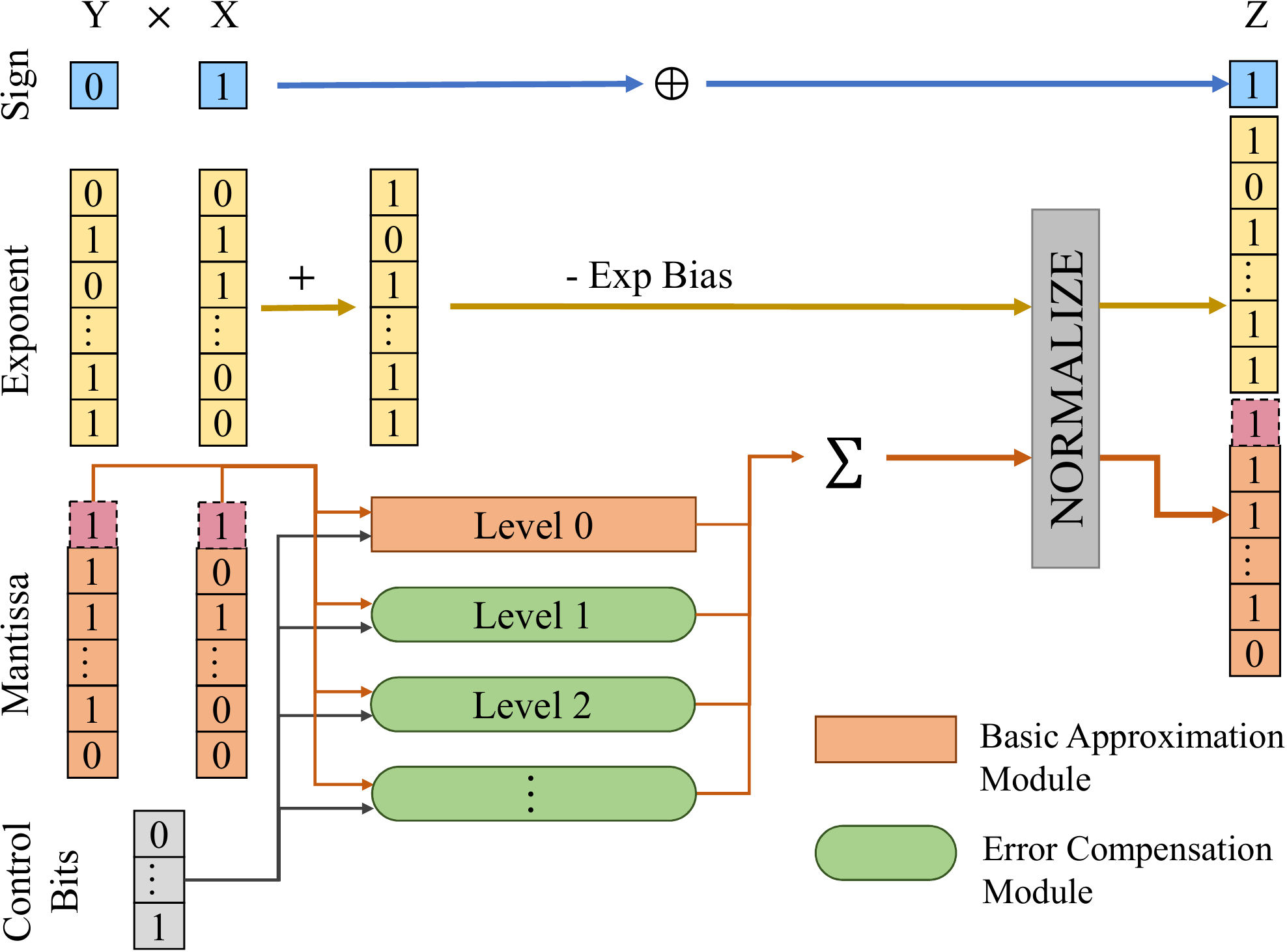}
    \hfil
    \caption{Architecture of the approximate multiplier OAM in \cite{chen2020optimally}.}
    \label{fig:oaum_struct}
\end{figure}

Fig.~\ref{fig:oaum_struct} demonstrates the multi-level approximate multiplier architecture of an optimally approximated multiplier (OAM) in \cite{chen2020optimally}. In the figure, Level 0 is denoted as the basic approximation module, which provides an initial estimation $f_{approx}^0$, while the deeper levels act as error compensation\wyadd{ (given in Eq.~\eqref{eq:optimized_output_of_each_level})} to gradually improve the overall accuracy. Thus, the run-time configurability can be easily realized by specifying the desired depth. Unlike ApproxLP~\cite{imani2019approxlp}, \hjdel{the }comparators are no longer needed for OAM~\cite{chen2020optimally}. Thus, the delay of OAM can be significantly reduced when compared to ApproxLP even for a similar number of sub-domains. 

Since the two coefficients of $a$ and $b$ \wadd{are}\qdel{lie in} the middle points of the intervals where the operands belong to, a circuit-friendly implementation can be achieved for the error compensation at each level $n$ as\hjdel{ in Eq.~\eqref{eq:optimized_output_of_each_level}}~\cite{chen2020optimally}:
\begin{align}\label{eq:optimized_output_of_each_level}
    \Delta f_n  = & \wyadd{f^{n}_{approx}-f^{n-1}_{approx} }\notag \\
                = & \bigg\{  \Big[\big(x[n]?(1):(-1)\big)\times (y-\hat{y_{n-1}})\Big]  \notag \\
                 + & \Big[\big(y[n]?(1):(-1)\big)\times (x-\hat{x_{n-1}})\Big] \bigg\} \gg (n+1) \\
                 + & \Big[\big(x[n]\oplus y[n]\big)?(1):(-1)\Big] \gg (2n+2), \notag
\end{align}
\qadd{where ``$?\ :$'' is the conditional operator, $\oplus$ is the XOR operator, $\gg$ is the right shift operator, $x[n]$ is the $n^{th}$ \qdel{bit}\wadd{most significant bit (MSB)} of mantissa $x$\qdel{ indexed from the most significant bit (MSB)}, and $\hat{x_{n-1}}$ preserves the $(n-1)$ MSBs of mantissa $x$ with an extra bit $1$ at the $n^{th}$\wadd{ MSB} position.}
Since the number of right shifts is pre-determined at each level, the right shift operation does not require\qadd{ an} additional circuit\qdel{s} to implement.
Thus, the number of operations at each level for the OAM is reduced to 5, including one XOR operation, two arithmetic negations, and two additions, which results in a constant area complexity, whereas ApproxLP has an area complexity of $O(4^n)$~\cite{chen2020optimally}. In addition, a hardware-friendly implementation of the OAM is provided in \cite{chen2021TC}.

The errors of the OAM~\cite{chen2020optimally} are \qdel{reported}\wadd{shown} below in terms of maximum \qdel{absolute error (MaxAE)}\wadd{error distance (MaxED)}, mean \qdel{absolute error (MAE)}\wadd{error distance (MED)}, and MSE for the approximation level~$n$:
\begin{equation}\label{eq:optimal_error_metric}
    \textrm{\qdel{MaxAE}\wadd{MaxED}} = \frac{1}{4^{n+1}},
    \textrm{\qdel{MAE}\wadd{MED}} = \frac{1}{4^{n+2}},
    \textrm{MSE} = \frac{1}{9\times 16^{n+1}}.
\end{equation}
The OAM~\cite{chen2020optimally} can produce a zero-mean error distribution, which is an appealing feature for applications with consecutive multiply-accumulate operations.

\subsection{Hybrid Approximation}
Several approximate designs combine the multipliers with different precisions together to adapt to varying accuracy requirements~\cite{imani2017cfpu, imani2018rmac}.
These designs are considered as using a hybrid approximation in this article.
For example, accurate and approximate multipliers are combined to adjust the computational accuracy by selecting the appropriate multiplier in \cite{imani2017cfpu}.
For the approximate multiplier, after detecting the number of consecutive 1’s or 0’s in the mantissa, the mantissa can then be rounded to 1 or 2, which converts the multiplication to a shift operation. If higher precision is required, the accurate multiplier is then invoked to conduct the calculation. Reference \cite{imani2018rmac} used the sum of two mantissas to approximate the multiplication. A tuning strategy is proposed to decide the working mode of the multiplier by detecting the number of consecutive bits in the inputs. However, such methods heavily rely on an accurate or high-precision multiplier, which significantly increases the circuit area.

\section{Approximate Multipliers with Architecture-level Approximation}\label{sec:arch}
As discussed in \qdel{section~\ref{sec:back}}\qadd{Section~\ref{sec:multiplier_binary}}, a conventional multiplier typically involves three stages, $i.e.$, data input, partial product generation, and accumulation. \qdel{To}\qadd{In addition}, an encoding stage is often deployed to further reduce the number of partial products, $e.g.$, Booth encoding~\cite{de1996low}. For approximate multipliers based on such an architecture, approximations can be introduced into any of the four aforementioned stages. 
\begin{figure}[!htbp]
\centering
\subfigure[]{
\begin{minipage}[h]{\linewidth}
\centering
\includegraphics[width=0.6\textwidth]{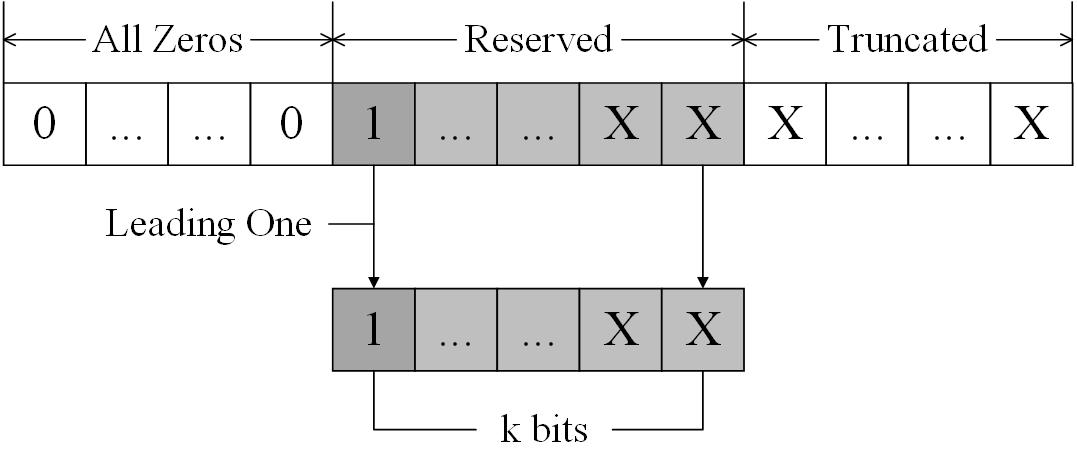}
\end{minipage}%
}%

\subfigure[]{
\begin{minipage}[h]{\linewidth}
\centering
\includegraphics[width=0.7\textwidth]{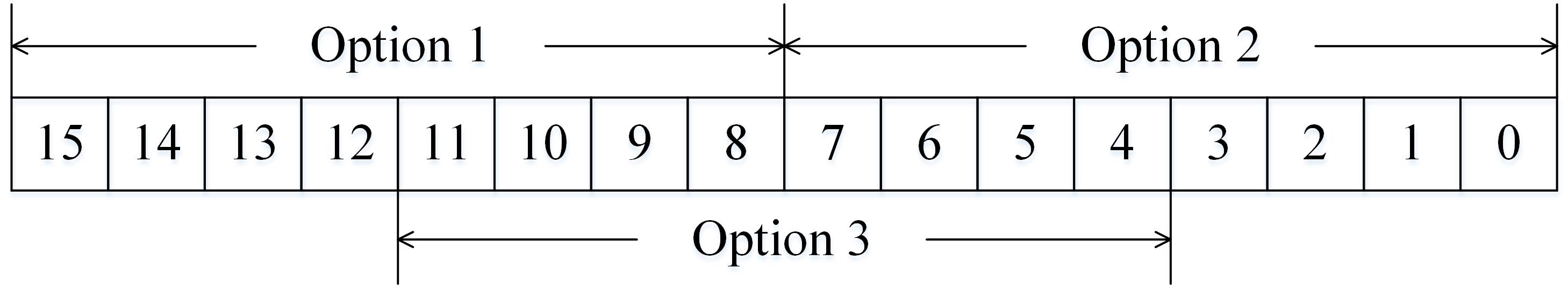}
\end{minipage}%
}%
\centering
\caption{(a) An example of DSM truncation; (b) An example of SSM truncation.}\label{fig:arch_DSM_SSM}
\end{figure}

\subsection{Approximation at Input }
It is simple yet effective to introduce approximation in the input for approximate multipliers. For example, a few least significant bits (LSBs) can be removed to reduce the input bit-width, which has a lower impact on the result than removing the MSBs~\cite{hashemi2015drum,vahdat2019tosam,vahdat2017letam,narayanamoorthy2014energy}. In general, there are two types of  input truncation schemes, the dynamic segment method (DSM) and the static segment method (SSM)~\cite{narayanamoorthy2014energy}. DSM segments the data according to the leading one, while SSM is based on a given segmentation option. 
For example, in Fig.~\ref{fig:arch_DSM_SSM}(a), DSM keeps $k$ consecutive bits from the \qdel{first non-zero bit}\qadd{leading one} of an unsigned number. The parameter $k$ determines the level of accuracy loss for an approximate multiplier. \qdel{On the other hand}\qadd{In contrast}, SSM in Fig.~\ref{fig:arch_DSM_SSM}(b) provides a few pre-determined ($i.e.$, \textit{static}) options when truncating the input\hjdel{ data}. The options can be either leading $k$ bits (option 1) or last $k$ bits (option 2), as suggested in~\cite{narayanamoorthy2014energy}. It is also possible to keep the bits in the middle (option 3) as a trade-off (Fig.~\ref{fig:arch_DSM_SSM}(b)). Unlike DSM, SSM requires fewer hardware resources but may include more redundant bits. As shown in the Dynamic Range Unbiased Multiplier (DRUM) in \cite{hashemi2015drum}, the additional support to DSM requires two extra Leading-One Detectors (LODs), two extra encoders, and one extra barrel shifter. \qdel{And}\qadd{Moreover,} the last bit of the reserved part is always set to one \qdel{aimed at}\qadd{for} unbiasedness.

\subsection{Approximation at Partial Product Generation}
\label{subsec:apx-ppg}
\begin{figure}[!htbp]
    \centering
    \includegraphics[scale=.7]{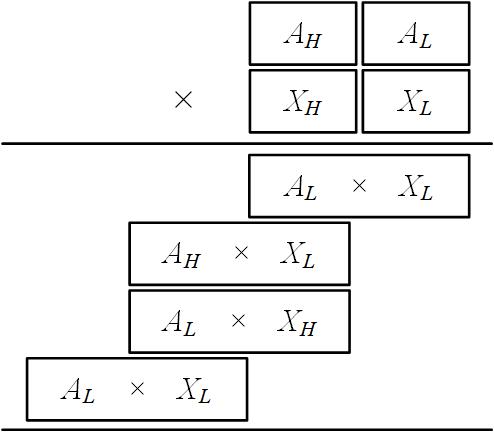}
    \caption{An example of UDM in \cite{kulkarni2011trading}.}
    \label{fig:arch3}       
\end{figure}

\wyadd{One of the techniques to bring approximation into the partial product generation stage is breaking large-sized multipliers down to smaller ones and then adopting approximate basic multiplication modules~\cite{kulkarni2011trading,AWTM_6783335,rehman2016architectural}.} Kulkarni \textit{et al.} proposed an under-designed multiplier (UDM) architecture, which partitioned both multiplier and multiplicand into 2 parts recursively until the partial products can be generated through $2 \times 2$ multiplication modules~\cite{kulkarni2011trading}. The basic partition scheme is presented in Fig.~\ref{fig:arch3}. 
\wyadd{In order to explore the design space of the UDM architecture, Rehman \textit{et al.} adopted a depth-first search (DFS) algorithm to find a solution within constraints based on basic modules including Pareto-optimal approximate $2\times 2$ multiplication modules and Pareto-optimal approximate 1-bit full adders~\cite{rehman2016architectural}.}

Another alternative to partial product generation is to introduce an intermediate variable to replace the partial products (\textit{a.k.a.} altered partial product (APP)) and then conduct approximations~\cite{venkatachalam2017design,yang2018low,yang2017low}. As discussed in \qdel{section~\ref{sec:back}}\qadd{Section~\ref{sec:multiplier_binary}}, a partial product can be generated using \qadd{an }AND gate\qdel{s}:
\begin{equation}
pp_{m,n}=x_m \cdot y_n\;,
\end{equation}
where $x_m$ and $y_n$ represent the $m^{th}$ and the $n^{th}$ bit of the two inputs $x$ and $y$, respectively. Similar to a carry look-ahead adder, the propagate and the generate signals can be defined as:
\begin{equation}
    p_{m,n}=pp_{m,n}+pp_{n,m}\;,
\end{equation}
\begin{equation}
    g_{m,n}=pp_{m,n}\cdot pp_{n,m}\;.
\end{equation}
Since the generate signals are possibly all 0's, they can be compressed column-wise using an OR gate. The propagate signals can be computed using approximate adders to achieve a more compact design than the original multiplier. Yang \textit{et al.} employed a similar idea of using two signals of approximate sum and error recovery vector to approximate the partial product~\cite{yang2018low}.

\subsection{Approximation at Accumulation}

In this section, we discuss the related works \qdel{about}\qadd{on} approximation at the accumulation stage. First\qadd{,} we focus on the basic modules \qdel{during this process}\qadd{used at this stage}, including approximate compressors and \qadd{approximate 1-bit }adders. Then\qadd{,} we review the works \qdel{about}\qadd{on} the allocation of different approximate compressors in the entire accumulation stage.

\subsubsection{Approximate Compressors}

\begin{figure}[!htbp]
\centering
\subfigure[]{\includegraphics[width=0.25\textwidth]{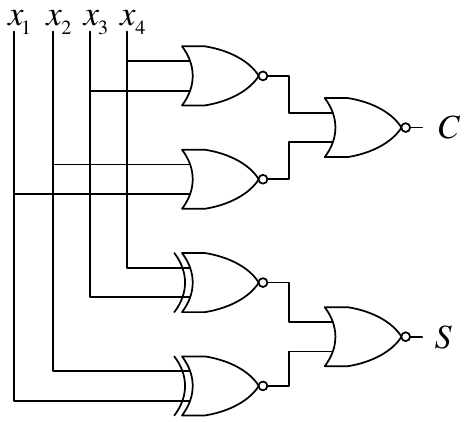}
\label{fig:Conventional AC_a}
}
\subfigure[]{
\includegraphics[width=0.3\textwidth]{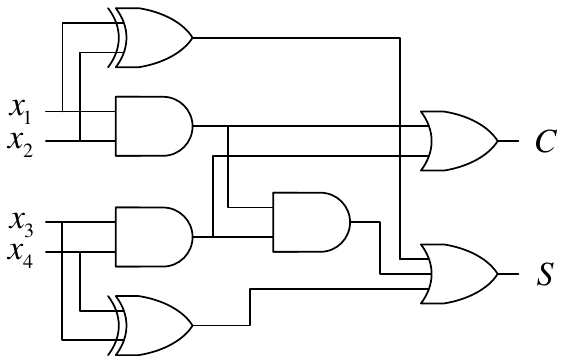}
\label{fig:Conventional AC_b}
}%
\centering
\caption{Two approximate 4-2 compressors proposed by (a) Momeni \textit{et al.}~\cite{Momeni2014}; and (b) Venkatachalam \textit{et al.}~\cite{venkatachalam2017design}, where $C$ is the carry signal\qdel{,}\qadd{ and} $S$ is the sum signal. \wdel{; (c) Strollo \textit{et al.}~\cite{Strollo2020}}}\label{fig:Conventional ACs}
\end{figure}

Instead of the exact compressors in Section 2.3, many \qdel{methods}\qadd{approximate compressor designs} have been proposed\qdel{ to design approximate compressors} to improve energy efficiency~\cite{Lin2013, Momeni2014, venkatachalam2017design, yang2015approximate, ha2017multipliers, Akbari2017, Ahmadinejad2019, Strollo2020, esposito2018approximate, yang2018low,tung2019low, marimuthu2016design, Xuan22, van2020fpga}.
\qdel{The approximate compressors can be divided}\qadd{We divide them} into two categories\qadd{, low-order and high-order approximate compressors}. 
\qdel{The first category are the}\qadd{The} low-order approximate compressors\qdel{, including}\qadd{ include} approximate 1-bit half adders, approximate 1-bit full adders, and approximate 4-2 compressors.
Taking the error rate into account, approximate 4-2 compressors with different approximation levels were designed to realize power-efficient Dadda multipliers~\cite{Momeni2014, yang2015approximate}.
One structure of the proposed approximate 4-2 compressors in ~\cite{Momeni2014} is shown in Fig.~\ref{fig:Conventional AC_a}.
Venkatachalam \textit{et al.} took the error difference into account to design an approximate 1-bit half adder, an approximate 1-bit full adder, and an approximate 4-2 compressor. The structure of the approximate 4-2 compressor is shown in Fig.~\ref{fig:Conventional AC_b}.
In order to provide more flexibility for \qadd{accuracy-power }trade-off, 
Akbari \textit{et al.} designed four approximate 4-2 compressors with different approximation levels, each of which can flexibly switch between the exact and the approximate operating modes~\cite{Akbari2017}.
\qadd{A survey on some approximate 4-2 compressors is presented in~\cite{Strollo2020} together with their performance comparison.}

The second category of approximate compressors is a high-order approximate compressor\qadd{, which has at least 5 inputs}. 
By modifying the truth table of the exact compressors, \qadd{an }approximate \qdel{4-2 and }5-2 compressor\qdel{s} with significantly fewer transistors \qdel{were}\qadd{was} proposed in~\cite{Ahmadinejad2019}. 
Tung \textit{et al.} proposed to design more general $n$-2 approximate compressors ($n \ge 5$) to further reduce the delay and power of the approximate multiplier~\cite{tung2019low}.
Instead of using a 2-bit output,
Marimuthu \textit{et al.}~\cite{marimuthu2016design} designed an approximate 15-4 compressor, where four different types of approximate 5-3 compressors were used as basic modules.
Esposito \textit{et al.} proposed more general approximate $n$-$\lceil n/2 \rceil$ compressors, where the outputs of these compressors are equally weighted~\cite{esposito2018approximate}.
Compared with the previously proposed unequally-weighted approximate compressors~\cite{Lin2013, Momeni2014, venkatachalam2017design, yang2015approximate, ha2017multipliers, Akbari2017, Ahmadinejad2019, Strollo2020,yang2018low,tung2019low, marimuthu2016design}, these equally-weighted approximate compressors have a smaller area, and their errors are easier to \qdel{estimate using mathematical model}\qadd{analyze\hjdel{ mathematically}}.

\begin{figure}[!htbp]
\centering
\subfigure[]{\includegraphics[width=0.22\textwidth]{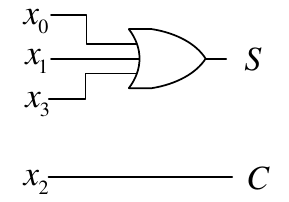}
\label{fig:MinAC_a}
}
\subfigure[]{
\includegraphics[width=0.28\textwidth]{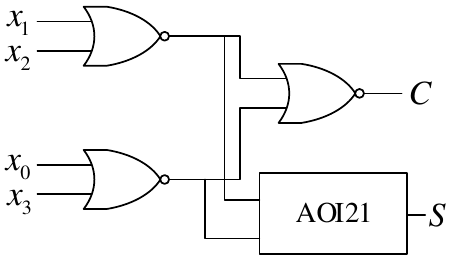}
\label{fig:MinAC_b}
}%
\centering
\caption{The optimized structures by MinAC for the two approximate 4-2 compressors in Fig.~\ref{fig:Conventional ACs}: (a) Optimized design for the approximate compressor in \cite{Momeni2014}; and (b) Optimized design for the approximate compressor in \cite{venkatachalam2017design}\wdel{; (c) MinAC design with the error bound as the MED of Strollo~\textit{et al.}~\cite{Strollo2020}}.}\label{fig:MinAC ACs}
\end{figure}

However, all the prior approximate compressors were manually designed. To provide a \hjdel{more }general solution to approximate multiplier design and optimize over a number of hardware cost metrics, \qadd{Wang \textit{et al.}} proposed\qdel{s} a method called MinAC to automatically generate minimal-area approximate 4-2 compressors~\cite{Xuan22}.
MinAC formulates the approximate circuit design as an exact synthesis problem~\cite{Haaswijk20s}, which \qdel{formulates}\qadd{introduces} proper encoding variables and constraints to determine the connections of the gates, the gate functions, and the output gates of the circuit.
For example, for any approximate 4-2 compressor in Fig.~\ref{fig:Conventional ACs}, its error metric such as \qdel{the mean error distance (MED)}\wadd{the MED} can be calculated and denoted as $e_b$. Then, MinAC can be applied to synthesize an area-optimal approximate 4-2 compressor with the MED no larger than $e_b$.
Fig.~\ref{fig:MinAC ACs} shows the structures of the two area-optimal approximate 4-2 compressors obtained by MinAC. Obviously, compared \qdel{with}\qadd{to} the original designs in Fig.~\ref{fig:Conventional ACs}, MinAC can synthesize \qdel{the corresponding }approximate 4-2 compressors with much fewer gates and smaller areas.

\subsubsection{Allocation of Approximate Compressors in the Compressor Tree}

\begin{figure}[!htbp]
    \centering
    \includegraphics[width=.8\textwidth]{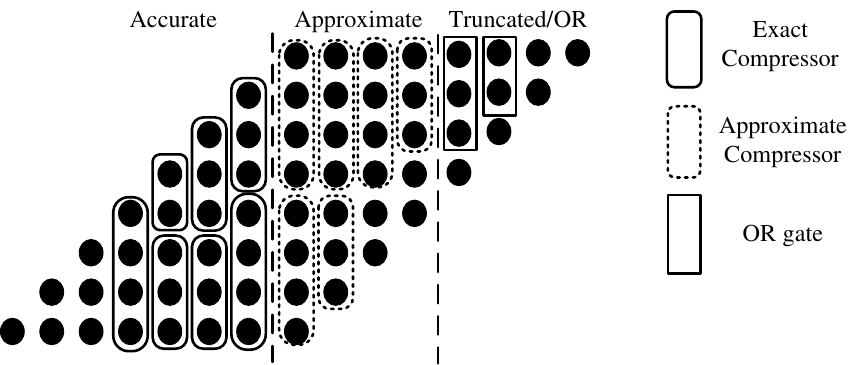}
    \caption{An example of partial product array that is divided into three groups with different levels of approximation.}
    \label{fig:arch5}
\end{figure}

\begin{figure}[!htbp]
    \centering
    \includegraphics[width=.6\textwidth]{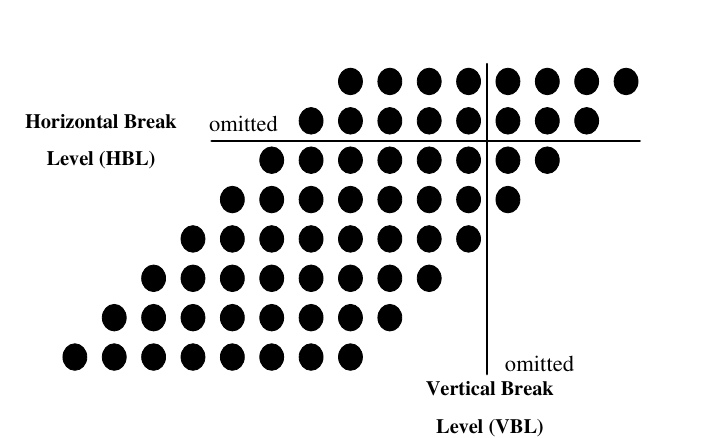}
    \caption{An example of broken-array multiplier (BAM)~\cite{mahdiani2009bio}.}
    \label{fig:BAM}       
\end{figure}

In the conventional multipliers, exact compressors, such as half adders, full adders, and\hjdel{\qadd{ exact}} 4-2 compressors, are applied to accumulate the partial product array.
Researchers have proposed \qdel{reduction schemes for allocating compressors efficiently}\qadd{efficient \hjdel{compressor }allocation schemes to use as few compressors as possible during the accumulation stage}\qdel{, such} in order to minimize area and delay.
However, the traditional \A{s} are not applicable to approximate multipliers as they do not \qdel{account}\qadd{consider} the accumulated errors in the tree structure.
\qdel{In~\cite{venkatachalam2017design}, it}\qadd{To address this issue, Venkatachalam~\textit{et al.}} proposed a reduction scheme \qdel{to completely take advantage of}\qadd{that uses} approximate compressors to compress the partial product array to two rows, which \qdel{were}\qadd{are} then added up by \qdel{using }an RCA~\cite{venkatachalam2017design}.
Jiang \textit{et al.} further proposed to ignore the carry signals in adders to reduce the critical path delay~\cite{Jiang19}. However, the error may become non-negligible if completely relying on approximate compressors during accumulation. Thus, many researchers proposed to separate the partial product array column-wise\hjdel{ly} to two or three groups, as shown in Fig.~\ref{fig:arch5}. Due to the \hjdel{different significance of different groups}difference in the significance of groups, each group can be \qdel{introduced into}\qadd{assigned with} different levels of approximation~\cite{esposito2018approximate,Wang20,yang2018low,ha2017multipliers,tung2019low,yang2015approximate,van2020fpga,cho2004design,song2007adaptive,wang2009high}. While the first group is always allocated with exact compressors, the second group can be allocated with either approximate compressors using equally weighted outputs~\cite{esposito2018approximate,Wang20}\hjdel{,} or high-order approximate compressors with error recovery~\cite{ha2017multipliers, tung2019low}. For the last group, one straightforward approach is to completely ignore \qdel{the group}\qadd{it}~\cite{yang2015approximate,yang2018low}. To improve the accuracy, \qadd{Tung \textit{et al.}} proposed to use OR gates for the last group~\cite{tung2019low}. To provide more flexibility, Mahdiani \textit{et al.} \hjdel{proposed \qdel{to}\qadd{the broken-array multiplier (BAM), which} }divided the partial product array into four groups through horizontal and vertical slicing in a broken-array multiplier (BAM), as shown in Fig.~\ref{fig:BAM}~\cite{mahdiani2009bio}. The partial products on the right of the vertical break level (VBL) or above the horizontal break level (HBL) are then ignored. In other words, only the partial products on the bottom left are used for calculation. Apparently, the approximation level can be adjusted by tuning VBL and HBL.

However, most prior works still allocate approximate compressors in an ad hoc manner or limited to some specific types of compressors. It is  desirable to automate the allocation for different kinds of approximate compressors. \qdel{Reference~\cite{Xiao22}}\qadd{For this purpose, Xiao \textit{et al.}} proposed a general framework for approximate compressor allocation called OPACT, which converts the allocation problem to \qadd{an }integer programming (IP)\qadd{ problem}~\cite{Xiao22}. The proposed framework not only accounts for the trade-off between area and accuracy, but also optimizes the connection order of different compressors.

\subsection{Approximation at Booth Encoding}
\label{sec:Booth-encoding}
Booth encoding is often used to reduce the number of partial products, which can be generated in parallel at the cost of additional area. The radix-4 Booth algorithm is a common option deployed for high bit-width multipliers~\cite{lin2004design}. Qian \textit{et al.} proposed an approximate Wallace-Booth multiplier with approximate modified Booth encoding (MBE), approximate 4-2 compressors, and approximate Wallace tree~\cite{qian2016design}.
In addition to the radix-4 algorithm, the radix-8 Booth algorithm is also widely used to further reduce the number of partial products. \wydel{However, Radix-8 algorithm demands odd multiples and hence needs additional adders.}However, the encoder in a radix-8 algorithm may generate odd values including 3 and $-3$ and hence needs additional adders to compute the odd multiples of multiplicands when generating the partial products. This step leads to an increased delay. \qdel{To reduce the increased partial product generation delay in Radix-8 algorithm}\qadd{To reduce the delay}, Jiang \textit{et al.} suggested an approximate adder to generate the odd multiples for multiplication~\cite{jiang2015approximate}, which can reduce the delay of carry propagation as a trade-off between speed and accuracy.
\section{Approximate Multipliers with Circuit-level Approximation}\label{sec:circuit}
This section reviews a few general circuit-level approximation techniques applicable to various architectures or algorithms, including Boolean rewriting, gate-level pruning, \wadd{evolutionary circuit design, }and voltage over-scaling (VOS)\hjdel{, which are applicable to various architectures or algorithms}.  

\subsection{Boolean Rewriting}

\begin{figure}[!htbp]
    \centering
    \includegraphics[scale=.6]{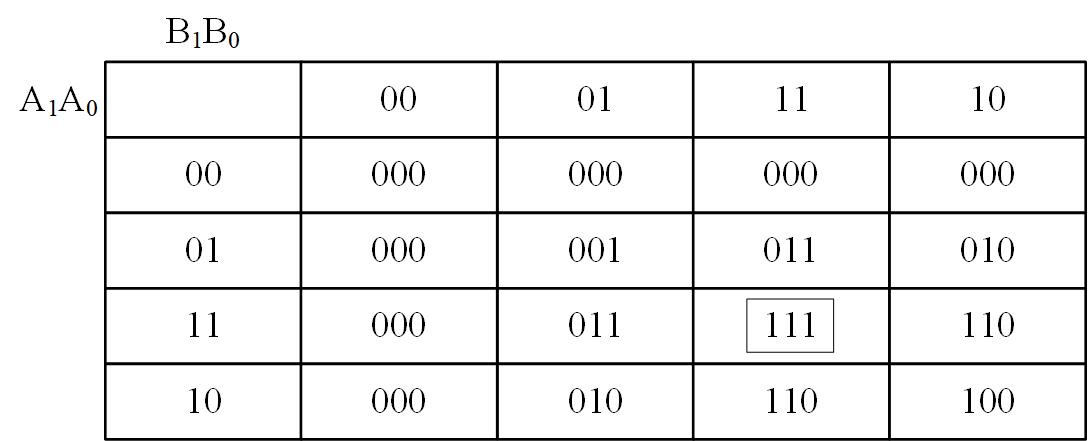}
    \caption{An example of modifying K-map to achieve more compact design~\cite{kulkarni2011trading}: the \qdel{original term of '1001' can be changed to '111' in the table}\qadd{accurate output `1001' at the cell $(A_1 A_0 B_1 B_0)=(1111)$ is changed to `111'}.}
    \label{fig:arch4}       
\end{figure}

\begin{figure}[!htbp]
\centering
\subfigure[]{
\begin{minipage}[t]{0.5\linewidth}
\centering
\includegraphics[width=0.7\textwidth]{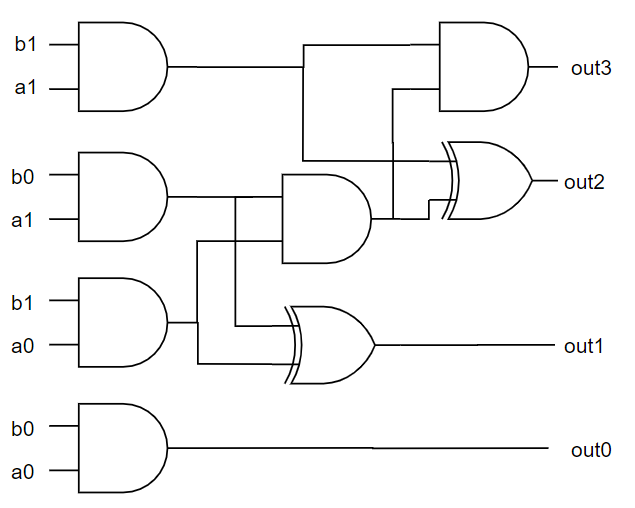}
\end{minipage}%
}%
\subfigure[]{
\begin{minipage}[t]{0.5\linewidth}
\centering
\includegraphics[width=0.7\textwidth]{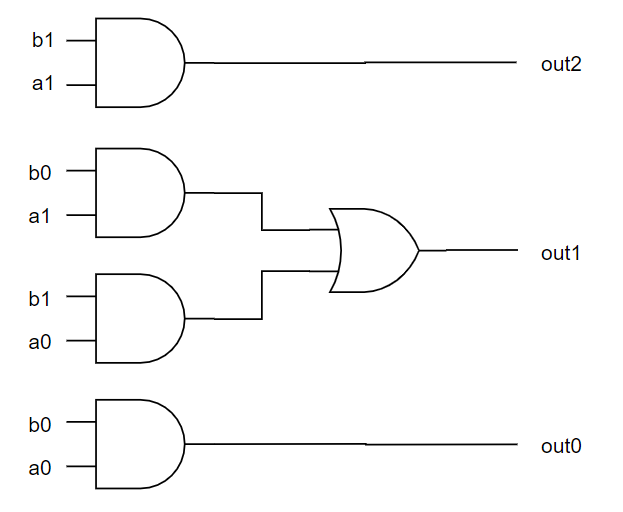}
\end{minipage}%
}%
\centering
\caption{Comparison on the implementations of the $2\times 2$ multiplier module: (a) Accurate logic implementation; (b) Approximate logic implementation~\cite{kulkarni2011trading}.}
\label{fig:arch4_circ} 
\end{figure}

Boolean rewriting modifies Boolean algebra expressions to simplify the circuit and is frequently utilized to approximate the basic modules in a multiplier.
Karnaugh map (K-map) modification is a commonly used technique in this category. The basic idea of the K-map is to group the adjacent \qdel{squares}\qadd{cells} with the same logic values as much as possible.
However, it is quite common in practice that one or more \qdel{squares}\qadd{cells} cannot be grouped, causing additional logic\qdel{s} and hence area. Thus, the approximation to the K-map can be introduced by modifying the adjacent \qdel{square}\qadd{cells} to the same value so \qdel{as to group the squares and}\qadd{that they can be grouped to} obtain a more compact representation.
For example, the approximate multiplier UDM discussed in \qdel{section~\ref{sec:arch}}\qadd{Section~\ref{subsec:apx-ppg}} is comprised of a $2\times 2$ multiplication module~\cite{kulkarni2011trading}, which can be designed through K-map modification. By modifying the K-map as in Fig.~\ref{fig:arch4}, the basic block can act as both a partial product generator and a compressor with an error rate of $1/16$~\cite{kulkarni2011trading}. As shown in Fig.~\ref{fig:arch4_circ}, when compared to the accurate logic implementation, the approximate implementation needs much fewer logic gates (37.5\% reduction) with a shorter critical path. 

\begin{figure}[!htbp]
    \centering
    \includegraphics[width=.6\textwidth]{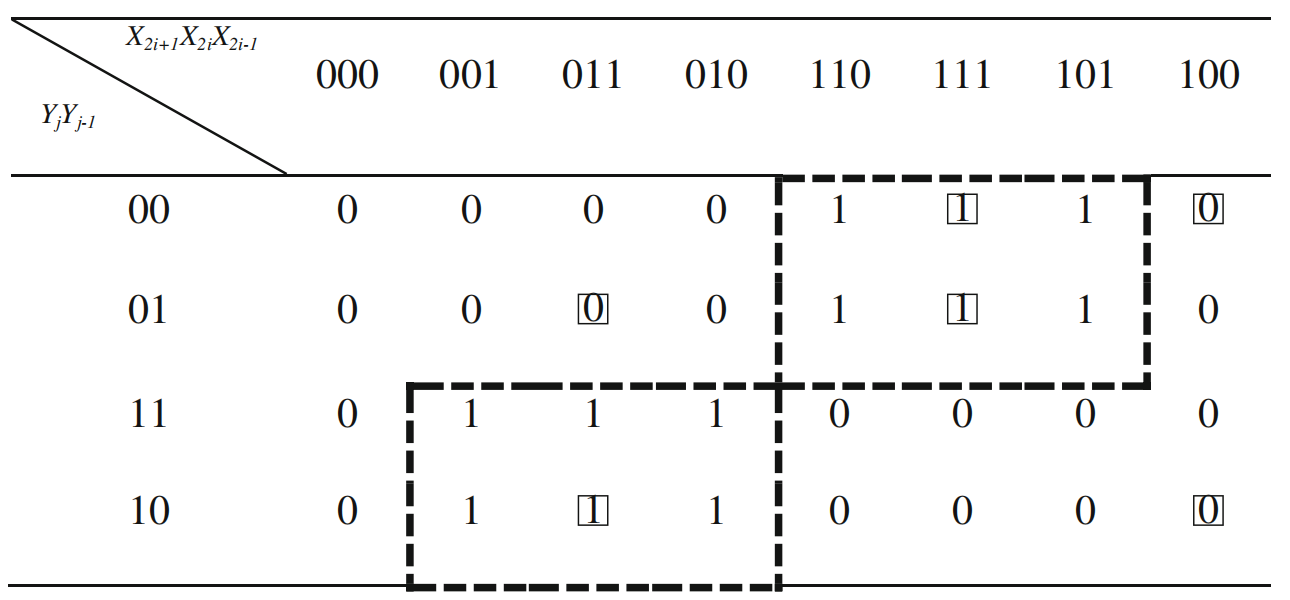}
    \caption{K-map modification of AMBE~\cite{yin2018designs}: The \qdel{original 0's and 1's}\qadd{circled values in the table} are flipped to enable\qadd{ a} more compact representation\qdel{ in the table}.}
    \label{fig:AMBE}       
\end{figure}

\wyadd{Apart from basic multiplication modules~\cite{kulkarni2011trading,rehman2016architectural}, the} K-map modification can also be applied to other \qdel{arithmetic functions}\qadd{basic modules} in multipliers, such as adders~\cite{pabithra2018analysis,rehman2016architectural,alouani2017novel}, compressors~\cite{van2020fpga,ha2017multipliers}, and \qdel{booth}\qadd{Booth} encoding modules~\cite{qian2016design,yin2018designs,liu2017design}. For example, Yin \textit{et al.} used K-map modification to design an approximate modified Booth encoding (AMBE) module~\cite{yin2018designs}. With the modified K-map in Fig.~\ref{fig:AMBE}, the original expression for the modified Booth encoding algorithm:
\begin{equation}
PP_{j}=(X_{2i}\oplus X_{2i-1})(X_{2i+1}\oplus Y_{j})+\overline{(X_{2i}\oplus X_{2i-1})}(X_{2i+1}\oplus X_{i})(X_{2i+1}\oplus Y_{j-1})\;,
\end{equation}
can be simplified to~\cite{yin2018designs}: 
\begin{equation}
PP_{j}^{'}=(X_{2i}\oplus X_{2i-1})(X_{2i+1}\oplus Y_{j})\;.
\end{equation}

In addition to K-map modification, \qadd{a }Boolean expression\qdel{s} can also be directly changed to create \qadd{a }more compact expression, such as the approximate adders SOA in \cite{liu2017design_log} and lower-part-or adder (LOA) in \cite{mahdiani2009bio}, \qdel{which either}\qadd{where the former} sets the least significant part of the addition to one\qdel{ or}\qadd{ and the latter} computes with OR gates\qdel{, respectively}.

\subsection{Gate-Level Pruning}
\begin{figure}[!htbp]
  \centering
\includegraphics[scale=.38]{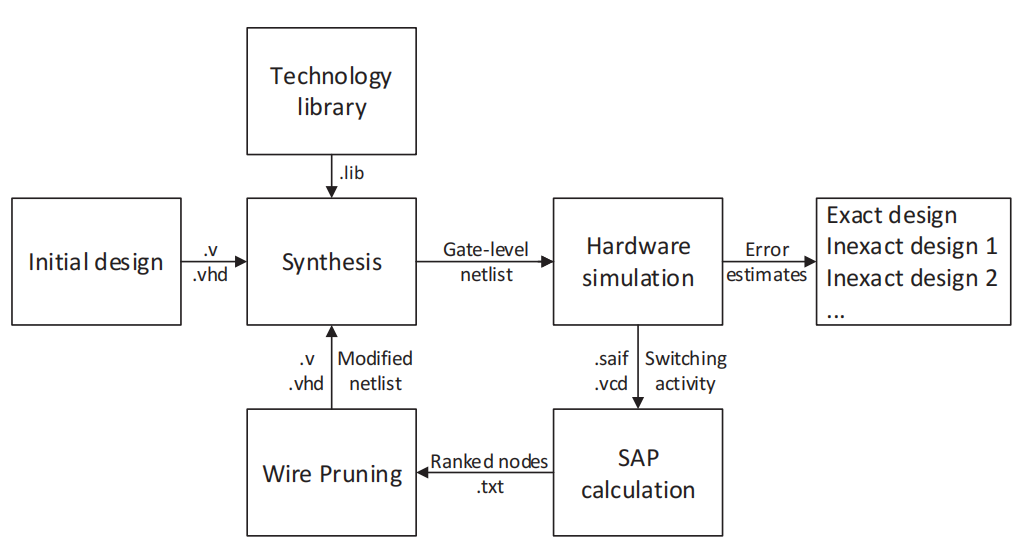}
\caption{An example of gate-level pruning in a digital design flow~\cite{schlachter2015automatic}.}
\label{fig:circuit1}       
\end{figure}

Gate-level pruning provides an alternative to simplify the netlist. It is based on the probabilistic pruning, which prunes less active gates from a circuit with a limited accuracy loss~\cite{lingamneni2011energy}. Schlachter \textit{et al.} proposed to transform the circuit to a graph and prune the nodes with the lowest significance-activity product (SAP) during synthesis~\cite{schlachter2015automatic}. The term ``significance” indicates the importance of each node/gate, while ``activity” refers to the toggling rate of the gate. The significance for the output nodes is user-defined and then backward propagated to calculate the significance of other gates. The activity of a node can be extracted from the \qdel{.SAIF file (Switching Activity Interchange Format)}\qadd{switching activity interchange format (SAIF) file}, which presents the toggle counts of wires. The digital design flow with gate-level pruning \qdel{was presented}\qadd{is shown} in Fig.~\ref{fig:circuit1}.

\subsection{Evolutionary Circuit Design}

Evolutionary algorithms, especially genetic programming, have been utilized to design digital circuits including approximate designs~\cite{sekanina2013approximate,vasicek2014evolutionary,hrbacek2016automatic,mrazek2017evoapprox8b}. Cartesian Genetic Programming (CGP) that uses graph representations is a flexible form of genetic programming~\cite{miller2008cartesian}. Based on CGP, circuits are represented by node arrays, where a node represents a basic \wadd{logic }function such as \qdel{adders and bit-wise operators}\wadd{AND and OR}. The design process starts with the circuit of an existing conventional multiplier. Then, a group of circuits is randomly generated from the initial one by applying \qdel{analogs of}\wadd{some operations analogy to the} naturally occurring genetic operations. Each circuit is evaluated \qdel{considering}\wadd{with} multiple objectives including error, delay, and power. Based on the evaluation, satisfactory circuits are selected \qdel{for}\wadd{as} a new generation of circuits. When the maximum number of generations is reached or the requirements of approximate multipliers are satisfied, the iteration is terminated and the connected nodes determine the multiplication implementation. Hrbacek \textit{et al.} followed the scheme and applied Non-dominated Sorting Genetic Algorithm II (NSGA-II) to sort out approximate multipliers among the Pareto front~\cite{hrbacek2016automatic}. Based on the previous work~\cite{hrbacek2016automatic}, reference \cite{mrazek2017evoapprox8b} constructed a library containing 471 8-bit approximate multipliers.

\subsection{Voltage Over-scaling}
Voltage scaling is another commonly used method to reduce power consumption~\cite{chandrakasan1995minimizing}. In general, the \qdel{operation }supply voltage needs to be higher than $V_{dd-crit}$, which is the minimum supply voltage to ensure the\qadd{ correct} timing of the critical path~\cite{lau2009energy,liu2009computation, mohapatra2011design, chen2012energy}.  
While voltage over-scaling (VOS) effectively reduces the power, timing violation induced errors are inevitably introduced~\cite{chen2012energy}. Thus, Lau \textit{et al.} provided different energy budgets for each column of the array multiplier in order to minimize the computation error~\cite{lau2009energy}. Since VOS mainly impacts the critical and near-critical paths, it is desired to adjust the architecture of each computation module to achieve a shorter critical path and mitigate the impact of low supply voltage~\cite{liu2009computation,chen2012energy}. Liu \textit{et al.} also proposed an analytical model to assess the VOS-induced computation errors, which can then be used to select the corresponding architecture and setup~\cite{liu2009computation}.

\section{Evaluation}\label{sec:evaluate}

In this section, we evaluate the representative approximate multipliers\qadd{ presented} in the previous sections in terms of accuracy and circuit characteristics (including delay, power, and area). Then, these approximate multipliers are applied in \wydel{two}\wyadd{three} machine learning applications by replacing the original exact multipliers\qadd{ with them}. The classification accuracy and energy consumption of the representative designs are presented for comparison.

\subsection{Quantitative Comparison Among Multipliers}

Given the above discussion\qdel{s} on various\qdel{ approximation} techniques introduced \qdel{to}\qadd{for} approximate multiplier design, it is necessary to further quantitatively compare \qdel{the}\qadd{their} performance under the same condition. Thus, the following subsections compare the designs under three categories, $i.e.$ fixed-point unsigned, fixed-point signed, and floating-point multipliers, where 16-bit is selected for fixed-point. It is noted that all the designs are evaluated with the same setup in the following experiment\qadd{s}. 
In particular, we use mean relative error distance (MRED) and normalized mean error distance (NMED) as accuracy metrics. The circuit measurements include delay, power, and area.
The approximate multipliers are implemented \qdel{with}\qadd{by} Verilog\qdel{,}\qadd{ and then} synthesized \qdel{in}\qadd{by Synopsys} Design Compiler~\cite{DC} using the UMC40 library~\cite{UMC40} \qdel{with}\qadd{under either} delay-optimized \qdel{and}\qadd{or} area-optimized \qdel{constraints}\qadd{mode}.
For each multiplier, we randomly generate 10-million input pairs following a uniform distribution and then run simulations in Verilator~\cite{verilator}.

\subsubsection{Fixed-Point Unsigned Multipliers}
\label{sec:fixed_unsigned}
We have implemented the aforementioned representative fixed-point unsigned multipliers\qadd{ listed} in Table~\ref{tab:work_summary}.
SSM~\cite{narayanamoorthy2014energy} and DRUM~\cite{hashemi2015drum} are implemented with the segment width 8. APP is implemented by altering partial products and globally applying the approximate reduction technique\hjdel{ globally}, \wydel{while APP(7,16,8) separates the partial product array \hjdel{column-wisely}\hjadd{in a column-wise manner} into three groups, \hjdel{\textit{i.e.},} accurate, approximate and truncated parts~\cite{venkatachalam2017design}.}while APP2 separates the partial product array in a column-wise manner into three groups, \textit{i.e.}, accurate, approximate, and truncated parts, with $7$, $16$, and $8$ columns, respectively~\cite{venkatachalam2017design}. \wydel{AW(7,16,8)}AW \qdel{takes}\qadd{adopts} the same partition scheme\qadd{ as APP2} for comparison\qdel{s using}\qadd{, but uses} an \qdel{AND-gate}\qadd{AND gate-based} partial product generation scheme~\cite{ha2017multipliers}.
\qadd{The approximate multiplier BAM is denoted as BAM($m,n$),}
\qdel{The two parameters of BAM}\qadd{where $m$ and $n$} denote the numbers of \qdel{dumped}\qadd{omitted} rows and columns when carrying out the accumulation~\cite{mahdiani2009bio}. OPACT~\cite{Xiao22}, UDM~\cite{kulkarni2011trading}, LM~\cite{mitchell1962computer}, Iterative LM (IterLM)~\cite{mitchell1962computer} and Improved LM (ImprLM)~\cite{ansari2020improved} are implemented following the original works.
\qadd{We also implement the approximate logarithm-based multiplier (ALM) and its iterative variant, IALM, both using approximate adders in the design.}
\qadd{The implementation of ALM uses the approximate adder SOA~\cite{liu2017design_log}\qdel{ configured by setting lower 11 bits approximate}\qadd{ that sets the 11 LSBs approximate}. We denote it as ALM\_SOA.}
\qadd{The implementation of IALM uses three approximate adders, including two SOAs and one LOA~\cite{mahdiani2009bio} configured by setting the numbers of approximate bits as 5, 11, and 16, respectively. We denote it as IALM\_SL.}
\wyadd{Finally, an approximate multiplier named mul16u\_1UG in the EvoApproxLib library~\cite{mrazek2017evoapprox8b}, an open library of approximate multipliers generated using evolutionary algorithms, is selected from the Pareto-optimal set based on power versus MRED with the MRED value comparable to the other designs.}

\begin{figure}[!htbp]
    \centering
    \includegraphics[width=\textwidth]{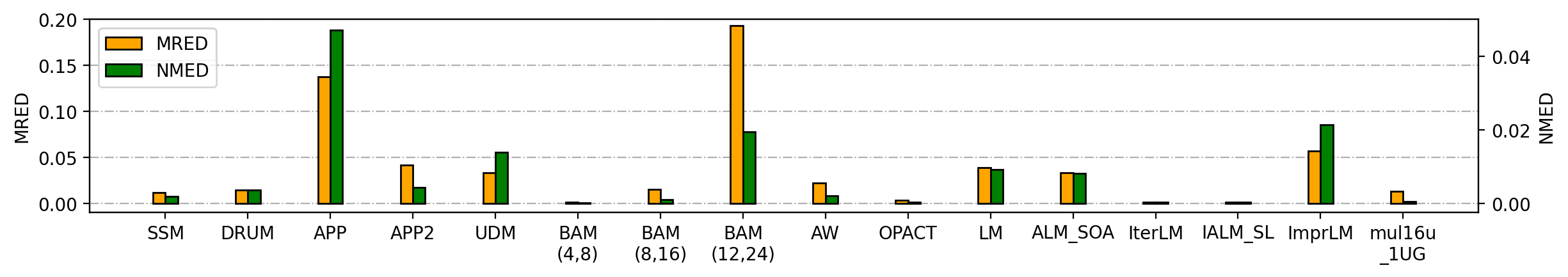}
    \caption{\qdel{Accuracy losses}\qadd{Error} comparison of approximate fixed-point unsigned multipliers.}
    \label{fig:acc_comp}       
\end{figure}

\begin{figure}[!htbp]
    \centering
    \includegraphics[width=\textwidth]{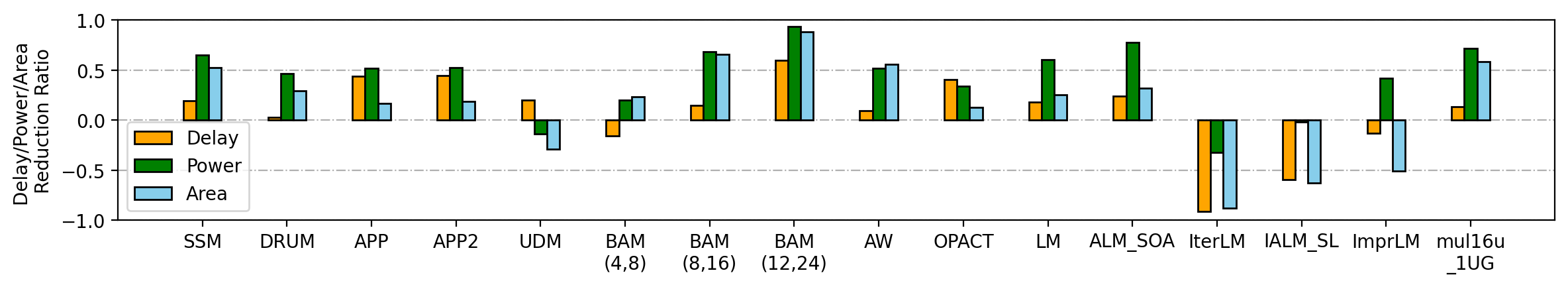}
    \caption{Circuit characteristics comparison of delay-optimized approximate multipliers.}
    \label{fig:circuit_DelayOpt_comp}       
\end{figure}

\begin{figure}[!htbp]
    \centering
    \includegraphics[width=\textwidth]{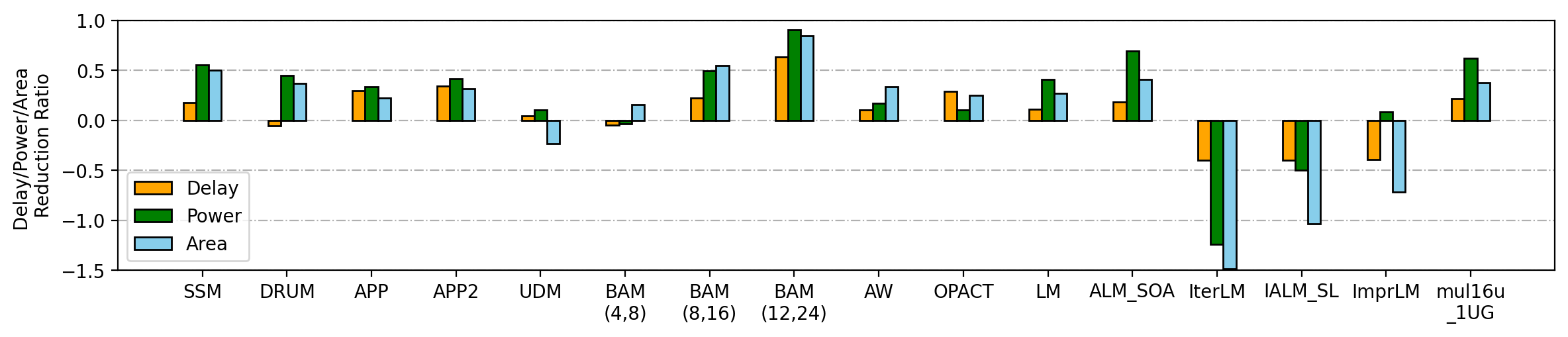}
    \caption{Circuit characteristics comparison of area-optimized approximate multipliers.}
    \label{fig:circuit_AreaOpt_comp}       
\end{figure}

Figs.~\ref{fig:acc_comp}--\ref{fig:circuit_AreaOpt_comp} show the accuracy and the circuit characteristics of multiple unsigned approximate multipliers when compared to an exact 16-bit multiplier IP obtained from DesignWare~\cite{DW}. \qdel{For Fig.~\ref{fig:acc_comp}}\qadd{In Fig.~\ref{fig:acc_comp}, we show the errors of the approximate multipliers, measured by MRED and NMED. Thus}, a lower bar indicates\qadd{ a} higher accuracy\qdel{, while for}\qadd{.} Figs.~\ref{fig:circuit_DelayOpt_comp} and~\ref{fig:circuit_AreaOpt_comp}\qdel{,}\qadd{ show the delay/power/area reduction ratio over the exact design under the delay-optimized and the area-optimized modes, respectively, where the corresponding exact design is synthesized under the same mode.} A higher positive bar indicates a larger delay, power, or area improvement.
\qdel{When using the delay-optimized constraints}\qadd{From Fig.~\ref{fig:circuit_DelayOpt_comp}}, we can observe that\qadd{ under the delay-optimized mode,} SSM shows better circuit characteristics than DRUM with comparable accuracy\qdel{ losses}. Due to the approximation in\qadd{ the MSBs}\qdel{ significant columns}, \qdel{APP has an accuracy loss over 13\% in terms of MRED and an accuracy loss about 5\% for NMED}\qadd{APP has an MRED over $13\%$ and an NMED about $5\%$}, while its variant APP2 shows much improved accuracy with very similar circuit characteristics. Comparing APP2 and AW, the altered partial products obviously help the efficiency of the reduction procedure but at the cost of area and power. Among all the BAM multipliers, BAM(8,16) provides \qdel{a better}\qadd{the best} trade-off between accuracy and circuit cost. OPACT, configured with automatically optimized compressor allocation and order connection, \qdel{generally }provides negligible \qdel{accuracy losses}\qadd{error} with moderate hardware savings. As for logarithm-based multipliers, the iterative schemes including IterLM and IALM\_SL consume too much hardware cost for error compensation to achieve the desired accuracy. ALM\_SOA \hjdel{then }improves the circuit characteristics with the approximate adder at the cost of moderate accuracy degradation.
\wyadd{The mul16u\_1UG achieves balanced circuits improvements in terms of area, delay, and power consumption with acceptable errors, showing a similar performance to BAM(8,16).}

\qdel{For the area-optimized constraints}\qadd{As shown in Fig.~\ref{fig:circuit_AreaOpt_comp}, under the area-optimized mode}, the results for most designs, \textit{e.g.}, DRUM, APP, APP2, UDM, OPACT, LM, ALM\_SOA, show an improvement in the area reduction ratio at the cost of a decreased delay reduction ratio, when compared to the delay-optimized cases. \qdel{While}\wyadd{In contrast, the multipliers BAMs, AW, IterLM, \wydel{and }IALM\_SL\wyadd{ and mul16u\_1UG} show improvements in delay reduction at the cost of smaller area and power reduction ratios.}\wydel{, the multipliers \qdel{of }UDM, IterLM, IALM\_SL(5,11,16) and ImprLM are found to have degradation in all the metrics when \qdel{synthesizing with the area constraint}\qadd{synthesized under the area-optimized mode}.}
The reason for the decrease in the area reduction ratio lies in the fact that the reduction ratio is measured against the exact multiplier synthesized under the area-optimized mode. In that case, the area\wadd{ of the exact multiplier} also reduces compared to the exact multiplier synthesized under the delay-optimized mode.
\textbf{Such a comparison interestingly \qdel{explore}\qadd{explores} the sensitivity of the various approximation techniques to the synthesis \qdel{constraints}\qadd{mode}}, which \qdel{have}\qadd{has} not been well illustrated in the prior work.

\subsubsection{Fixed-Point Signed Multipliers}
For fixed-point signed approximate multiplication, Booth multipliers are widely used by adding an extra stage of Booth encoding to the conventional binary multiplier. Thus, the aforementioned architecture-level approximation techniques are applicable to Booth multipliers as well~\cite{cho2004design,wang2009high,chen2011high,song2007adaptive,farshchi2013new,jiang2015approximate}. Given the comprehensive comparison of the above techniques in Section~\ref{sec:fixed_unsigned}, here we only focus on the comparison of the approximate techniques in Booth encoding\qadd{ (see Section~\ref{sec:Booth-encoding})}. In particular, we choose R4SA, and its variants R4ABE1 and R4ABE2, which neglect the\qadd{ signal} neg\_cin in the last partial product line in comparison to the exact Booth radix-4 multiplier~\cite{liu2017design}.
Here, \qadd{the signal }neg\_cin \qdel{denotes ``1''}\qadd{is the additional logic value 1} added to the LSB when \qdel{computing 2's complement due to the negative encoding}\qadd{handling a negative number in 2's complement\hjdel{ form}}.
R4ABE1 and R4ABE2 utilize different approximate encoders designed by K-map modification and are configured by setting the number of approximate LSBs as 15.
The prefix ``R4'' refer\qadd{s} to the radix-4 approximate multiplier, which is compared to the exact radix-4 multiplier. Similarly, R8ATM is a radix-8 approximate multiplier by introducing an approximate re-coding adder for \qdel{"Y+2Y" format computations}\qadd{the computation of the expression $(Y+2Y)$} and hence\qadd{, it is} compared to the exact radix-8 multiplier~\cite{jiang2015approximate}. 
\wyadd{Finally, an approximate multiplier named mul16s\_GRU is selected from the Pareto-optimal set based on power versus MRED in the EvoApproxLib library~\cite{mrazek2017evoapprox8b} with the MRED value comparable to the other designs.}

\begin{figure}[!htbp]
    \centering
    \includegraphics[scale=.5]{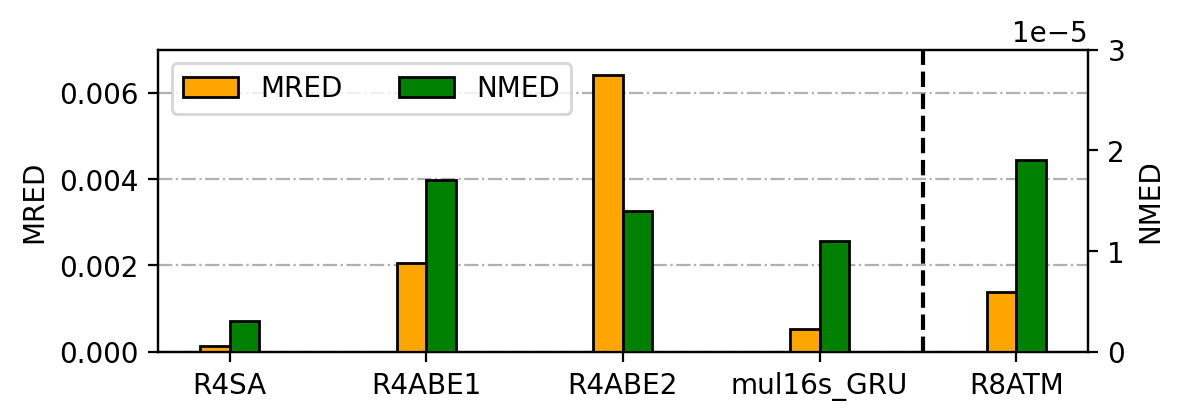}
    \caption{\qdel{Accuracy losses}\qadd{Error} comparison of approximate fixed-point signed multipliers.}
    \label{fig:booth_acc}       
\end{figure}

\begin{figure}[!htbp]
    \centering
    \includegraphics[scale=.5]{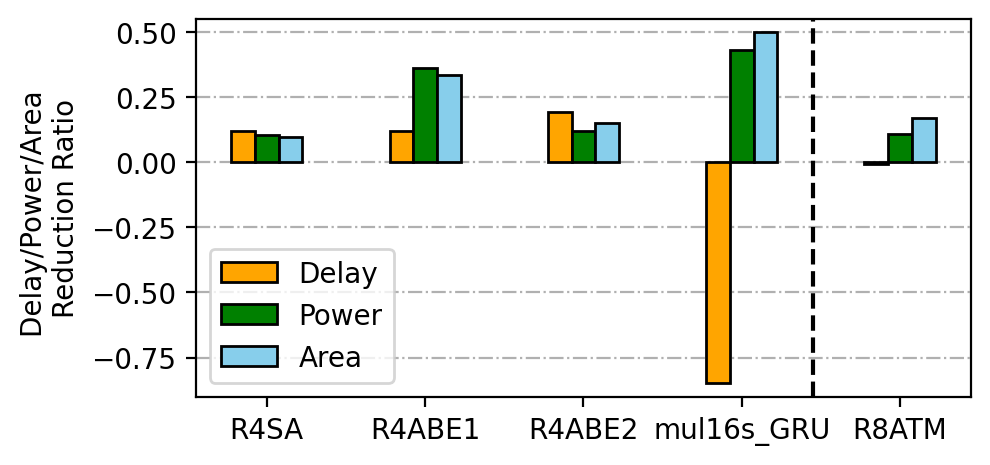}
    \caption{Circuit characteristics comparison of delay-optimized approximate fixed-point signed multipliers.}
    \label{fig:Booth_circuit_DelayOpt_comp}       
\end{figure}

\begin{figure}[!htbp]
    \centering
    \includegraphics[scale=.5]{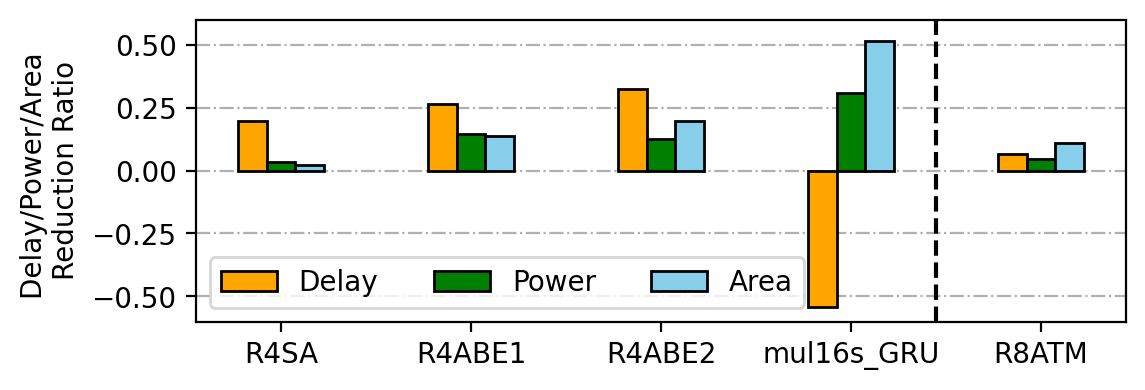}
    \caption{Circuit characteristics comparison of area-optimized approximate fixed-point signed multipliers.}
    \label{fig:Booth_circuit_AreaOpt_comp}       
\end{figure}

As we can see in Figs.~\ref{fig:booth_acc}--\ref{fig:Booth_circuit_AreaOpt_comp}, R4SA has a small error but distinct improvement in circuit characteristics. The ignorance of neg\_cin in the last partial product line reduces the extra hardware cost for accumulation. Since R4ABE1 and R4ABE2 introduce additional approximation\qdel{s} in the encoders based on R4SA, we mainly focus on the differences among them.
Both R4ABE1 and R4ABE2 have \qdel{larger accuracy losses}\qadd{a larger MRED and NMED} than R4SA, but they produce mixed results for different \qdel{accuracy}\qadd{error} measures\qadd{ (i.e., MRED and NMED)} when compared to each other. When the delay is optimized, R4ABE2 presents a higher improvement \qdel{of}\qadd{in} delay and R4ABE1 is superior in power and area. When the area is optimized, R4ABE1 and R4ABE2 demonstrate improvement\qdel{s} in all the three circuit \hjdel{characteristics }metrics. 
\wyadd{The mul16s\_GRU outperforms the other circuits in terms of area and power consumption under both the delay-optimized and the area-optimized mode. However, it has a notable degradation in delay, which prioritizes improvements in area and power consumption at the expense of increased delay.}
On the other hand, R8ATM shows a negligible delay improvement over the exact radix-8 multiplier \qdel{with delay-optimized constraints}\qadd{under the delay-optimized mode}. \qdel{When using area-optimized synthesis}\qadd{Under the area-optimized mode}, R8ATM is found to have more balanced improvements in circuit characteristics\hjdel{ improvement\qdel{s}}.

\subsubsection{Floating-Point Multipliers}
Many floating-point multipliers introduce approximation\qdel{s} at\qadd{ the} algorithm level, \textit{e.g.}, through linearization or hybrid approximation\qdel{s}. We implement and compare the \hjdel{most }representative approximation techniques in this section, \textit{i.e.}, ApproxLP~\cite{imani2019approxlp}, OAM~\cite{chen2020optimally}, and RMAC~\cite{imani2018rmac}. An exact 32-bit floating-point multiplier IP from DesignWare~\cite{DW} is used as the baseline.
\qadd{We denote the multipliers ApproxLP and OAM with the approximation level $n$ as ApproxLP($n$) and OAM($n$), respectively. A smaller $n$ indicates a higher approximation.}

\begin{figure}[!htbp]
    \centering
    \includegraphics[width=\textwidth]{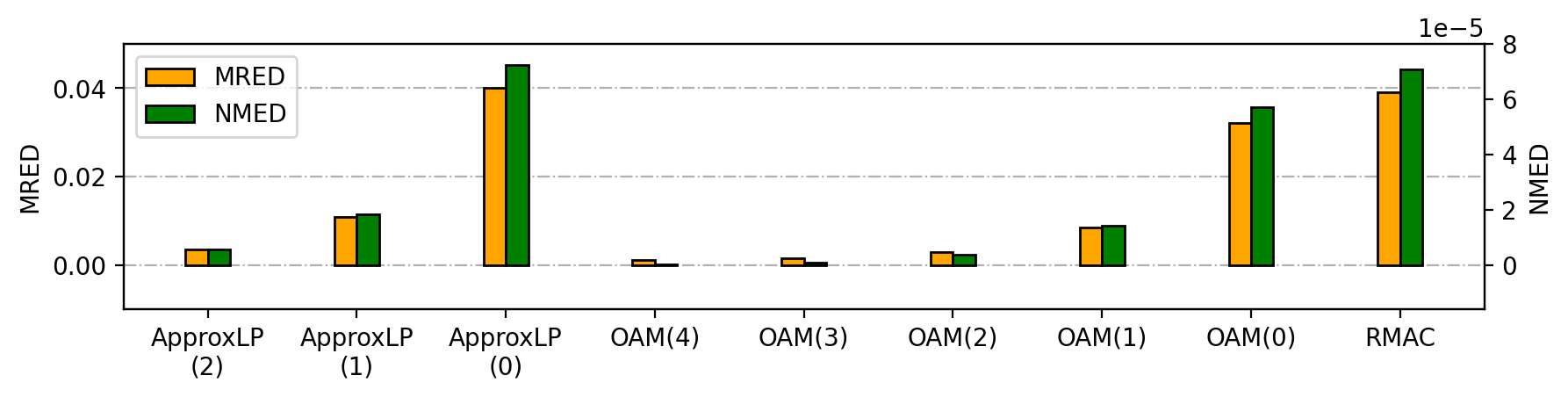}
    \caption{\qdel{Accuracy losses}\qadd{Error} comparison of approximate floating-point multipliers.}
    \label{fig:float_acc}       
\end{figure}

\begin{figure}[!htbp]
    \centering
    \includegraphics[width=\textwidth]{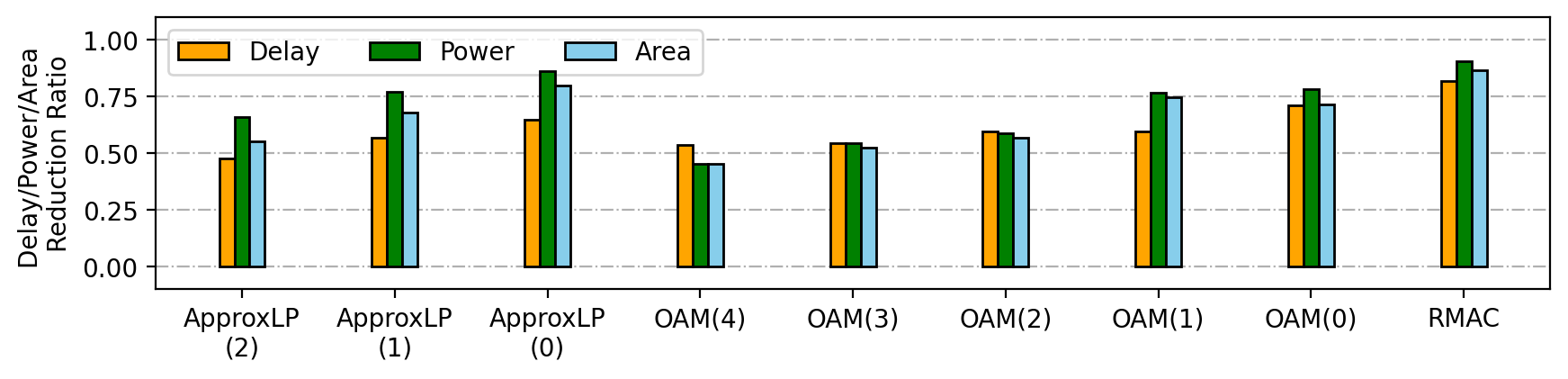}
    \caption{Circuit characteristics comparison of delay-optimized approximate floating-point multipliers.}
    \label{fig:Float_circuit_DelayOpt_comp}       
\end{figure}

\begin{figure}[!htbp]
    \centering
    \includegraphics[width=\textwidth]{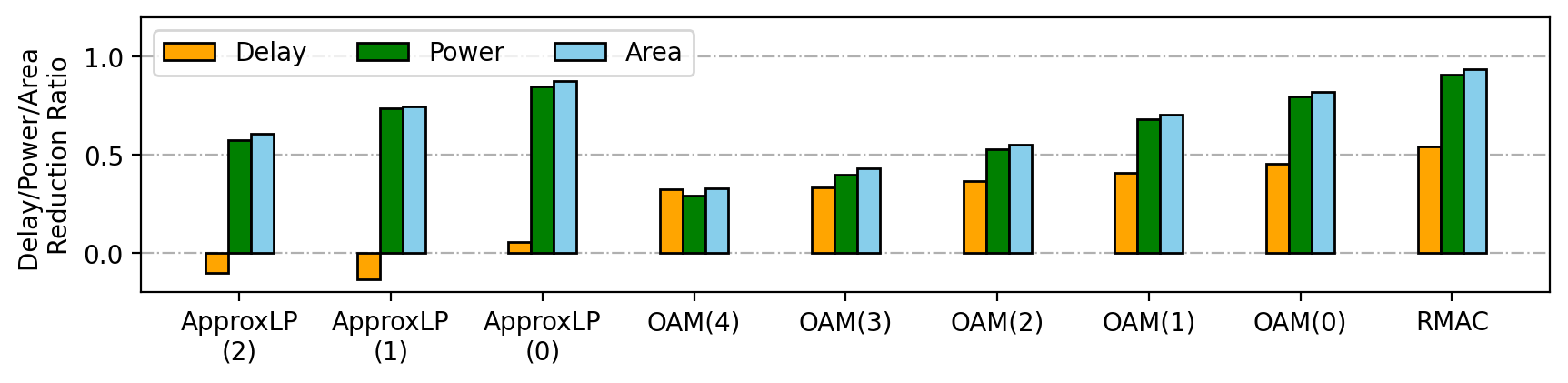}
    \caption{Circuit characteristics comparison of area-optimized approximate floating-point multipliers.}
    \label{fig:Float_circuit_AreaOpt_comp}       
\end{figure}

As shown in Figs.~\ref{fig:float_acc}--\ref{fig:Float_circuit_AreaOpt_comp}, when the approximation level decreases, both ApproxLP and OAM show a larger error with smaller hardware costs.
In general, \qadd{at the same approximation level, OAM has a smaller error, area, and delay than ApproxLP, while ApproxLP has \wadd{a }lower power dissipation}.\qdel{OAM compared with ApproxLP at the same level is superior in terms of accuracy, delay and area while ApproxLP shows advantages in power reduction.}
In Fig.~\ref{fig:Float_circuit_DelayOpt_comp}, 
unlike ApproxLP showing stable improvements in circuit characteristics \hjdel{improvement\qdel{s} }with an increased approximation level, OAM shows sharper changes in \wydel{both \qdel{accuracy loss}\qadd{error} and}circuit characteristics from level-(2, 3, 4) to level-(0, 1). When considering area-optimized synthesis in Fig.~\ref{fig:Float_circuit_AreaOpt_comp}, the circuit \hjdel{characteristics }metrics show steady improvement\qdel{s} with the increased approximation level\qdel{s} for both ApproxLP and OAM. However, ApproxLP shows a negative delay improvement, simply indicating a longer delay than the exact multiplier design. RMAC maps the mantissa multiplication to the addition between the two inputs and shows a similar performance \qdel{to}\qadd{as} ApproxLP(0) \qdel{while with better delay optimization}\wydel{\qadd{except \hjadd{for a higher}\hjdel{more} delay reduction}}with a larger delay reduction ratio. 

\subsection{Application-Level Comparison}
\label{sec:app_comp}
In addition to the generic evaluations, we further \qdel{explore}\qadd{study} the performance of the approximate multipliers in machine learning applications, which \qdel{is}\qadd{are} typically considered multiplication-intensive but error-\qdel{tolerance}\qadd{tolerant} by its nature. \wyadd{\wydel{Two}\wyadd{Three} representative machine learning applications are selected for comparison: \wyadd{the first }one is a multi-layer perceptron (MLP) network~\cite{gardner1998artificial} with one hidden layer on the MNIST dataset~\cite{deng2012mnist}, which is denoted as MNIST (MLP); the \wydel{other}\wyadd{second one} is the convolutional neural network AlexNet~\cite{krizhevsky2017imagenet} on the CIFAR10 dataset~\cite{krizhevsky2009learning}, which is denoted as CIFAR10 (AlexNet); \wyadd{ the third one is a lightweight convolutional neural network SqueezeNet~\cite{iandola2016squeezenet} on the ImageNet~\cite{deng2009imagenet}, which is denoted as ImageNet (SqueezeNet). }}

\wydel{Both}\wyadd{All the three} neural networks are pre-trained with accurate floating-point training infrastructure to obtain the trained weights in double precision. The baselines have been implemented in Verilog using the exact multipliers to find out the baseline classification accuracy, power, and delay of executing the \wydel{two}\wyadd{three} tasks, which can be measured by Prime-Time PX. During the evaluation, the original exact multipliers are replaced by the approximate multipliers to evaluate their impact on both classification accuracy and energy efficiency. The accuracy loss is defined as the classification accuracy change due to the use of the approximate multipliers, while the energy efficiency is given by the power-delay-product (PDP). For the evaluations using fixed-point multipliers, post-training quantization has been applied to obtain the fixed-point weights~\cite{jacob2018quantization}. For example, the weights and inputs \qdel{can be}\qadd{are} mapped to the range of $[0,65535]$ for 16-bit fixed-point unsigned multipliers, and $[-32767,32767]$ for 16-bit fixed-point signed multipliers. 

\subsubsection{Fixed-Point Unsigned Multipliers}
\label{sec:fixed-point-unsigned}

\begin{figure}[!htbp]
    \centering
    \includegraphics[width=\textwidth]{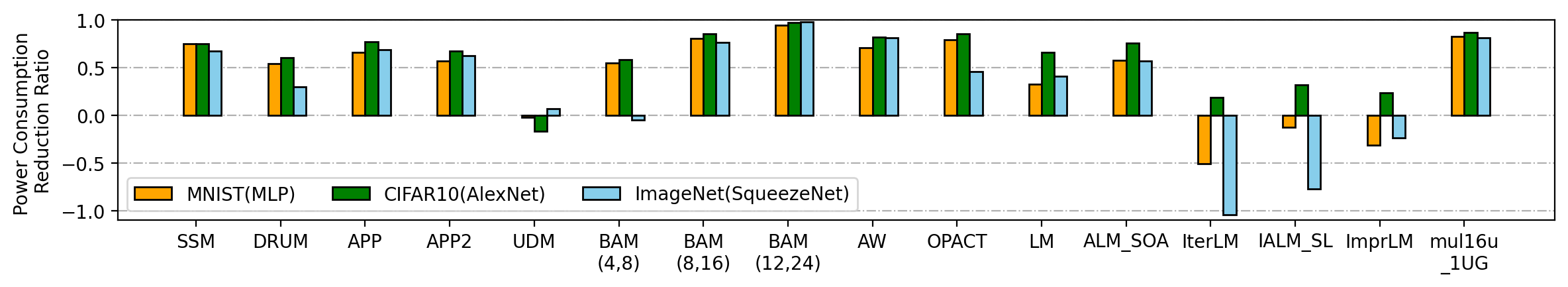}
    \caption{Power reduction of deploying approximate fixed-point unsigned multipliers in MNIST (MLP)\wydel{ and}\wyadd{,} CIFAR10 (AlexNet) \wyadd{and ImageNet (SqueezeNet) }when compared to the case of using the exact multiplier.}
    \label{fig:application_power}       
\end{figure}

\begin{figure}[!htbp]
    \centering
    \includegraphics[width=\textwidth]{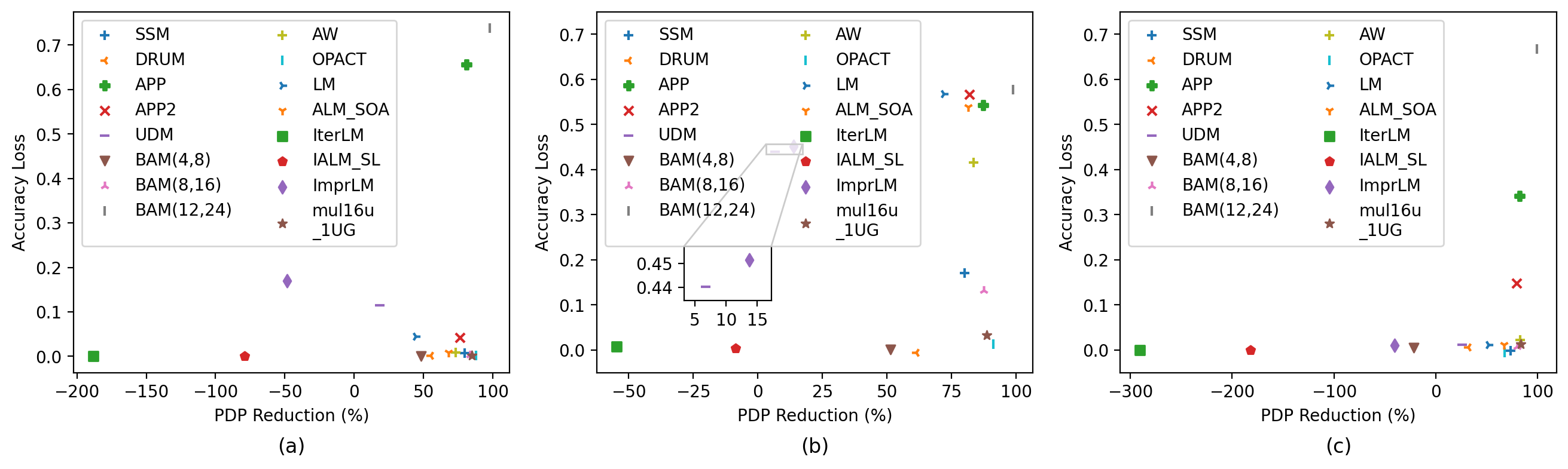}
    \caption{Classification accuracy loss versus PDP reduction ratio for the delay-optimized approximate \qdel{16 × 16}\qadd{$16 \times 16$} unsigned multipliers on (a) MNIST (MLP)\wydel{ and}\wyadd{,} (b) CIFAR10 (AlexNet)\wyadd{ and (c) ImageNet (SqueezeNet)}.}
    \label{fig:appUnsigned_loss_PDP}       
\end{figure}

\qadd{This section studies the performance of various approximate fixed-point unsigned multipliers.}
\wyadd{Fig.~\ref{fig:application_power} presents the power reduction ratio when deploying the approximate multiplier in MNIST (MLP)\wydel{ and}\wyadd{,} CIFAR10 (AlexNet) \wyadd{and ImageNet (SqueezeNet) }in comparison to the cases of using the exact multiplier. Other than a few logarithm-based designs requiring \qdel{the }extra power consumption for error compensation, most designs can achieve quite large power reductions with the largest power saving above 95\%.} Thus, the application-level accuracy needs to be further considered to determine the most suitable approximate multipliers for the machine learning applications.
Fig.~\ref{fig:appUnsigned_loss_PDP} shows the classification accuracy loss\qdel{es} versus the energy efficiency trade-off for various approximate multipliers on the \wydel{two}\wyadd{three} applications\hjdel{, which provides us more insights}. The PDP reduction ratio is defined as the relative PDP change over the original PDP, where a positive PDP reduction ratio indicates improved energy efficiency and a negative ratio indicates reduced energy efficiency compared to the original design. On the other hand, the higher accuracy loss indicates a \qdel{more poor}\qadd{worse} classification performance. Thus, the designs that are closer to the lower right corner are \hjdel{more }preferred.
\wyadd{It can be observed that OPACT, BAM(8,16), SSM, ALM\_SOA, AW, APP2, BAM(4,8), DRUM, \wydel{and }LM\wyadd{ and mul16u\_1UG} can achieve good improvements in energy efficiency with a limited or almost negligible accuracy loss on the MNIST (MLP). OPACT, DRUM, \wydel{and }BAM(4,8)\wyadd{ and mul16u\_1UG} can keep good accuracy with large improvements in energy efficiency \wydel{improvement }on the CIFRA10 (AlexNet).
\wyadd{Besides, APP2 and BAM(4,8) become suboptimal choices on the ImageNet (SqueezeNet) due to the accuracy loss and power consumption overhead, respectively.} }
\wydel{This simply indicates}\wyadd{These indicate} that the potential approximation design space actually changes for different algorithms and demands more in-depth understanding of the approximation error propagation when executing the algorithm. As discussed earlier, iterative log-based designs do not perform well in terms of energy efficiency in \wydel{both}\wyadd{these} cases, which are located close to the lower left\qadd{ corner of the plots}.

Moreover, if we compare Fig.~\ref{fig:acc_comp} and Fig.~\ref{fig:appUnsigned_loss_PDP}(a), it is found that, in general, the multipliers with smaller MREDs \qdel{intend}\qadd{tend} to cause smaller accuracy losses on MNIST (MLP). Both APP and BAM(12,24) are located at the upper right corner of Fig.~\ref{fig:appUnsigned_loss_PDP}(a) due to their large MREDs in the generic evaluation\hjdel{s}. However, when the application (or algorithm) becomes more complex, as shown in Fig.~\ref{fig:appUnsigned_loss_PDP}(b), more designs even with moderate MREDs in Fig.~\ref{fig:acc_comp} show very poor performances in terms of accuracy on CIFAR10 (AlexNet). The difference is partly due to the shallowness of the MLP network, where the approximation errors can hardly get accumulated. In addition, some multipliers, like ALM\_SOA and LM, introduce one-sided errors that can become more difficult to be cancelled out during the approximation error propagation within the AlexNet. Finally, the multipliers such as BAM(8,16), AW, and APP2 tend to introduce larger relative errors for small inputs with their aggressive approximations in the least-significant bits. \textbf{Thus, when deploying an approximate multiplier, instead of simply selecting the one based on its MRED, \qdel{the approximate multiplier needs to}\qadd{we should choose one with} a balance among error bias, approximation aggressiveness, and its fitness to a particular algorithm.}

\begin{figure}[!htbp]
    \centering
    \includegraphics[width=\textwidth]{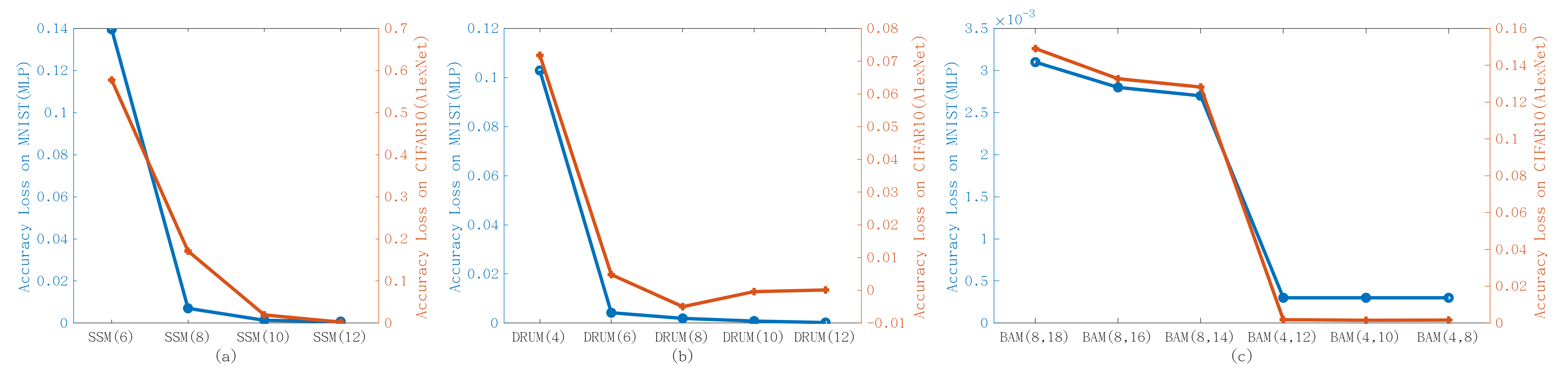}
    \caption{Classification accuracy loss under different approximation configurations of (a) SSM, (b) DRUM and (c) BAM.}
    \label{fig:acc_vs_app}       
\end{figure}

\wyadd{Furthermore, we present the classification accuracy loss of SSM, DRUM, and BAM under different approximation configurations on MNIST (MLP) and CIFAR10 (AlexNet). As shown in Fig.~\ref{fig:acc_vs_app}, the accuracy loss increases as the approximation degree increases, with the left side representing a higher degree of approximation and the right side representing a lower degree of approximation. When configured with a larger truncation width, both DRUM and SSM exhibit minimal accuracy losses. However, as the truncation width decreases, DRUM demonstrates smaller accuracy losses than SSM because DRUM aims to preserve more data within the limited bitwidth. In cases of smaller truncation width, SSM may need to provide more truncation choices to retain data information better. The accuracy loss curves of BAM($4,n$) and BAM($8,n$) show a significant difference, whereas the adjustment of the discarded column numbers in BAM($4,n$) and BAM($8,n$) has a relatively smaller impact. This is because each row of partial products covers 16 partial product columns, and discarding the same partial product rows has a greater impact than discarding the same partial product columns.}

\subsubsection{Fixed-Point Signed Multipliers}
\label{sec:fixed-point-signed}

\begin{figure}[!htbp]
    \centering
    \includegraphics[scale=.6]{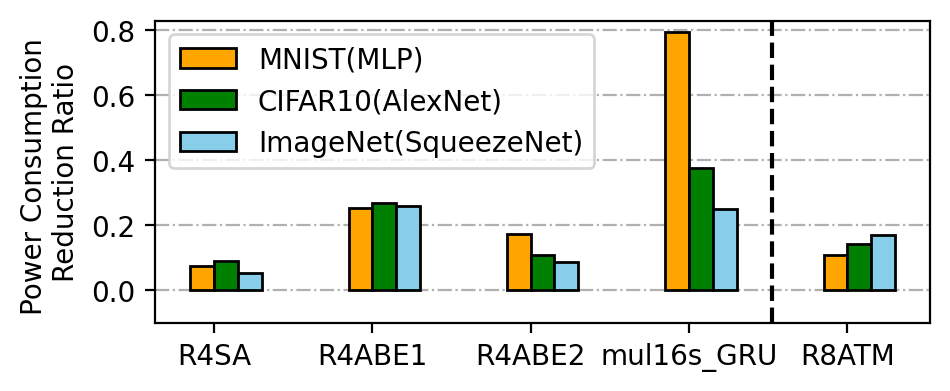}
    \caption{Power reduction of deploying approximate fixed-point signed multipliers in MNIST (MLP)\wydel{ and}\wyadd{,} CIFAR10 (AlexNet) \wyadd{and ImageNet (SqueezeNet) }when compared to the case of using the exact multiplier.}
    \label{fig:application_booth_power}       
\end{figure}

\begin{figure}[!htbp]
    \centering
    \includegraphics[width=\textwidth]{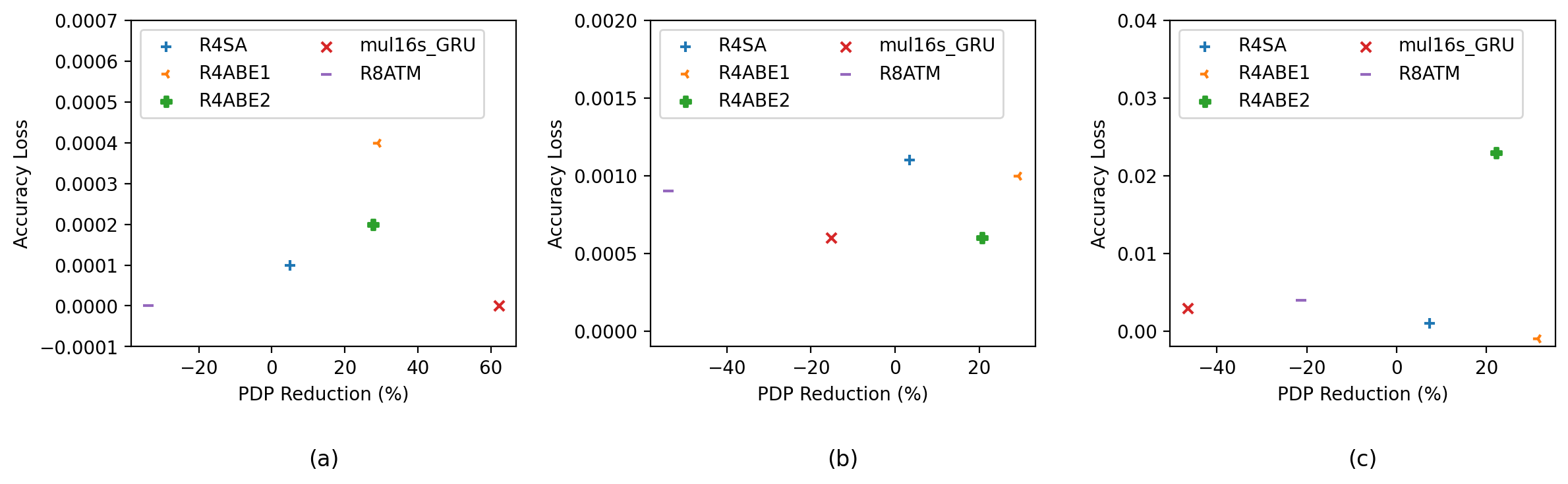}
    \caption{Classification accuracy loss versus PDP reduction ratio for the delay-optimized approximate 16 × 16 signed multipliers on (a) MNIST (MLP)\wydel{ and}\wyadd{,} (b) CIFAR10 (AlexNet)\wyadd{ and (c) ImageNet (SqueezeNet)}.}
    \label{fig:appBooth_loss_PDP}       
\end{figure}

Fig.~\ref{fig:application_booth_power} presents the comparison of power reduction among different approximate fixed-point signed multipliers, with the baseline design using the exact signed multipliers. Similar as in Fig.~\ref{fig:Booth_circuit_DelayOpt_comp}, all the designs show power reductions over the exact multipliers, where \wydel{R4ABE1}\wyadd{mul16s\_GRU} can achieve the most power saving. Fig.~\ref{fig:appBooth_loss_PDP} \hjdel{also }compares the trade-off between classification accuracy loss and PDP reduction ratio for the \wydel{two}\wyadd{three} applications using different approximate signed fixed-point \hjdel{signed }multipliers. \wyadd{When comprehensively considering the accuracy loss and PDP, \wyadd{the mul16s\_GRU exhibits the best performance on MNIST (MLP), and }R4ABE1 is \hjdel{obviously }superior to the other multipliers\wyadd{ on CIFAR10 (AlexNet) and ImageNet (SqueezeNet)}. \wyadd{The mul6s\_GRU performs well on MNIST (MLP) due to the significant power saving. However, its performance is not as favorable in the other two applications, primarily due to its inherently high delay, which results in high energy requirements. }}R8ATM has the worst performance in terms of PDP reduction due to its more complicated encoding. 

\begin{figure}[!htbp]
    \centering
    \includegraphics[width=\textwidth]{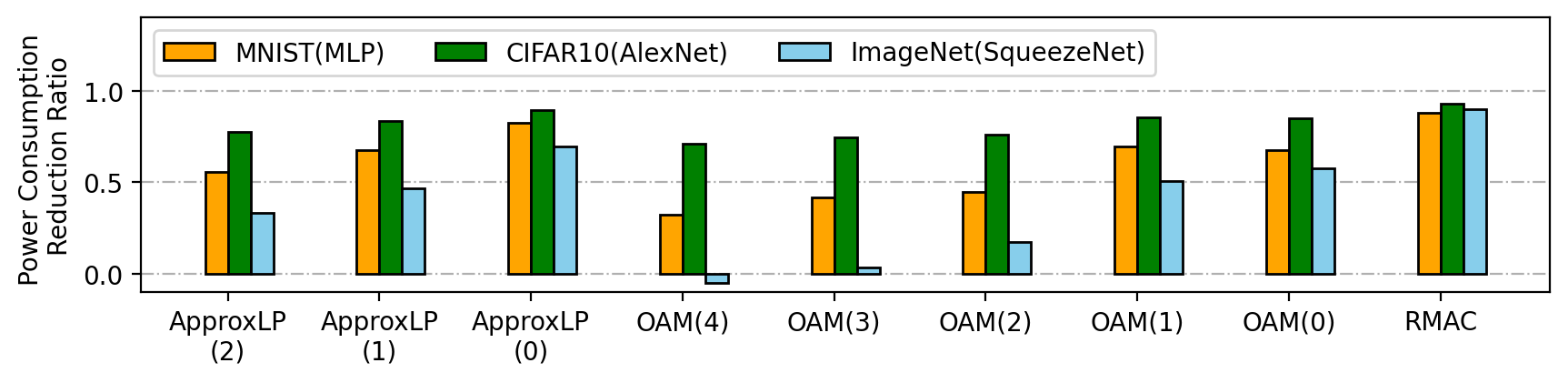}
    \caption{Power reduction of deploying approximate floating-point multipliers in MNIST (MLP)\wydel{ and}\wyadd{,} CIFAR10 (AlexNet) \wyadd{and ImageNet (SqueezeNet) }when compared to the case of using the exact multiplier.}
    \label{fig:application_float_power}       
\end{figure}

\begin{figure}[!htbp]
    \centering
    \includegraphics[width=\textwidth]{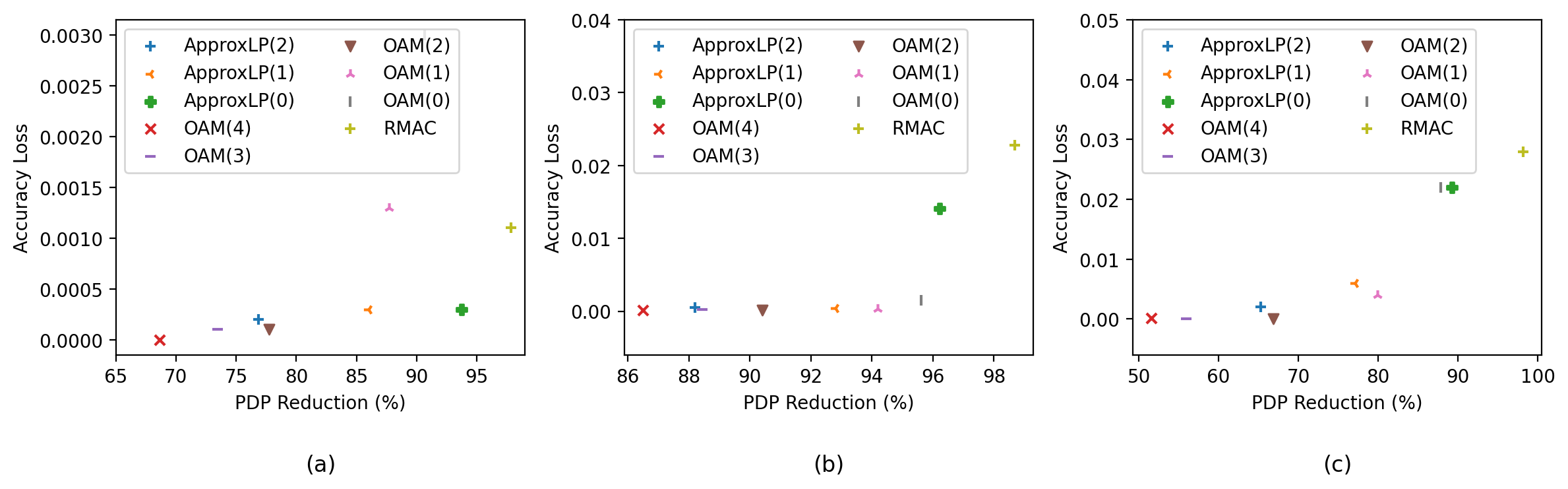}
    \caption{Classification accuracy loss versus PDP reduction ratio for the delay-optimized approximate floating-point multipliers on (a) MNIST (MLP)\wydel{ and}\wyadd{,} (b) CIFAR10 (AlexNet)\wyadd{ and (c) ImageNet (SqueezeNet)}.}
    \label{fig:appFloat_loss_PDP}       
\end{figure}

\subsubsection{Floating-point Multipliers}
\label{sec:floating-point} 
\wyadd{Finally, we investigate the results of designs using approximate floating-point multipliers. As shown in
Fig.~\ref{fig:application_float_power}, \wyadd{almost }all the approximate floating-point multipliers can achieve dramatic power reduction\wyadd{ except the OAM(4) and OAM(3) on ImageNet(SqueezeNet)}. Such power saving eventually contributes to the large PDP reduction in Fig.~\ref{fig:appFloat_loss_PDP}, in which the smallest energy efficiency improvement is more than \wydel{65\% for OAM(4) on MNIST (MLP)}\wyadd{50\% for OAM(4) on ImageNet (SqueezeNet)}. Due to the error tolerance \hjdel{nature }of the \wydel{two}\wyadd{three} applications and the large dynamic range of the floating-point multipliers, all the applications adopting the multipliers in this category have negligible accuracy losses, where the worst case is \wydel{CIFAR10 (AlexNet)}\wyadd{ImageNet (SqueezeNet)} using RMAC with an accuracy loss of only \wydel{0.024}\wyadd{0.028}.} Moreover, the impact of the approximation level is very limited when comparing OAM or ApproxLP with different approximation levels. Thus, compared to the fixed-point counterparts, \textbf{the larger dynamic range in approximate floating-point multipliers actually contributes to a better trade-off of accuracy and energy efficiency on the \wydel{two}\wyadd{three} machine learning applications.}

\section{Discussions}\label{sec:discussion}
Both the reviews and evaluations in the \qdel{last few}\qadd{previous} sections demonstrate the growing popularity and maturity of the research on approximate multipliers. However, there are still various challenges to be overcome, ranging from the design methodology, tools, to applications. This section\qdel{ would discuss}\qadd{ discusses} a few potential research \qdel{\& development }directions that may shed light on easier and more efficient integration of approximate multipliers in the digital design flow in the future.
\begin{itemize}
    \item \textbf{Top-down design methodology}: The demand for approximate multipliers is application driven to achieve higher energy efficiency at the cost of accuracy. On the other hand, as shown in the last section, the accuracy of approximate multipliers is not equivalent to the accuracy at the application level. For example, the metrics like MRED and NMED measuring the error distance between approximate and exact results do not \qdel{easily meet the requirements at the application level}\wadd{correlate well with the application-level quality measures}, such as \qdel{the measurement by using }the peak signal-to-noise ratio (PSNR) for image processing and the average precision (AP) \qdel{measure }for object detection. \qdel{In other words, the specifications of applications do not always fully exploit the accuracy of approximate multipliers. }Such a mismatch inevitably incurs additional design effort or over-design when deploying approximate multipliers. Thus, for a top-down design methodology, it is necessary to build a link between the application-level specification and\qadd{ error metrics of} approximate multiplier\qdel{ metric}s. In addition, when both exact and approximate arithmetic circuits are used, the data flow needs to be revised to support such mixed precision~\cite{zhou2020convolutional}. Both processes demand a general design methodology to support the top-down design flow.
    
    \item \textbf{Design \qdel{synthesis and optimization}\qadd{automation methods}}:
    \qadd{Most existing approximate multipliers were designed manually.}
    \qdel{With}\qadd{However, with the} various approximation options \qdel{and deviated behavior for different synthesis options, it becomes increasingly important to have an optimization tool to further exploit the design space and automatically select the appropriate approximation techniques and synthesis options.}at different design levels, the design space for approximate multipliers is huge. Thus, the design of an optimized approximate multiplier calls for \hjdel{design }automation methods that systematically explore the design space.
    \qadd{In this direction, some methods have been proposed for the architecture-level and circuit-level approximation, such as the automatic synthesis of approximate 4-2 compressors~\cite{Xuan22} and optimized allocation of approximate compressors in the compression stage~\cite{Xiao22}. These automatic synthesis methods are able to obtain better designs than the existing manual designs.}
    \qadd{However, given the large and complicated design space of approximate multipliers, there is still room for the development of more efficient and powerful design automation methods.}
    \item \textbf{Algorithm-circuit co-optimization}: Finally, \wydel{most applications deploy approximate multipliers as the algorithm itself is often error-tolerant with \qdel{heavy}\qadd{a large number of} multiplications.}many applications such as machine learning are error-tolerant\qdel{ as nature}. They can then \qdel{deploy}\wadd{utilize} approximate multipliers to achieve energy efficiency improvements without much accuracy \qdel{compromises}\wadd{loss}. However, as shown in the last section, there are still many unknowns on the interplay between the algorithm and the underlying approximate multiplier. In particular, it is highly desired to conduct research on designing approximation-friendly algorithms to fully utilize the benefits of \hjdel{underlying }approximate multipliers. Such a study calls for an understanding of the approximation theory, error propagation, and error representation in the algorithms to define the appropriate \hjdel{approximation }design space. It is clear that the design space \qdel{is highly dependent}\wadd{highly depends} on the underlying approximate multiplier circuit. Thus, more research efforts need to be placed upon the interplay between the approximate algorithm design space and the characteristics of the underlying approximate multiplier circuit.
\end{itemize}

\section{Conclusions}\label{sec:con}
Widespread use of multiplication in error-tolerant applications enables significant improvements in energy efficiency when \qdel{adopting}\wadd{using} approximate multipliers. In this article, we review approximate multipliers at the algorithm, architecture, and circuit levels. Detailed experimental results are presented on both generic and machine learning applications to help understand the advantages and disadvantages of various techniques at different design levels.

At the algorithm level, logarithm-based, linearization-based, and hybrid approximations are discussed. 
Mitchell's logarithmic multiplier shows significant hardware savings but with one-sided errors.
The iterative logarithmic ones \qdel{focus on error alleviation}\wadd{reduce the error, but suffer from hardware overhead} \qdel{while suffering from hardware costs }for error compensation. The linearization scheme saves substantial energy and area with negligible errors, \qdel{providing a zero-mean}\wadd{and also provides an} error distribution\wadd{ with zero mean}. The hybrid scheme collaboratively utilizes highly approximated multipliers and more accurate ones, which may increase the circuit area.
Although the reviewed works of the first two techniques focus on fixed-point and floating-point \qdel{designs}\wadd{multiplications,} respectively, there are some similarities between them.
Both the logarithm-based and the linearization-based algorithms transform the multiplication to simpler \wadd{linear }computations, \qdel{with a difference in the domain of linear computing}\wadd{despite their different mechanisms}.
The \qdel{first one}\wadd{logarithm-based method} \qdel{makes use of the linear computation of multiplication on a logarithmic scale}\wadd{exploits the fact that in the logarithmic domain, a multiplication becomes a linear addition,} while the \qdel{second one makes use of}\wadd{linearization-based method directly performs} piecewise-linear approximation\wadd{ for the non-linear multiplication}.
With the same need for locating the leading one, both of them fit floating-point multipliers well due to the normalized mantissa, but it may take \qdel{relatively }extra \qdel{efforts}\wadd{hardware costs} for their utilization in fixed-point \qdel{designs}\wadd{multipliers}.

At the architecture level, various approximation strategies are presented and discussed at different stages of a conventional exact multiplier. Truncating input operands or partial products is a simple \qdel{but}\wadd{yet} effective way to reduce \qdel{required }hardware\wadd{ cost}\qdel{, while}\wadd{. However,} static truncation may cause \wadd{a }large relative error \qdel{provided}\wadd{for} small operands, which \qdel{could bring impacts on}\wadd{can affect} the solution quality of complex applications.
Applying the altered partial products enables the acceleration of accumulation, while the altering process increases the circuit area to some extent.
\wadd{Using approximate compressors in the accumulation stage is another effective way to design approximate multipliers.}
\qadd{However, t}he allocation of approximate compressors in the \qdel{tree-based }accumulation\wadd{ stage} and\wadd{ the determination of} their connection orders are \qdel{not studied well in previous works and need}\wadd{complicated problems that deserve} \qdel{future investigations}\wadd{further study}.

At the circuit level, approximation techniques can be applied with \hjdel{any of }the aforementioned approximation methods at the architecture and the algorithm levels. Boolean rewriting is often utilized to simplify the circuit of basic modules adopted in approximated multipliers. Gate-level pruning and evolutionary circuit design are both objective-oriented, and thus the trade-off between hardware and accuracy can be adjusted to meet different requirements. Voltage over-scaling \qdel{could}\wadd{can} reduce energy consumption effectively, while the induced errors due to timing violations are often non-negligible, since the \qdel{significant part basically covers the critical path}\wadd{timing violations occur on the critical paths, which typically involve computation over the MSBs}. 

\qdel{What's more}\wadd{Furthermore}, the generic and application-level evaluations give two insights: (1) various approximation techniques show different sensitivity to the synthesis mode, and thus may present circuit improvements \qdel{at}\wadd{of} different degrees \qdel{when synthesized in various}\wadd{under different synthesis} modes; (2) the selection of approximate multipliers for specific applications is concerned with not only \wadd{the basic error metrics such as NMED and }MRED, but also error bias, approximation aggressiveness, and its fitness to the algorithm.

\begin{acks}
This work was supported in part by National Key R\&D Program of China (Grant No. 2021ZD0114703), National Natural Science Foundation of China (Grant No. 62034007, 61974133, and 62141404) and SGC Cooperation Project (Grant No. M-0612).
\end{acks}
\bibliographystyle{ACM-Reference-Format}
\bibliography{00_Main}

\begin{thebibliography}{100}

\bibitem{atzori2010internet}
Luigi Atzori, Antonio Iera, and Giacomo Morabito.
\newblock {The Internet of Things: A survey}.
\newblock {\em Computer networks}, 54(15):2787--2805, 2010.

\bibitem{jiang2020approximate}
Honglan Jiang, Francisco Javier~Hernandez Santiago, Hai Mo, Leibo Liu, and Jie
  Han.
\newblock Approximate arithmetic circuits: A survey, characterization, and
  recent applications.
\newblock {\em Proceedings of the IEEE}, 108(12):2108--2135, 2020.

\bibitem{5993675}
Vaibhav Gupta, Debabrata Mohapatra, Sang~Phill Park, Anand Raghunathan, and
  Kaushik Roy.
\newblock {IMPACT}: Imprecise adders for low-power approximate computing.
\newblock In {\em IEEE/ACM International Symposium on Low Power Electronics and
  Design}, pages 409--414. IEEE, 2011.

\bibitem{6569370}
Jie Han and Michael Orshansky.
\newblock Approximate computing: An emerging paradigm for energy-efficient
  design.
\newblock In {\em 2013 18th IEEE European Test Symposium (ETS)}, pages 1--6,
  2013.

\bibitem{10.5555/2971808.2972118}
Mohsen Imani, Abbas Rahimi, and Tajana~S. Rosing.
\newblock Resistive configurable associative memory for approximate computing.
\newblock In {\em Proceedings of the 2016 Conference on Design, Automation \&
  Test in Europe}, DATE '16, page 1327–1332, San Jose, CA, USA, 2016. EDA
  Consortium.

\bibitem{10.1145/2540708.2540710}
Swagath Venkataramani, Vinay~K. Chippa, Srimat~T. Chakradhar, Kaushik Roy, and
  Anand Raghunathan.
\newblock {Quality Programmable Vector Processors for Approximate Computing}.
\newblock In {\em Proceedings of the 46th Annual IEEE/ACM International
  Symposium on Microarchitecture}, MICRO-46, page 1–12, New York, NY, USA,
  2013. Association for Computing Machinery.

\bibitem{liu2020retrospective}
Weiqiang Liu, Fabrizio Lombardi, and Michael Shulte.
\newblock A retrospective and prospective view of approximate computing.
\newblock {\em Proceedings of the IEEE}, 108(3):394--399, 2020.

\bibitem{deng2019energy}
Jianing Deng, Zhiguo Shi, and Cheng Zhuo.
\newblock Energy-efficient real-time {UAV} object detection on embedded
  platforms.
\newblock {\em IEEE Transactions on Computer-Aided Design of Integrated
  Circuits and Systems}, 39(10):3123--3127, 2019.

\bibitem{1303135}
Gokul Govindu, Ling Zhuo, Seonil Choi, and Viktor Prasanna.
\newblock Analysis of high-performance floating-point arithmetic on {FPGAs}.
\newblock In {\em 18th International Parallel and Distributed Processing
  Symposium, 2004. Proceedings.}, page 149, 2004.

\bibitem{465364}
Robert~K Yu and Gregory~B Zyner.
\newblock 167 {MHz} radix-4 floating point multiplier.
\newblock In {\em Proceedings of the 12th Symposium on Computer Arithmetic},
  pages 149--154, 1995.

\bibitem{mitchell1962computer}
John~N Mitchell.
\newblock Computer multiplication and division using binary logarithms.
\newblock {\em IRE Transactions on Electronic Computers}, EC-11(4):512--517,
  1962.

\bibitem{lim1992single}
Yong~Ching Lim.
\newblock Single-precision multiplier with reduced circuit complexity for
  signal processing applications.
\newblock {\em IEEE transactions on Computers}, 41(10):1333--1336, 1992.

\bibitem{schulte1993truncated}
Michael~J Schulte and Earl~E Swartzlander.
\newblock {Truncated multiplication with correction constant [for DSP]}.
\newblock In {\em Proceedings of IEEE Workshop on VLSI Signal Processing},
  pages 388--396. IEEE, 1993.

\bibitem{jou1999design}
Jer~Min Jou, Shiann~Rong Kuang, and Ren Der~Chen.
\newblock Design of low-error fixed-width multipliers for {DSP} applications.
\newblock {\em IEEE Transactions on Circuits and Systems II: Analog and Digital
  Signal Processing}, 46(6):836--842, 1999.

\bibitem{van2000design}
Lan-Da Van, Shuenn-Shyang Wang, and Wu-Shiung Feng.
\newblock Design of the lower error fixed-width multiplier and its application.
\newblock {\em IEEE Transactions on Circuits and Systems II: Analog and Digital
  Signal Processing}, 47(10):1112--1118, 2000.

\bibitem{jon2000fixed}
Shyh-Jye Jon and Hui-Hsuan Wang.
\newblock Fixed-width multiplier for {DSP} application.
\newblock In {\em Proceedings 2000 International Conference on Computer
  Design}, pages 318--322. IEEE, 2000.

\bibitem{cho2004design}
Kyung-Ju Cho, Kwang-Chul Lee, Jin-Gyun Chung, and Keshab~K Parhi.
\newblock Design of low-error fixed-width modified booth multiplier.
\newblock {\em IEEE Transactions on Very Large Scale Integration (VLSI)
  Systems}, 12(5):522--531, 2004.

\bibitem{van2005generalized}
L-D Van and Chih-Chyau Yang.
\newblock Generalized low-error area-efficient fixed-width multipliers.
\newblock {\em IEEE Transactions on Circuits and Systems I: Regular Papers},
  52(8):1608--1619, 2005.

\bibitem{courbariaux2014training}
Matthieu Courbariaux, Yoshua Bengio, and Jean-Pierre David.
\newblock Training deep neural networks with low precision multiplications.
\newblock {\em arXiv preprint arXiv:1412.7024}, 2014.

\bibitem{10.1109/TCSVT.2013.2243658}
Ku~He, Andreas Gerstlauer, and Michael Orshansky.
\newblock {Circuit-Level Timing-Error Acceptance for Design of Energy-Efficient
  DCT/IDCT-Based Systems}.
\newblock {\em Circuits and Systems for Video Technology, IEEE Transactions
  on}, 23:961--974, 06 2013.

\bibitem{Imani_ACAM_10.1145/2934583.2934595}
Mohsen Imani, Yeseong Kim, Abbas Rahimi, and Tajana Rosing.
\newblock {ACAM: Approximate Computing Based on Adaptive Associative Memory
  with Online Learning}.
\newblock In {\em Proceedings of the 2016 International Symposium on Low Power
  Electronics and Design}, ISLPED '16, page 162–167, New York, NY, USA, 2016.
  Association for Computing Machinery.

\bibitem{10.5555/2971808.2971893}
Mohsen Imani, Shruti Patil, and Tajana~S. Rosing.
\newblock {MASC: Ultra-Low Energy Multiple-Access Single-Charge TCAM for
  Approximate Computing}.
\newblock In {\em Proceedings of the 2016 Conference on Design, Automation \&
  Test in Europe}, DATE '16, page 373–378, San Jose, CA, USA, 2016. EDA
  Consortium.

\bibitem{imani2018rapidnn}
Mohsen Imani, Mohammad Samragh, Yeseong Kim, Saransh Gupta, Farinaz Koushanfar,
  and Tajana Rosing.
\newblock Rapidnn: In-memory deep neural network acceleration framework.
\newblock {\em arXiv preprint arXiv:1806.05794}, 2018.

\bibitem{kulkarni2011trading}
Parag Kulkarni, Puneet Gupta, and Milos Ercegovac.
\newblock Trading accuracy for power with an underdesigned multiplier
  architecture.
\newblock In {\em 2011 24th Internatioal Conference on VLSI Design}, pages
  346--351. IEEE, 2011.

\bibitem{7544342}
Muhammad Shafique, Rehan Hafiz, Semeen Rehman, Walaa El-Harouni, and J{\"o}rg
  Henkel.
\newblock Invited: Cross-layer approximate computing: From logic to
  architectures.
\newblock In {\em 2016 53nd ACM/EDAC/IEEE Design Automation Conference (DAC)},
  pages 1--6, 2016.

\bibitem{6637825}
Alexander Suhre, Furkan Keskin, Tulin Ersahin, Rengul Cetin-Atalay, Rashid
  Ansari, and A~Enis Cetin.
\newblock A multiplication-free framework for signal processing and
  applications in biomedical image analysis.
\newblock In {\em 2013 IEEE International Conference on Acoustics, Speech and
  Signal Processing}, pages 1123--1127, 2013.

\bibitem{7927254}
Shaghayegh Vahdat, Mehdi Kamal, Ali Afzali-Kusha, Massoud Pedram, and
  Zainalabedin Navabi.
\newblock {TruncApp}: A truncation-based approximate divider for energy
  efficient {DSP} applications.
\newblock In {\em Design, Automation Test in Europe Conference Exhibition
  (DATE), 2017}, pages 1635--1638, 2017.

\bibitem{10.1109/TCAD.2018.2834438}
Cheng Zhuo, Kassan Unda, Yiyu Shi, and Wei-Kai Shih.
\newblock From layout to system: Early stage power delivery and architecture
  co-exploration.
\newblock {\em IEEE Transactions on Computer-Aided Design of Integrated
  Circuits and Systems}, 38(7):1291--1304, 2018.

\bibitem{chandrakasan1995minimizing}
Anantha~P Chandrakasan and Robert~W Brodersen.
\newblock Minimizing power consumption in digital {CMOS} circuits.
\newblock {\em Proceedings of the IEEE}, 83(4):498--523, 1995.

\bibitem{liu2009computation}
Yang Liu, Tong Zhang, and Keshab~K Parhi.
\newblock Computation error analysis in digital signal processing systems with
  overscaled supply voltage.
\newblock {\em IEEE transactions on very large scale integration (VLSI)
  systems}, 18(4):517--526, 2009.

\bibitem{imani2019approxlp}
Mohsen Imani, Alice Sokolova, Ricardo Garcia, Andrew Huang, Fan Wu, Baris
  Aksanli, and Tajana Rosing.
\newblock {ApproxLP}: Approximate multiplication with linearization and
  iterative error control.
\newblock In {\em Proceedings of the 56th Annual Design Automation Conference
  2019}, pages 1--6, 2019.

\bibitem{chen2020optimally}
Chuangtao Chen, Sen Yang, Weikang Qian, Mohsen Imani, Xunzhao Yin, and Cheng
  Zhuo.
\newblock Optimally approximated and unbiased floating-point multiplier with
  runtime configurability.
\newblock In {\em Proceedings of the 39th International Conference on
  Computer-Aided Design}, pages 1--9, 2020.

\bibitem{chen2021TC}
Chuangtao Chen, Weikang Qian, Mohsen Imani, Xunzhao Yin, and Cheng Zhuo.
\newblock {PAM: A piecewise-linearly-approximated floating-point multiplier
  with unbiasedness and configurability}.
\newblock {\em IEEE Transactions on Computers}, 71(10):2473--2486, 2021.

\bibitem{imani2019searchd}
Mohsen Imani, Xunzhao Yin, John Messerly, Saransh Gupta, Michael Niemier,
  Xiaobo~Sharon Hu, and Tajana Rosing.
\newblock Searchd: A memory-centric hyperdimensional computing with stochastic
  training.
\newblock {\em IEEE Transactions on Computer-Aided Design of Integrated
  Circuits and Systems}, 39(10):2422--2433, 2019.

\bibitem{Ni_2019}
Kai Ni, Xunzhao Yin, Ann~Franchesca Laguna, Siddharth Joshi, Stefan D{\"u}nkel,
  Martin Trentzsch, Johannes M{\"u}ller, Sven Beyer, Michael Niemier,
  Xiaobo~Sharon Hu, et~al.
\newblock Ferroelectric ternary content-addressable memory for one-shot
  learning.
\newblock {\em Nature Electronics}, 2(11):521--529, 2019.

\bibitem{vahdat2019tosam}
Shaghayegh Vahdat, Mehdi Kamal, Ali Afzali-Kusha, and Massoud Pedram.
\newblock {TOSAM}: An energy-efficient truncation-and rounding-based scalable
  approximate multiplier.
\newblock {\em IEEE Transactions on Very Large Scale Integration (VLSI)
  Systems}, 27(5):1161--1173, 2019.

\bibitem{zendegani2016roba}
Reza Zendegani, Mehdi Kamal, Milad Bahadori, Ali Afzali-Kusha, and Massoud
  Pedram.
\newblock {RoBA} multiplier: A rounding-based approximate multiplier for
  high-speed yet energy-efficient digital signal processing.
\newblock {\em IEEE Transactions on Very Large Scale Integration (VLSI)
  Systems}, 25(2):393--401, 2016.

\bibitem{sarwar2016multiplier}
Syed~Shakib Sarwar, Swagath Venkataramani, Anand Raghunathan, and Kaushik Roy.
\newblock Multiplier-less artificial neurons exploiting error resiliency for
  energy-efficient neural computing.
\newblock In {\em 2016 Design, Automation \& Test in Europe Conference \&
  Exhibition (DATE)}, pages 145--150. IEEE, 2016.

\bibitem{liu2018design}
Weiqiang Liu, Jiahua Xu, Danye Wang, Chenghua Wang, Paolo Montuschi, and
  Fabrizio Lombardi.
\newblock Design and evaluation of approximate logarithmic multipliers for low
  power error-tolerant applications.
\newblock {\em IEEE Transactions on Circuits and Systems I: Regular Papers},
  65(9):2856--2868, 2018.

\bibitem{ansari2020improved}
Mohammad~S Ansari, Bruce~F Cockburn, and Jie Han.
\newblock An improved logarithmic multiplier for energy-efficient neural
  computing.
\newblock {\em IEEE Transactions on Computers}, 70(4):614--625, 2020.

\bibitem{ansari2019hardware}
Mohammad~S Ansari, Bruce~F Cockburn, and Jie Han.
\newblock A hardware-efficient logarithmic multiplier with improved accuracy.
\newblock In {\em 2019 Design, Automation \& Test in Europe Conference \&
  Exhibition (DATE)}, pages 928--931. IEEE, 2019.

\bibitem{saadat2018minimally}
Hassaan Saadat, Haseeb Bokhari, and Sri Parameswaran.
\newblock Minimally biased multipliers for approximate integer and
  floating-point multiplication.
\newblock {\em IEEE Transactions on Computer-Aided Design of Integrated
  Circuits and Systems}, 37(11):2623--2635, 2018.

\bibitem{hashemi2015drum}
Soheil Hashemi, R~Iris Bahar, and Sherief Reda.
\newblock {DRUM}: A dynamic range unbiased multiplier for approximate
  applications.
\newblock In {\em 2015 IEEE/ACM International Conference on Computer-Aided
  Design (ICCAD)}, pages 418--425. IEEE, 2015.

\bibitem{vahdat2017letam}
Shaghayegh Vahdat, Mehdi Kamal, Ali Afzali-Kusha, and Massoud Pedram.
\newblock {LETAM}: A low energy truncation-based approximate multiplier.
\newblock {\em Computers \& Electrical Engineering}, 63:1--17, 2017.

\bibitem{narayanamoorthy2014energy}
Srinivasan Narayanamoorthy, Hadi~Asghari Moghaddam, Zhenhong Liu, Taejoon Park,
  and Nam~Sung Kim.
\newblock Energy-efficient approximate multiplication for digital signal
  processing and classification applications.
\newblock {\em IEEE transactions on very large scale integration (VLSI)
  systems}, 23(6):1180--1184, 2014.

\bibitem{venkatachalam2017design}
Suganthi Venkatachalam and Seok-Bum Ko.
\newblock Design of power and area efficient approximate multipliers.
\newblock {\em IEEE Transactions on Very Large Scale Integration (VLSI)
  Systems}, 25(5):1782--1786, 2017.

\bibitem{yang2018low}
Tongxin Yang, Tomoaki Ukezono, and Toshinori Sato.
\newblock A low-power high-speed accuracy-controllable approximate multiplier
  design.
\newblock In {\em 2018 23rd Asia and South Pacific Design Automation Conference
  (ASP-DAC)}, pages 605--610. IEEE, 2018.

\bibitem{yang2017low}
Tongxin Yang, Tomoaki Ukezono, and Toshinori Sato.
\newblock Low-power and high-speed approximate multiplier design with a tree
  compressor.
\newblock In {\em 2017 IEEE International Conference on Computer Design
  (ICCD)}, pages 89--96. IEEE, 2017.

\bibitem{esposito2018approximate}
Darjn Esposito, Antonio Giuseppe~Maria Strollo, Ettore Napoli, Davide De~Caro,
  and Nicola Petra.
\newblock Approximate multipliers based on new approximate compressors.
\newblock {\em IEEE Transactions on Circuits and Systems I: Regular Papers},
  65(12):4169--4182, 2018.

\bibitem{Wang20}
Manzhen Wang, Yuanyong Luo, Mengyu An, Yuou Qiu, Muhan Zheng, Zhongfeng Wang,
  and Hongbing Pan.
\newblock {An Optimized Compression Strategy for Compressor-Based Approximate
  Multiplier}.
\newblock In {\em 2020 IEEE International Symposium on Circuits and Systems
  (ISCAS)}, 2020.

\bibitem{ha2017multipliers}
Minho Ha and Sunggu Lee.
\newblock Multipliers with approximate 4--2 compressors and error recovery
  modules.
\newblock {\em IEEE Embedded Systems Letters}, 10(1):6--9, 2017.

\bibitem{tung2019low}
Che-Wei Tung and Shih-Hsu Huang.
\newblock Low-power high-accuracy approximate multiplier using approximate
  high-order compressors.
\newblock In {\em 2019 2nd International Conference on Communication
  Engineering and Technology (ICCET)}, pages 163--167. IEEE, 2019.

\bibitem{yang2015approximate}
Zhixi Yang, Jie Han, and Fabrizio Lombardi.
\newblock Approximate compressors for error-resilient multiplier design.
\newblock In {\em 2015 IEEE International Symposium on Defect and Fault
  Tolerance in VLSI and Nanotechnology Systems (DFTS)}, pages 183--186. IEEE,
  2015.

\bibitem{van2020fpga}
Nguyen Van~Toan and Jeong-Gun Lee.
\newblock {FPGA}-based multi-level approximate multipliers for high-performance
  error-resilient applications.
\newblock {\em IEEE Access}, 8:25481--25497, 2020.

\bibitem{liu2017design}
Weiqiang Liu, Liangyu Qian, Chenghua Wang, Honglan Jiang, Jie Han, and Fabrizio
  Lombardi.
\newblock Design of approximate radix-4 {Booth} multipliers for error-tolerant
  computing.
\newblock {\em IEEE Transactions on Computers}, 66(8):1435--1441, 2017.

\bibitem{jiang2015approximate}
Honglan Jiang, Jie Han, Fei Qiao, and Fabrizio Lombardi.
\newblock Approximate radix-8 {Booth} multipliers for low-power and
  high-performance operation.
\newblock {\em IEEE Transactions on Computers}, 65(8):2638--2644, 2015.

\bibitem{pabithra2018analysis}
S~Pabithra and S~Nageswari.
\newblock Analysis of approximate multiplier using 15--4 compressor for error
  tolerant application.
\newblock In {\em 2018 International Conference on Control, Power,
  Communication and Computing Technologies (ICCPCCT)}, pages 410--415. IEEE,
  2018.

\bibitem{yin2018designs}
Peipei Yin, Chenghua Wang, Weiqiang Liu, Earl~E Swartzlander, and Fabrizio
  Lombardi.
\newblock Designs of approximate floating-point multipliers with variable
  accuracy for error-tolerant applications.
\newblock {\em Journal of Signal Processing Systems}, 90(4):641--654, 2018.

\bibitem{ashtaputre1985using}
Sunil Ashtaputre, Carla~D Savage, and Wesley~E Snyder.
\newblock Using an approximate multiplier in a one-dimensional array
  architecture for real-time convolution.
\newblock Technical report, North Carolina State University. Center for
  Communications and Signal Processing, 1985.

\bibitem{lingamneni2013improving}
Avinash Lingamneni, Arindam Basu, Christian Enz, Krishna~V Palem, and Christian
  Piguet.
\newblock Improving energy gains of inexact {DSP} hardware through
  reciprocative error compensation.
\newblock In {\em 2013 50th ACM/EDAC/IEEE Design Automation Conference (DAC)},
  pages 1--8. IEEE, 2013.

\bibitem{brandalero2018approximate}
Marcelo Brandalero, Antonio Carlos~S Beck, Luigi Carro, and Muhammad Shafique.
\newblock Approximate on-the-fly coarse-grained reconfigurable acceleration for
  general-purpose applications.
\newblock In {\em 2018 55th ACM/ESDA/IEEE Design Automation Conference (DAC)},
  pages 1--6. IEEE, 2018.

\bibitem{akbari2018px}
Omid Akbari, Mehdi Kamal, Ali Afzali-Kusha, Massoud Pedram, and Muhammad
  Shafique.
\newblock {PX-CGRA}: Polymorphic approximate coarse-grained reconfigurable
  architecture.
\newblock In {\em 2018 Design, Automation \& Test in Europe Conference \&
  Exhibition (DATE)}, pages 413--418. IEEE, 2018.

\bibitem{yoo20206}
Byoung-Joo Yoo, Dong-Hyuk Lim, Hyonguk Pang, June-Hee Lee, Seung-Yeob Baek,
  Naxin Kim, Dong-Ho Choi, Young-Ho Choi, Hyeyeon Yang, Taehun Yoon, et~al.
\newblock {6.4 A 56Gb/s 7.7 mW/Gb/s PAM-4 wireline transceiver in 10nm FinFET
  using MM-CDR-Based ADC timing skew control and low-power DSP with approximate
  multiplier}.
\newblock In {\em 2020 IEEE International Solid-State Circuits
  Conference-(ISSCC)}, pages 122--124. IEEE, 2020.

\bibitem{snigdha2016optimal}
Farhana~Sharmin Snigdha, Deepashree Sengupta, Jiang Hu, and Sachin~S
  Sapatnekar.
\newblock Optimal design of {JPEG} hardware under the approximate computing
  paradigm.
\newblock In {\em 2016 53nd ACM/EDAC/IEEE Design Automation Conference (DAC)},
  pages 1--6. IEEE, 2016.

\bibitem{du2014leveraging}
Zidong Du, Krishna Palem, Avinash Lingamneni, Olivier Temam, Yunji Chen, and
  Chengyong Wu.
\newblock Leveraging the error resilience of machine-learning applications for
  designing highly energy efficient accelerators.
\newblock In {\em 2014 19th Asia and South Pacific design automation conference
  (ASP-DAC)}, pages 201--206. IEEE, 2014.

\bibitem{zhang2015approxann}
Qian Zhang, Ting Wang, Ye~Tian, Feng Yuan, and Qiang Xu.
\newblock {ApproxANN}: An approximate computing framework for artificial neural
  network.
\newblock In {\em 2015 Design, Automation \& Test in Europe Conference \&
  Exhibition (DATE)}, pages 701--706. IEEE, 2015.

\bibitem{mrazek2016design}
Vojtech Mrazek, Syed~Shakib Sarwar, Lukas Sekanina, Zdenek Vasicek, and Kaushik
  Roy.
\newblock Design of power-efficient approximate multipliers for approximate
  artificial neural networks.
\newblock In {\em Proceedings of the 35th International Conference on
  Computer-Aided Design}, pages 1--7, 2016.

\bibitem{hammad2018impact}
Issam Hammad and Kamal El-Sankary.
\newblock Impact of approximate multipliers on {VGG} deep learning network.
\newblock {\em IEEE Access}, 6:60438--60444, 2018.

\bibitem{ansari2019improving}
Mohammad~S Ansari, Vojtech Mrazek, Bruce~F Cockburn, Lukas Sekanina, Zdenek
  Vasicek, and Jie Han.
\newblock Improving the accuracy and hardware efficiency of neural networks
  using approximate multipliers.
\newblock {\em IEEE Transactions on Very Large Scale Integration (VLSI)
  Systems}, 28(2):317--328, 2019.

\bibitem{hammad2019deep}
Issam Hammad, Kamal El-Sankary, and Jason Gu.
\newblock Deep learning training with simulated approximate multipliers.
\newblock In {\em 2019 IEEE International Conference on Robotics and
  Biomimetics (ROBIO)}, pages 47--51. IEEE, 2019.

\bibitem{leon2019cooperative}
Vasileios Leon, Konstantinos Asimakopoulos, Sotirios Xydis, Dimitrios Soudris,
  and Kiamal Pekmestzi.
\newblock Cooperative arithmetic-aware approximation techniques for
  energy-efficient multipliers.
\newblock In {\em Proceedings of the 56th Annual Design Automation Conference
  2019}, pages 1--6, 2019.

\bibitem{hammad2021cnn}
Issam Hammad, Ling Li, Kamal El-Sankary, and W~Martin Snelgrove.
\newblock {CNN} inference using a preprocessing precision controller and
  approximate multipliers with various precisions.
\newblock {\em IEEE Access}, 9:7220--7232, 2021.

\bibitem{ullah2018area}
Salim Ullah, Semeen Rehman, Bharath~Srinivas Prabakaran, Florian Kriebel,
  Muhammad~Abdullah Hanif, Muhammad Shafique, and Akash Kumar.
\newblock Area-optimized low-latency approximate multipliers for {FPGA}-based
  hardware accelerators.
\newblock In {\em Proceedings of the 55th annual design automation conference},
  pages 1--6, 2018.

\bibitem{guo2020small}
Yi~Guo, Heming Sun, and Shinji Kimura.
\newblock Small-area and low-power {FPGA}-based multipliers using approximate
  elementary modules.
\newblock In {\em 2020 25th Asia and South Pacific Design Automation Conference
  (ASP-DAC)}, pages 599--604. IEEE, 2020.

\bibitem{ullah2021high}
Salim Ullah, Semeen Rehman, Muhammad Shafique, and Akash Kumar.
\newblock High-performance accurate and approximate multipliers for
  {FPGA}-based hardware accelerators.
\newblock {\em IEEE Transactions on Computer-Aided Design of Integrated
  Circuits and Systems}, 41(2):211--224, 2021.

\bibitem{ullah2018smapproxlib}
Salim Ullah, Sanjeev~Sripadraj Murthy, and Akash Kumar.
\newblock {SMApproxLib: Library of FPGA-based approximate multipliers}.
\newblock In {\em Proceedings of the 55th Annual Design Automation Conference},
  pages 1--6, 2018.

\bibitem{prabakaran2020approxfpgas}
Bharath~Srinivas Prabakaran, Vojtech Mrazek, Zdenek Vasicek, Lukas Sekanina,
  and Muhammad Shafique.
\newblock {ApproxFPGAs: Embracing ASIC-based approximate arithmetic components
  for FPGA-based systems}.
\newblock In {\em 2020 57th ACM/IEEE Design Automation Conference (DAC)}, pages
  1--6. IEEE, 2020.

\bibitem{waris2021axbms}
Haroon Waris, Chenghua Wang, Weiqiang Liu, and Fabrizio Lombardi.
\newblock {AxBMs}: Approximate radix-8 booth multipliers for high-performance
  {FPGA}-based accelerators.
\newblock {\em IEEE Transactions on Circuits and Systems II: Express Briefs},
  68(5):1566--1570, 2021.

\bibitem{ullah2022appaxo}
Salim Ullah, Siva~Satyendra Sahoo, Nemath Ahmed, Debabrata Chaudhury, and Akash
  Kumar.
\newblock {AppAxO}: Designing application-specific approximate operators for
  {FPGA}-based embedded systems.
\newblock {\em ACM Transactions on Embedded Computing Systems (TECS)},
  21(3):1--31, 2022.

\bibitem{sebastian2020memory}
Abu Sebastian, Manuel Le~Gallo, Riduan Khaddam-Aljameh, and Evangelos
  Eleftheriou.
\newblock Memory devices and applications for in-memory computing.
\newblock {\em Nature nanotechnology}, 15(7):529--544, 2020.

\bibitem{am_lib}
Ying Wu and Cheng Zhuo.
\newblock Verilog implementation of approximate multipliers.
\newblock \url{https://github.com/skycrapers/AM-Lib}, 2022.

\bibitem{IEEE_754_2019_8766229}
{IEEE Standard for Floating-Point Arithmetic}.
\newblock {\em IEEE Std 754-2019 (Revision of IEEE 754-2008)}, pages 1--84,
  2019.

\bibitem{wallace1964suggestion}
Christopher~S Wallace.
\newblock A suggestion for a fast multiplier.
\newblock {\em IEEE Transactions on electronic Computers}, EC-13(1):14--17,
  1964.

\bibitem{dadda1965some}
Luigi Dadda.
\newblock Some schemes for parallel multipliers.
\newblock {\em Alta frequenza}, 34:349--356, 1965.

\bibitem{Chang2004ultra}
Chip-Hong Chang, Jiangmin Gu, and Mingyan Zhang.
\newblock Ultra low-voltage low-power {CMOS} 4-2 and 5-2 compressors for fast
  arithmetic circuits.
\newblock {\em IEEE Transactions on Circuits and Systems I: Regular Papers},
  51(10):1985--1997, 2004.

\bibitem{pishvaie2012improved}
Abdoreza Pishvaie, Ghassem Jaberipur, and Ali Jahanian.
\newblock Improved {CMOS} (4; 2) compressor designs for parallel multipliers.
\newblock {\em Computers \& Electrical Engineering}, 38(6):1703--1716, 2012.

\bibitem{baran2010energy}
Dursun Baran, Mustafa Aktan, and Vojin~G Oklobdzija.
\newblock Energy efficient implementation of parallel {CMOS} multipliers with
  improved compressors.
\newblock In {\em Proceedings of the 16th ACM/IEEE international symposium on
  Low power electronics and design}, pages 147--152, 2010.

\bibitem{arasteh2018energy}
Armineh Arasteh, Mohammad~H Moaiyeri, MohammadReza Taheri, Keivan Navi, and
  Nader Bagherzadeh.
\newblock An energy and area efficient 4:2 compressor based on finfets.
\newblock {\em Integration}, 60:224--231, 2018.

\bibitem{veeramachaneni2007novel}
Sreehari Veeramachaneni, Kirthi~M Krishna, Lingamneni Avinash, Sreekanth~Reddy
  Puppala, and MB~Srinivas.
\newblock Novel architectures for high-speed and low-power 3-2, 4-2 and 5-2
  compressors.
\newblock In {\em 20th International Conference on VLSI Design held jointly
  with 6th International Conference on Embedded Systems (VLSID'07)}, pages
  324--329, 2007.

\bibitem{fritz2017fast}
Christopher Fritz and Adly~T Fam.
\newblock Fast binary counters based on symmetric stacking.
\newblock {\em IEEE Transactions on Very Large Scale Integration (VLSI)
  Systems}, 25(10):2971--2975, 2017.

\bibitem{Momeni2014}
Amir Momeni, Jie Han, Paolo Montuschi, and Fabrizio Lombardi.
\newblock {Design and analysis of approximate compressors for multiplication}.
\newblock {\em IEEE Transactions on Computers}, 64(4):984--994, 2014.

\bibitem{ahmed2016iterative}
Syed~E Ahmed, Sanket Kadam, and MB~Srinivas.
\newblock An iterative logarithmic multiplier with improved precision.
\newblock In {\em 2016 IEEE 23nd Symposium on Computer Arithmetic (ARITH)},
  pages 104--111. IEEE, 2016.

\bibitem{Approx_53_7598342}
Vincent Camus, Jeremy Schlachter, Christian Enz, Michael Gautschi, and Frank~K
  Gurkaynak.
\newblock Approximate 32-bit floating-point unit design with 53\% power-area
  product reduction.
\newblock In {\em ESSCIRC Conference 2016: 42nd European Solid-State Circuits
  Conference}, pages 465--468, 2016.

\bibitem{AWTM_6783335}
Kartikeya Bhardwaj, Pravin~S Mane, and J{\"o}rg Henkel.
\newblock {Power- and area-efficient Approximate Wallace Tree Multiplier for
  error-resilient systems}.
\newblock In {\em Fifteenth International Symposium on Quality Electronic
  Design}, pages 263--269, 2014.

\bibitem{liu2014low}
Cong Liu, Jie Han, and Fabrizio Lombardi.
\newblock A low-power, high-performance approximate multiplier with
  configurable partial error recovery.
\newblock In {\em 2014 Design, Automation \& Test in Europe Conference \&
  Exhibition (DATE)}, pages 1--4. IEEE, 2014.

\bibitem{guo2020reconfigurable}
Chuliang Guo, Li~Zhang, Xian Zhou, Weikang Qian, and Cheng Zhuo.
\newblock A reconfigurable approximate multiplier for quantized {CNN}
  applications.
\newblock In {\em 2020 25th Asia and South Pacific Design Automation Conference
  (ASP-DAC)}, pages 235--240. IEEE, 2020.

\bibitem{qian2016design}
Liangyu Qian, Chenghua Wang, Weiqiang Liu, Fabrizio Lombardi, and Jie Han.
\newblock Design and evaluation of an approximate {Wallace-Booth} multiplier.
\newblock In {\em 2016 IEEE international symposium on circuits and systems
  (ISCAS)}, pages 1974--1977. IEEE, 2016.

\bibitem{lingamneni2011energy}
Avinash Lingamneni, Christian Enz, Jean-Luc Nagel, Krishna Palem, and Christian
  Piguet.
\newblock Energy parsimonious circuit design through probabilistic pruning.
\newblock In {\em 2011 Design, Automation \& Test in Europe}, pages 1--6. IEEE,
  2011.

\bibitem{schlachter2015automatic}
Jeremy Schlachter, Vincent Camus, Christian Enz, and Krishna~V Palem.
\newblock Automatic generation of inexact digital circuits by gate-level
  pruning.
\newblock In {\em 2015 IEEE International Symposium on Circuits and Systems
  (ISCAS)}, pages 173--176. IEEE, 2015.

\bibitem{imani2017cfpu}
Mohsen Imani, Daniel Peroni, and Tajana Rosing.
\newblock {CFPU}: Configurable floating point multiplier for energy-efficient
  computing.
\newblock In {\em 2017 54th ACM/EDAC/IEEE Design Automation Conference (DAC)},
  pages 1--6. IEEE, 2017.

\bibitem{imani2018rmac}
Mohsen Imani, Ricardo Garcia, Saransh Gupta, and Tajana Rosing.
\newblock Rmac: Runtime configurable floating point multiplier for approximate
  computing.
\newblock In {\em Proceedings of the International Symposium on Low Power
  Electronics and Design}, pages 1--6, 2018.

\bibitem{jiang2018low}
Honglan Jiang, Cong Liu, Fabrizio Lombardi, and Jie Han.
\newblock Low-power approximate unsigned multipliers with configurable error
  recovery.
\newblock {\em IEEE Transactions on Circuits and Systems I: Regular Papers},
  66(1):189--202, 2018.

\bibitem{song2007adaptive}
Min-An Song, Lan-Da Van, and Sy-Yen Kuo.
\newblock Adaptive low-error fixed-width {Booth} multipliers.
\newblock {\em IEICE Transactions on Fundamentals of Electronics,
  Communications and Computer Sciences}, 90(6):1180--1187, 2007.

\bibitem{wang2009high}
Jiun-Ping Wang, Shiann-Rong Kuang, and Shish-Chang Liang.
\newblock High-accuracy fixed-width modified {Booth} multipliers for lossy
  applications.
\newblock {\em IEEE transactions on very large scale integration (VLSI)
  systems}, 19(1):52--60, 2009.

\bibitem{chen2011high}
Yuan-Ho Chen and Tsin-Yuan Chang.
\newblock A high-accuracy adaptive conditional-probability estimator for
  fixed-width booth multipliers.
\newblock {\em IEEE Transactions on Circuits and Systems I: Regular Papers},
  59(3):594--603, 2011.

\bibitem{farshchi2013new}
Farzad Farshchi, Muhammad~S Abrishami, and Sied~M Fakhraie.
\newblock New approximate multiplier for low power digital signal processing.
\newblock In {\em The 17th CSI International Symposium on Computer Architecture
  \& Digital Systems (CADS 2013)}, pages 25--30. IEEE, 2013.

\bibitem{mahdiani2009bio}
Hamid~Reza Mahdiani, Ali Ahmadi, Sied~Mehdi Fakhraie, and Caro Lucas.
\newblock Bio-inspired imprecise computational blocks for efficient {VLSI}
  implementation of soft-computing applications.
\newblock {\em IEEE Transactions on Circuits and Systems I: Regular Papers},
  57(4):850--862, 2009.

\bibitem{Jiang19}
Honglan Jiang, Cong Liu, Fabrizio Lombardi, and Jie Han.
\newblock {Low-Power Approximate Unsigned Multipliers With Configurable Error
  Recovery}.
\newblock {\em IEEE Transactions on Circuits and Systems I: Regular Papers},
  66:189--202, 2019.

\bibitem{Xiao22}
Weihua Xiao, Cheng Zhuo, and Weikang Qian.
\newblock {OPACT}: Optimization of approximate compressor tree for approximate
  multiplier.
\newblock In {\em 2022 Design, Automation, and Test in Europe Conference
  (DATE)}, 2022.

\bibitem{mrazek2017evoapprox8b}
Vojtech Mrazek, Radek Hrbacek, Zdenek Vasicek, and Lukas Sekanina.
\newblock Evoapprox8b: Library of approximate adders and multipliers for
  circuit design and benchmarking of approximation methods.
\newblock In {\em Design, Automation \& Test in Europe Conference \& Exhibition
  (DATE), 2017}, pages 258--261. IEEE, 2017.

\bibitem{hrbacek2016automatic}
Radek Hrbacek, Vojtech Mrazek, and Zdenek Vasicek.
\newblock Automatic design of approximate circuits by means of multi-objective
  evolutionary algorithms.
\newblock In {\em 2016 International Conference on Design and Technology of
  Integrated Systems in Nanoscale Era (DTIS)}, pages 1--6. IEEE, 2016.

\bibitem{chen2012energy}
Jienan Chen and Jianhao Hu.
\newblock Energy-efficient digital signal processing via
  voltage-overscaling-based residue number system.
\newblock {\em IEEE Transactions on Very Large Scale Integration (VLSI)
  Systems}, 21(7):1322--1332, 2012.

\bibitem{lau2009energy}
Mark~SK Lau, Keck-Voon Ling, and Yun-Chung Chu.
\newblock Energy-aware probabilistic multiplier: design and analysis.
\newblock In {\em Proceedings of the 2009 international conference on
  Compilers, architecture, and synthesis for embedded systems}, pages 281--290,
  2009.

\bibitem{WOS000287716800004}
Zdenka Babi{\'c}, Aleksej Avramovi{\'c}, and Patricio Buli{\'c}.
\newblock An iterative logarithmic multiplier.
\newblock {\em Microprocessors and Microsystems}, 35(1):23--33, 2011.

\bibitem{kim2019cost}
HyunJin Kim, Min~Soo Kim, Alberto~A Del~Barrio, and Nader Bagherzadeh.
\newblock A cost-efficient iterative truncated logarithmic multiplication for
  convolutional neural networks.
\newblock In {\em 2019 IEEE 26th symposium on computer arithmetic (ARITH)},
  pages 108--111. IEEE, 2019.

\bibitem{ahmed2019improved}
Syed~Ershad Ahmed and MB~Srinivas.
\newblock An improved logarithmic multiplier for media processing.
\newblock {\em Journal of Signal Processing Systems}, 91:561--574, 2019.

\bibitem{yin2020design}
Peipei Yin, Chenghua Wang, Haroon Waris, Weiqiang Liu, Yinhe Han, and Fabrizio
  Lombardi.
\newblock Design and analysis of energy-efficient dynamic range approximate
  logarithmic multipliers for machine learning.
\newblock {\em IEEE Transactions on Sustainable Computing}, 6(4):612--625,
  2020.

\bibitem{liu2017design_log}
Weiqiang Liu, Jiahua Xu, Danye Wang, and Fabrizio Lombardi.
\newblock Design of approximate logarithmic multipliers.
\newblock In {\em Proceedings of the on Great Lakes Symposium on VLSI 2017},
  pages 47--52, 2017.

\bibitem{de1996low}
Edwin de~Angel and EE~Swartzlander.
\newblock Low power parallel multipliers.
\newblock In {\em VLSI Signal Processing, Ix}, pages 199--208. IEEE, 1996.

\bibitem{rehman2016architectural}
Semeen Rehman, Walaa El-Harouni, Muhammad Shafique, Akash Kumar, Jorg Henkel,
  and J{\"o}rg Henkel.
\newblock Architectural-space exploration of approximate multipliers.
\newblock In {\em 2016 IEEE/ACM International Conference on Computer-Aided
  Design (ICCAD)}, pages 1--8. IEEE, 2016.

\bibitem{Lin2013}
Chia-Hao Lin and Chao Lin.
\newblock {High accuracy approximate multiplier with error correction}.
\newblock In {\em International Conference on Computer Design}, pages 33--38,
  2013.

\bibitem{Akbari2017}
Omid Akbari, Mehdi Kamal, Ali Afzali-Kusha, and Massoud Pedram.
\newblock {Dual-quality 4:2 compressors for utilizing in dynamic accuracy
  configurable multipliers}.
\newblock {\em IEEE Transactions on Very Large Scale Integration (VLSI)
  Systems}, 25(4):1352--1361, 2017.

\bibitem{Ahmadinejad2019}
Mohammad Ahmadinejad, Mohammad~H Moaiyeri, and Farnaz Sabetzadeh.
\newblock {Energy and area efficient imprecise compressors for approximate
  multiplication at nanoscale}.
\newblock {\em AEU-International Journal of Electronics and Communications},
  110:152859, 2019.

\bibitem{Strollo2020}
Antonio G.~M. Strollo, Ettore Napoli, Davide De~Caro, Nicola Petra, and Gennaro
  Di~Meo.
\newblock {Comparison and extension of approximate 4-2 compressors for
  low-power approximate multipliers}.
\newblock {\em IEEE Transactions on Circuits and Systems I: Regular Papers},
  67(9):3021--3034, 2020.

\bibitem{marimuthu2016design}
R~Marimuthu, Y~Elsie Rezinold, and Partha~Sharathi Mallick.
\newblock Design and analysis of multiplier using approximate 15-4 compressor.
\newblock {\em IEEE Access}, 5:1027--1036, 2016.

\bibitem{Xuan22}
Xuan Wang and Weikang Qian.
\newblock {MinAC}: Minimal-area approximate compressor design based on exact
  synthesis for approximate multipliers.
\newblock In {\em 2022 IEEE International Symposium on Circuits and Systems
  (ISCAS), to be published}, 2022.

\bibitem{Haaswijk20s}
Winston Haaswijk, Mathias Soeken, Alan Mishchenko, and Giovanni De~Micheli.
\newblock {SAT-Based Exact Synthesis: Encodings, Topology Families, and
  Parallelism}.
\newblock {\em IEEE TCAD}, 39(4):871--884, 2020.

\bibitem{lin2004design}
Hsin-Lei Lin, Robert~C Chang, and Ming-Tsai Chan.
\newblock Design of a novel radix-4 booth multiplier.
\newblock In {\em The 2004 IEEE Asia-Pacific Conference on Circuits and
  Systems}, volume~2, pages 837--840. Citeseer, 2004.

\bibitem{alouani2017novel}
Ihsen Alouani, Hamzeh Ahangari, Ozcan Ozturk, and Smail Niar.
\newblock A novel heterogeneous approximate multiplier for low power and high
  performance.
\newblock {\em IEEE embedded Systems letters}, 10(2):45--48, 2017.

\bibitem{sekanina2013approximate}
Lukas Sekanina and Zdenek Vasicek.
\newblock Approximate circuit design by means of evolvable hardware.
\newblock In {\em 2013 IEEE International Conference on Evolvable Systems
  (ICES)}, pages 21--28. IEEE, 2013.

\bibitem{vasicek2014evolutionary}
Zdenek Vasicek and Lukas Sekanina.
\newblock Evolutionary design of approximate multipliers under different error
  metrics.
\newblock In {\em 17th International Symposium on Design and Diagnostics of
  Electronic Circuits \& Systems}, pages 135--140. IEEE, 2014.

\bibitem{miller2008cartesian}
Julian~Francis Miller and Simon~L Harding.
\newblock Cartesian genetic programming.
\newblock In {\em Proceedings of the 10th annual conference companion on
  Genetic and evolutionary computation}, pages 2701--2726, 2008.

\bibitem{mohapatra2011design}
Debabrata Mohapatra, Vinay~K Chippa, Anand Raghunathan, and Kaushik Roy.
\newblock Design of voltage-scalable meta-functions for approximate computing.
\newblock In {\em 2011 Design, Automation \& Test in Europe}, pages 1--6. IEEE,
  2011.

\bibitem{DC}
Synopsys.
\newblock Design compiler.
\newblock \url{https://www.synopsys.com/}, 2022.

\bibitem{UMC40}
UMC.
\newblock Umc40.
\newblock \url{https://www.umc.com}, 2022.

\bibitem{verilator}
Wilson Snyder.
\newblock Verilator.
\newblock \url{https://github.com/verilator/verilator}, 2003-2022.

\bibitem{DW}
Synopsys.
\newblock Designware.
\newblock \url{https://www.synopsys.com/designware-ip.html}, 2022.

\bibitem{gardner1998artificial}
Matt~W Gardner and SR~Dorling.
\newblock Artificial neural networks (the multilayer perceptron)—a review of
  applications in the atmospheric sciences.
\newblock {\em Atmospheric environment}, 32(14-15):2627--2636, 1998.

\bibitem{deng2012mnist}
Li~Deng.
\newblock The mnist database of handwritten digit images for machine learning
  research.
\newblock {\em IEEE Signal Processing Magazine}, 29(6):141--142, 2012.

\bibitem{krizhevsky2017imagenet}
Alex Krizhevsky, Ilya Sutskever, and Geoffrey~E Hinton.
\newblock Imagenet classification with deep convolutional neural networks.
\newblock {\em Communications of the ACM}, 60(6):84--90, 2017.

\bibitem{krizhevsky2009learning}
Alex Krizhevsky, Geoffrey Hinton, et~al.
\newblock Learning multiple layers of features from tiny images.
\newblock 2009.

\bibitem{iandola2016squeezenet}
Forrest~N Iandola, Song Han, Matthew~W Moskewicz, Khalid Ashraf, William~J
  Dally, and Kurt Keutzer.
\newblock Squeezenet: Alexnet-level accuracy with 50x fewer parameters and< 0.5
  mb model size.
\newblock {\em arXiv preprint arXiv:1602.07360}, 2016.

\bibitem{deng2009imagenet}
Jia Deng, Wei Dong, Richard Socher, Li-Jia Li, Kai Li, and Li~Fei-Fei.
\newblock Imagenet: A large-scale hierarchical image database.
\newblock In {\em 2009 IEEE conference on computer vision and pattern
  recognition}, pages 248--255. Ieee, 2009.

\bibitem{jacob2018quantization}
Benoit Jacob, Skirmantas Kligys, Bo~Chen, Menglong Zhu, Matthew Tang, Andrew
  Howard, Hartwig Adam, and Dmitry Kalenichenko.
\newblock Quantization and training of neural networks for efficient
  integer-arithmetic-only inference.
\newblock In {\em Proceedings of the IEEE conference on computer vision and
  pattern recognition}, pages 2704--2713, 2018.

\bibitem{zhou2020convolutional}
Xian Zhou, Li~Zhang, Chuliang Guo, Xunzhao Yin, and Cheng Zhuo.
\newblock A convolutional neural network accelerator architecture with
  fine-granular mixed precision configurability.
\newblock In {\em 2020 IEEE International Symposium on Circuits and Systems
  (ISCAS)}, pages 1--5. IEEE, 2020.

\end{thebibliography}


\end{document}